\documentclass[twocolumn,epjc3]{svjour3}  

\journalname{Eur. Phys. J. A}

\usepackage{lipsum}
\usepackage{xcolor}
\usepackage{hyperref}
\usepackage{amsmath,amssymb} 
\usepackage{widetext}
\usepackage{hhline}
\usepackage{bbm}
\usepackage{diagbox}
\usepackage{rotating}
\usepackage{float}
\restylefloat{table}

\usepackage{cite}

\usepackage{graphicx}

\usepackage{subfigure}

\usepackage{tikz}
\usepackage{lipsum}

\hypersetup{colorlinks=true,citecolor= blue}
\newcommand{\etal}{\textit{et al}. }

\begin{document}
\title{S-shell $\Lambda\Lambda$ hypernuclei based on chiral interactions}
\author{Hoai Le\thanksref{addr1,e1}
\and Johann Haidenbauer\thanksref{addr1,e2}
\and Ulf-G. Mei{\ss}ner\thanksref{addr2,addr1,addr3,e3}
\and Andreas Nogga\thanksref{addr1,e4}
}
\thankstext{e1}{e-mail: h.le@fz-juelich.de}
\thankstext{e2}{e-mail: j.haidenbauer@fz-juelich.de}
\thankstext{e3}{e-mail: meissner@hiskp.uni-bonn.de}
\thankstext{e4}{e-mail: a.nogga@fz-juelich.de}

\institute{IAS-4, IKP-3 and JHCP, Forschungszentrum J\"ulich, D-52428 J\"ulich, Germany \label{addr1}
           \and
           HISKP and BCTP, Universit\"at Bonn, D-53115 Bonn, Germany \label{addr2}
           \and 
           Tbilisi State University, 0186 Tbilisi, Georgia \label{addr3}
}

\date{March 14, 2021}

\maketitle

\begin{abstract}
We generalize the Jacobi no-core shell model (J-NCSM) to study double-strangeness hypernuclei. All particle conversions in  the strangeness $S=-1,-2$  sectors are explicitly taken into account. In two-body space, such transitions may lead to the coupling between states of identical particles and of non-identical
ones. Therefore, a careful consideration is required when 
determining the  combinatorial factors that connect the many-body potential matrix elements and the free-space two-body potentials. Using second quantization, we systematically derive the combinatorial factors in question for $S=0,-1,-2$ sectors.  As a first application, we use the J-NCSM to investigate $\Lambda \Lambda$
s-shell hypernuclei based on hyperon-hyperon (YY) potentials derived within chiral
effective field theory at leading order (LO) and up to 
next-to-leading order (NLO). 
We find that the LO potential overbinds $^{\text{ }\text{ }\text{ } \text{}6}_{\Lambda \Lambda}\text{He}$ while the prediction of the NLO
interaction is close to experiment. Both interactions also yield a bound state for $^{\text{ }\text{ }\text{ } \text{}5}_{\Lambda \Lambda}\text{He}$. The $^{\text{}\text{ }\text{ }\text{}4}_{\Lambda \Lambda}\text{H}$ system is predicted to be unbound.

\end{abstract}


\section{Introduction}
The scarcity of hyperon-nucleon (YN; Y=$\Lambda$, $\Sigma$) data 
and the almost complete lack of direct empirical information on 
the hyperon-hyperon (YY) and $\Xi N$
systems poses an enormous challenge for theorists in the 
  attempt to derive baryon-baryon (BB) interactions in the strangeness sector on a microscopic level. By exploiting SU(3) flavor
  symmetry, several sophisticated YN potentials have been derived \cite{Rijken:1998yy,PhysRevC.72.044005,Haidenbauer:2013oca,Haidenbauer:2019boi,PhysRevC.99.044003}, which all describe the available YN data on an adequate quantitative level. The situation however remains largely  unsatisfactory for the strangeness
  $S=-2$ sector, at least for the foreseeable future, because it is  practically impossible to perform direct
  YY scattering experiments
  and so far there have been no two-body $S=-2$ bound states observed. 
  Data on 
  $\Lambda\Lambda$- and $\Xi$ hypernuclei are therefore an 
  indispensable source of information that can provide valuable additional constraints for constructing YY interactions. 
  The latter requires solving the exact $S=-2$ many-body Hamiltonian with microscopic two- and higher-body BB interactions as input.
  
  In the present work, we utilize the Jacobi no-core shell model 
  (J-NCSM) \cite{Navratil:1999pw,Liebig:2015kwa}
  to study double-strangeness hypernuclei. Historically, 
  since the first observations of $\Lambda\Lambda$  hypernuclei,  $^{\text{ }\text{ } \text{}10}_{\Lambda \Lambda}\text{Be}$  \cite{Danysz:1963zza}, $^{\text{ }\text{ }\text{ } \text{}6}_{\Lambda \Lambda}\text{He}$ \cite{PhysRevLett.17.782} and especially 
  after publication of the so-called Nagara event 
  \cite{PhysRevLett.87.212502,Nakazawa:2010zza}, various approaches 
  have been employed to study 
  doubly-strange hypernuclei 
  \cite{Nemura:1999qp,Nemura:2004xb,10.1143/PTP.97.881,PhysRevC.66.024007,Filikhin:2002wm,PhysRevLett.89.172502,PhysRevC.68.024002,Fujiwara:2004na,Vidana:2003ic,Lanskoy:2003ia,Usmani:2004vs, Hiyama:2010eca, Richard:2014pwa, CONTESSI2019134893,Hiyama:2018ivm}. 
  For example, Nemura \etal used the so-called  stochastic variational method in combination with phenomenological effective central 
  $\Lambda N$ and
  $\Lambda \Lambda$ potentials to investigate $\Lambda \Lambda$ s-shell hypernuclei \cite{Nemura:1999qp}. Thereby, 
  it was assumed that the effects of tensor forces, three-body forces and $\Lambda-\Sigma$ conversion are effectively included in such central potentials by fitting the binding energies of the s-shell (core)  $\Lambda$-hypernuclei. 
  In their later work \cite{Nemura:2004xb}, the channel coupling, e.g., $\Lambda N - \Sigma N$ or $\Lambda \Lambda - \Xi N$ was fully taken into account, but again the YY interaction consisted of only central potentials. 
  Hiyama and co-workers  have successfully  applied the Jacobian-coordinate Gaussian expansion method to $\Lambda\Lambda$ hypernuclei with $A=6-11$, which treats hypernuclear systems as three-, four-, or five-cluster structures \cite{10.1143/PTP.97.881,PhysRevC.66.024007,Hiyama:2010eca}. 
  The authors have advanced the approach in order to allow for all possible  rearrangement channels so that any  changes in the  dynamic structure due to $\Lambda$ interactions can be taken into account. The interactions between a $\Lambda$ and a cluster are approximated using 
   the simulated G-matrix YN potentials that are  derived from a set of one-boson-exchange potentials. Here, 
  the $\Lambda N -\Sigma N$ and $\Lambda \Lambda - \Xi N$ couplings were not treated  explicitly but the tuning parameters of the simulated  potentials are chosen to reproduce   some experimental separation energies such as those of 
  $^{\,5}_{\Lambda}\text{He}$ and  
  $^{\text{ }\text{ }\text{ } \text{}6}_{\Lambda \Lambda}\text{He}$.  It is therefore rather difficult to relate the properties of the employed potentials  to the free-space BB interactions. Filikhin and Gal have solved the Faddeev-Yakubovsky equations formulated
  for three- ($\Lambda \Lambda \alpha $), four-cluster 
  ($\Lambda\Lambda \alpha \alpha$)  and $\Lambda \Lambda n p$ components \cite{Filikhin:2002wm,PhysRevLett.89.172502,PhysRevC.68.024002}. 
  Their calculations are also based on simulated potentials similar to those used in the works of Hiyama
  \etal but the $\Lambda \Lambda$  interactions were mainly restricted to the $s$-wave. Lately, Contessi and co-workers  \cite{CONTESSI2019134893} have combined the stochastic variational method with pionless effective field theory (EFT) interactions at LO to investigate the consistency of  $A=4-6$ $\Lambda\Lambda$ hypernuclei. 
  
 Recently, the Jacobi NCSM has been successfully employed by us in
 studies of single-$\Lambda$ hypernuclei up to $A=7$ \cite{Le:2019gjp,Le:2020zdu}.  
 In these investigations, the full complexity of the
 underlying nucleon-nucleon (NN) and YN interactions 
 (tensor forces, channel coupling) could 
 be incorporated. Now we extend the Jacobi NCSM to $S=-2$ systems. Also here all channel couplings, i.e.,  $\Lambda N -\Sigma N$, and  $ \Lambda\Lambda - \Lambda\Sigma - \Sigma\Sigma - \Xi N$ are explicitly considered.  As first application, we use the approach to obtain predictions of the chiral leading order (LO) \cite{Polinder:2007mp} and next-to-leading order (NLO) \cite{Haidenbauer:2015zqb,Haidenbauer:2018gvg} YY interactions for 
 $\Lambda \Lambda$ s-shell hypernuclei. 
  Chiral EFT \cite{Epelbaum:2008ga} is a very successful tool for describing the NN interaction (see \cite{Reinert:2017usi} and references therein) and allows for accurate calculations of nuclear observables \cite{Epelbaum:2018ogq,Piarulli:2017dwd,Epelbaum:2019zqc,Maris:2020qne}. The YN 
  interaction derived within the chiral EFT approach up to NLO 
  likewise leads to realistic results for ($s$- and p-shell) hypernuclei \cite{Gazda:2013jia,Wirth:2014ko,Wirth:2018kh,LE2020135189,Le:2020zdu} 
  and for nuclear matter \cite{Haidenbauer:2018gvg,HAIDENBAUER201529}. 
  It is therefore of great interest to study the predictions of the chiral YY potentials for $\Lambda \Lambda$ hypernuclei.
  
  The paper is structured in the following way.
  We present in Section~\ref{sec:Numerical} details of the technical realization. Some relevant formulas are also provided in the appendix. First  results for 
   the $^{\text{ }\text{ }\text{ } \text{}6}_{\Lambda \Lambda}\text{He}$,  $^{\text{ }\text{ }\text{ } \text{}5}_{\Lambda \Lambda}\text{He}$ and  $^{\text{ }\text{ }\text{ } \text{}4}_{\Lambda \Lambda}\text{H}$ hypernuclei are discussed in Section~\ref{sec:result}. Finally, conclusions and an outlook are given in Section~\ref{sec:concl}.

\section{Numerical realization}\label{sec:Numerical}
\subsection{Jacobi basis for $ S=-2$ systems}\label{sec:Jbasis}
In this section, we generalize our Jacobi no-core shell model (J-NCSM) formalism \cite{Le:2020zdu} to $S=-2$ hypernuclei. 
Adding a second $\Lambda$ hyperon to  single-strangeness  systems   complicates the numerical realization in many  ways. All  particle conversions
 that involve  a $\Xi$ hyperon,   for instance $\Lambda \Lambda \leftrightarrow  \Xi N, \,  \Sigma \Sigma \leftrightarrow  \Xi N \, \text{or} \,\,  \Lambda \Sigma \leftrightarrow  \Xi N  $   change the total number of nucleons in   the system by one. The latter must be explicitly   taken into account  for the many-body Hamiltonian and for the  basis  states.  Furthermore, 
 particle conversions in both $S=-1$ and $S=-2$ sectors can also  lead to  couplings between states of identical and non-identical hyperons. Because of that,  special
 attention is required  when evaluating the Hamiltonian matrix elements.  These issues  will thoroughly be  addressed   in  this section and in 
 \ref{appendix:appendixA}.  We start with  the construction of   many-body basis states first.
Since the total number of  nucleons in the system can change depending on the strange particles,  we  split the basis functions into two orthogonal sets: 
 one set that involves two singly strange hyperons  referred
to as $|\alpha^{*{(Y_1 Y_2)}}\rangle$, and the other that contains a doubly strange $\Xi$ hyperon denoted as  $|\alpha^{*{(\Xi)}}\rangle$. 
 The former    is  constructed  by    coupling the  antisymmetrized  states 
 of  $A-2$ nucleons, 
 $|\alpha_{(A-2)N}\rangle$, 
 to the  states describing a system of two hyperons, $| Y_1 Y_2 \rangle$
 \begin{eqnarray}\label{eq:basisS=Y1Y2}
& &   |\alpha^{*(Y_1 Y_2)}\rangle     =  |\alpha_{(A-2)N} \rangle  \otimes | Y_1  Y_2\rangle\nonumber\\[2pt]
&       & = | \mathcal{N}{J}{T}, \alpha_{(A-2)N}\, \alpha_{Y_1 Y_2}\, n_{\lambda} \lambda;\nonumber \\[2pt]
 & & \qquad   (((l_{Y_1 Y_2} (s_{Y_1} s_{Y_2}) S_{Y_1 Y_2}) J_{Y_1 Y_2} 
(\lambda J_{A-2})I_{\lambda}) J,\nonumber\\[2pt]
&  & \qquad  ((t_{Y_1} t_{Y_2})T_{Y_1 Y_2} T_{A-2}) T \rangle  \equiv \big| \,\begin{tikzpicture}[baseline={([yshift=-0.5ex]current bounding box.center)},scale=0.65]
                \filldraw[color=black, ultra thick, fill=gray ]  (0.,0.) circle(0.20cm);
                \filldraw[red] (-0.7,0.24)   circle(0.7mm) node[anchor=base, above=0.07em, text=black] {\scriptsize{$Y_1$}} ;
                \filldraw[red] (-0.7,-0.24)  circle(0.7mm) node[anchor=base, below=0.07em, text=black] {\scriptsize{$Y_2$}} ;
                \draw[baseline,thick, -]  (-0.7,-0.19) -- (-0.7,0.17);
                \draw[baseline,thick, -]  (-0.7,0.) -- (-0.2,0.0);
                \end{tikzpicture} \big\rangle,
\end{eqnarray}
 with $Y_1, Y_2 =\Lambda, \Sigma$  and $Y_1 \leq Y_2$.  Here the  inequality $Y_{1} \le Y_2 $  indicates the  fact that we  distinguish among  the three two-hyperon states   $|\Lambda \Lambda\rangle$, 
 $|\Lambda \Sigma \rangle$  and $|\Sigma \Sigma\rangle$  but do not   consider   the $|\Sigma \Lambda\rangle$ state explicitly. The notations in Eq.~(\ref{eq:basisS=Y1Y2}) are the same  as introduced in Ref.~\cite{Liebig:2015kwa,Le:2020zdu}.  For example,  the symbol $\alpha_{(A-2)N}$
  stands for all  quantum numbers characterizing the  antisymmetrized  states of $A-2$ nucleons:
the total number of  oscillator  quanta $\mathcal{N}_{A-2}$, total angular momentum $J_{A-2}$, isospin $T_{A-2}$ 
and state index $\zeta_{A-2} $ as well. Similarly,  $\alpha_{Y_1 Y_2}$ stands for a complete set of  quantum numbers describing the subcluster of two hyperons $Y_1$ and $Y_2$.  The principal quantum  number $n_{\lambda}$ of the
harmonic oscillator (HO) together with the orbital 
 angular $\lambda$  describe the  relative motion of the $(A-2)$N core with respect
to the center-of-mass (C.M.)~of the $Y_1 Y_2$ subcluster.   The orders, in which these quantum numbers are coupled, are  shown after the semicolon. As for the transition coefficients of for standard nuclei and single $\Lambda$ hypernuclei \cite{Liebig:2015kwa,Le:2020zdu}, the corresponding momenta or position vectors point to $Y_1$ and the $A-2$ cluster, respectively.  
   
  Analogously, in order to construct the basis  $|\alpha^{*(\Xi)}\rangle$, one  combines
 the  antisymmetrized  states of an $(A-1)$N system, $|\alpha_{(A-1)N}\rangle$,  with  the HO states,  $|\Xi \rangle$, describing the relative motion of   a $\Xi$ hyperon  with respect to  the C.M.~of the (A-1)N subcluster
  \begin{eqnarray}\label{eq:basisS=Xi}
&        &     |\alpha^{* (\Xi)}\rangle=  |\alpha_{(A-1)N}\rangle \otimes  |\Xi\rangle\nonumber\\[2pt]
&   & = | \mathcal{N}{J}{T}, \alpha_{(A-1)N}\, n_{\Xi}\, I_{\Xi} \,t_{\Xi} ; \nonumber\\
& & \quad  (J_{A-1} (l_{\Xi}\, s_{\Xi})\,I_{\Xi}){J}, 
(T_{A-1}\, t_{\Xi}) {T} \rangle
  \equiv \big| \begin{tikzpicture}[baseline={([yshift=-.5ex]current bounding box.center)},scale=0.65]
                \filldraw[color=black, ultra thick, fill=gray ]  (0.,0.) circle(0.22cm);
                \filldraw[red]  (0.6,0.) circle (0.7mm)  node[anchor=base, above=0.02em, text=black] {\scriptsize{$\Xi$}} ;
                \filldraw[red]  (0.6,0.) circle (0.7mm)  node[anchor=base, below=0.02em, text=white] {\scriptsize{$\Xi$}} ;
                \draw[baseline,thick, -]  (0.22,0.) -- (0.53,0.);
                \end{tikzpicture}  \big\rangle.
\end{eqnarray}
Here,  also  $\alpha_{(A-1)N}$  denotes a set of quantum numbers representing  
an antisymmetrized state of $A-1$ nucleons.     The relative motion of  a $\Xi$  hyperon is labelled by the HO principal quantum number $n_{\Xi}$, the orbital angular momentum $l_{\Xi}$ and
 spin $s_{\Xi}=\frac{1}{2}$ which combine together  to form the total angular momentum $I_{\Xi}$, and  the isospin $t_\Xi=\frac{1}{2}$. Again, following the definition of our coefficients of fractional parantage (CFPs) \cite{Liebig:2015kwa}, the momentum or position vector points towards the spectator particel, i.e. the $\Xi$. Finally,  the last lines in Eqs.~(\ref{eq:basisS=Y1Y2},\ref{eq:basisS=Xi})  also show the graphical  representations of the states. 
 
With the basis states defined in Eq.~(\ref{eq:basisS=Y1Y2},\ref{eq:basisS=Xi}),
the  $S=-2$ hypernuclear wave function $|\Psi(\pi J T)\rangle$ can be expanded as 
 \begin{eqnarray}\label{eq:expandS=-2wave}
&  &  \!\! \big |\Psi(\pi J T) \big \rangle = \sum_{\alpha^{*{(Y_1 Y_2)}}}  C_{\alpha^{*(Y_1 Y_2)}}  \, \big |\alpha^{*(Y_1 Y_2)}(\mathcal{N} J T)\big\rangle \nonumber\\[2pt]
&  & \qquad \qquad  \quad + \sum_{\alpha^{*(\Xi)}}  C_{\alpha^{*(\Xi)}}  \,\big | \alpha^{*(\Xi)}(\mathcal{N}J T)\big\rangle
\end{eqnarray}
The expansion coefficients are obtained when diagonalizing the   $S=-2$ Hamiltonian   in the basis   Eq.~(\ref{eq:basisS=Y1Y2},\ref{eq:basisS=Xi}).   For practical   calculations, the model space is truncated 
by limiting the total HO energy quantum number 
$\mathcal{N} = \mathcal{N}_{A-2} + 2n_{\lambda} + l_{\lambda} + \mathcal{N}_{Y_1 Y_2 } = \mathcal{N}_{A-1}  + l_{\Xi} + 2n_{\Xi}  \leq \mathcal{N}_{max}$. 
 Of course,  by doing so,  the  computed binding energies will be $\mathcal{N}_{max}$- and  HO $\omega$-dependent. 
 For extracting the converged results,   we  follow    the  two-step  extrapolation procedure  that has been extensively  employed for 
  nuclear and  single-$\Lambda$ hypernuclear calculations within the J-NCSM approach \cite{Liebig:2015kwa,Le:2020zdu}.
  For energies, we first define $E_{\cal N}$ for a given 
  $\mathcal{N}_max = \mathcal{N}$ by minimizing 
  the energy with respect to $\omega$. Then we perform an exponential fit to $E_\mathcal{N}$ to extrapolate to $\mathcal{N} \to \infty$.

  \subsection{$S=-2$  many-body Hamiltoninan}
  
 For the solution of the $A$-body Schrödinger equation, 
 \begin{equation}
    H \ | \Psi \rangle = E \ | \Psi \rangle
 \end{equation}
 we use a standard, iterative Lanczos solver. In order to 
 introduce the pertinent ingredients, we will in the following
present the evaluation of an expectation value of $H$. The extension 
to the evalution of $H \ | \Psi \rangle$ is then straightforward. 
 Using the 
   wave function in Eq.~(\ref{eq:expandS=-2wave}),  
 one can write down  the final expression for the  energy expectation value     as follows
 \begin{eqnarray}\label{eq:breakupHamiltonian}
 &  & \langle  \Psi(\pi J T) \,| \,H \, |\,   \Psi(\pi J T)  \rangle  =\nonumber\\[2pt]
 & &  \sum_{\substack{\alpha^{*(Y_1 Y_2)} \\[3pt]  \alpha^{\prime*(Y_1 Y_2)}}}  C_{\alpha^{*(Y_1 Y_2)}}
  C_{\alpha^{\prime*(Y_1 Y_2)}} \,    \langle \alpha^{*(Y_1 Y_2)} |\,  H \,|\,
 \alpha^{\prime*(Y_1 Y_2)} \rangle\nonumber\\[3pt]
 &    & +\sum_{\alpha^{*(\Xi)}, \,\alpha^{\prime*(\Xi)}}  C_{\alpha^{*(\Xi)}}
  C_{\alpha^{\prime*(\Xi)}}    \langle \alpha^{*(\Xi)} \, | H \,|\,
 \alpha^{\prime*(\Xi)} \rangle\nonumber\\[3pt]
& & +2 \sum_{\substack{\alpha^{*(Y_1 Y_2)} \\[3pt]  \alpha^{\prime*(\Xi)}}}  C_{\alpha^{*(Y_1 Y_2)}}
  C_{\alpha^{\prime*(\Xi)}} \,    \langle \alpha^{*(Y_1 Y_2)} |\,  H  \,|\,
 \alpha^{\prime*(\Xi)} \rangle\,.
\end{eqnarray}
\newline
\noindent{The} last line in Eq.~(\ref{eq:breakupHamiltonian}) is obtained
by exploiting the hermiticity of the Hamiltonian. 
It should be clear from Eq.~(\ref{eq:breakupHamiltonian}) that
 the part of the Hamiltonian that only involves the doubly-strange hyperon $\Xi$ does not contribute to the matrix element $ \langle \alpha^{*(Y_1 Y_2)} |\,  H \,|\,
 \alpha^{\prime*(Y_1 Y_2)}\rangle$  (in the first line).  Likewise,  $\langle \alpha^{*(\Xi)} \, | H \,|\,
 \alpha^{\prime*(\Xi)} \rangle$ will not receive  any contributions from the part of the Hamiltonian that contains two singly-strange hyperons $Y_1 $ and $Y_2$,  whereas   the last term is  nonzero  only for the  transition potentials in the $S=-2$ channels. Therefore,  
 in order to  write down the  explicit form of the   $S=-2$  $A$-body  Hamiltonian, we   distinguish  three parts of the Hamiltonian, namely $H_{Y_1 Y_2}, H_{\Xi}$ and $H^{S=-2}_{Y_1 Y_2 -\Xi N}$,  which   contributes   to the matrix elements in the first, second and third lines in Eq.~(\ref{eq:breakupHamiltonian}), respectively. 
 The first part of the Hamiltonian $H_{Y_1 Y_2}$  corresponds to  a system consisting  of   $A-2$ nucleons and two singly-strange  hyperons   $Y_1 $ and $Y_2$, and has  the following form, 
 \begin{widetext}
\begin{eqnarray}  \label{eq:hamiltonian2Y}
&  & \,H_{Y_1 Y_2}   = H^{S=0}_{Y_1 Y_2}  \,+ H^{S=-1}_{Y_1 Y_2} \,+ H^{S=-2}_{Y_1 Y_2} \nonumber\\
&  &  \qquad\quad= \sum_{i < j=1}^{A-2} \Big( \frac{2p^{2}_{ij}}{M(t_{Y_1}, t_{Y_2})} \,+ V^{S=0}_{ij} \Big)\nonumber\\[2pt]
 &   &  \qquad \qquad +  \sum_{i=1}^{A-2}
\Big( \frac{m_N + m(t_{Y_1})}{M(t_{Y_1}, t_{Y_2})} \,\frac{p^2_{iY_1}}{2\mu_{iY_1}} \,+ V^{S=-1}_{iY_1}
+   \frac{m_N + m(t_{Y_2})}{M(t_{Y_1}, t_{Y_2})} \,\frac{p^2_{iY_2}}{2\mu_{iY_2}} \,+ V^{S=-1}_{iY_2} \Big)\nonumber\\[3pt]
& &  \qquad \qquad + \frac{m(t_{Y_1}) + m(t_{Y_2})}{M(t_{Y_1}, {t_{Y2}})}\, \frac{p^{2}_{Y_1 Y_2}}{2\mu_{Y_1 Y_2}}
 \, +V^{S=-2}_{Y_1 Y_2} 
  + \big(m(t_{Y_1}) + m({t_{Y_2}}) -  2m_{\Lambda}\big) +\cdots,
\end{eqnarray}
\end{widetext}
with  $Y_1, Y_2 = \Lambda, \Sigma$  and \,  $Y_1 \leq Y_2$.   
Here,  $m({t_{Y_1}}),  m(t_{Y_2})$  and $m_N$  are the  $Y_1$, $Y_2$  hyperon   and nucleon rest masses, respectively.  $M(t_{Y_1},t_{Y_2})$ denotes the total rest mass of the system 
 $M(t_{Y_1},t_{Y_2}) = m(t_{Y_1}) + m(t_{Y_2}) + (A-2)m_N$, while  $\mu_{i Y_{1}}$ and $\mu_{Y_{1} Y_{2}}$ are the YN and YY reduced masses, respectively.   The  rest  mass differences  within the nucleon-  and  hyperon-isospin multiplets are neglected. $V_{ij}^{S=0}$,\break $V_{iY}^{S=-1}$, and 
 $V_{YY}^{S=-2}$ are the nucleon-nucleon (NN), YN and YY potentials. Finally, the last term in Eq.~(\ref{eq:hamiltonian2Y}) accounts for
the difference in the  rest masses of   the hyperons  arising due to  particle conversions.

Likewise, the second Hamiltonian, $H_{\Xi}$  (involving a $\Xi$ hyperon)  corresponds to a system composed of a $\Xi$ hyperon and $A-1$ nucleons. Hence,  
\begin{eqnarray} \label{eq:hamiltonianXi}
 & & \,\,H_{\Xi}   = H^{S=0}_{\Xi}  \,   +  \,    H^{S=-2}_{\Xi} \nonumber\\
& & \qquad = \sum_{i < j=1}^{A-1} \Big( \frac{2p^{2}_{ij}}{M({\Xi})} \,+ V^{S=0}_{ij} \Big)\nonumber\\
 &  &\qquad  +  \sum_{i=1}^{A-1}
\Big( \frac{m_N + m_{\Xi}}{M({\Xi})} \,\frac{p^2_{\Xi i}}{2\mu_{\Xi i}} \,+ V^{S=-2}_{\Xi i }\Big)\nonumber\\
  &  & \qquad\qquad\quad + \big(m_{\Xi} + m_N -  2m_{\Lambda}\big)+ \cdots,
  \end{eqnarray}
where $m_{\Xi}$  is the  $\Xi$ hyperon rest mass and  $\mu_{i \Xi}$ is  the reduced mass of a $\Xi$ and a nucleon. The total mass of the system is now given by 
$M(\Xi) = m_{\Xi} + (A-1)m_N$. 
$V^{S=-2}_{\Xi i }$ is the $\Xi N$ potential. 
The ellipses in 	Eqs.~(\ref{eq:hamiltonian2Y},\ref{eq:hamiltonianXi}) stand for those  higher-body forces that are omitted here. The transition Hamiltonian $H^{S=-2}_{Y_1 Y_2 ,\Xi N}$ is simply  given by the YY-$\Xi$N transition potential 
  \begin{align} \label{eq:hamiltonian_transition}
\begin{split}
   H^{S=-2}_{Y_1 Y_2 ,\Xi N} = \sum_{i=1}^{A-1}   V^{S=-2}_{Y_{1} Y_{2} , \Xi i}\,.
\end{split}
\end{align}
 \subsection{Evaluation of the $S=-2$ Hamiltonian matrix elements}\label{sec:S=-2matrix}
 Now,  taking into account the explicit forms of the  $A$-body Hamiltonian in  Eqs.~(\ref{eq:hamiltonian2Y}-\ref{eq:hamiltonian_transition}),  all possible  contributions 
 to the matrix element $\langle  \Psi(\pi J T) \,| \,H \, |\,   \Psi(\pi J T)  \rangle$ can then be  split  into three groups involving
 the  non-strange $H^{S=0}$, single-strange  $H^{S=-1}$ and double-strange  $H^{S=-2}$ parts of the total Hamiltonian, 
 \begin{eqnarray}\label{eq:matrixE_S=-2}
& & \langle  \Psi(\pi J T) \,|\, H  \,| \, \Psi(\pi J T)  \rangle  =   \langle  \Psi(\pi J T) \,|\, H^{S=0}\,| \,  \Psi(\pi J T)  \rangle\nonumber\\[4pt]
&  &  \qquad\qquad+  \langle  \Psi(\pi J T) \, |\, H^{S=-1} \, |  \, \Psi(\pi J T)  \rangle\nonumber\\[4pt]
&  &  \qquad\qquad +  \langle  \Psi(\pi J T) \,| \,H^{S=-2} \, |\,   \Psi(\pi J T)  \rangle.
\end{eqnarray}
The evaluation of the  non-strange  part,
\begin{eqnarray}\label{eq:non-strangepart}
 &  & \langle  \Psi(\pi J T) \,|\, H^{S=0}\,| \,  \Psi(\pi J T)  \rangle =\nonumber\\[4pt] 
 &  &   \sum_{\substack{\alpha^{*(Y_1 Y_2)} \\[3pt]  \alpha^{\prime*(Y_1 Y_2)}}}  C_{\alpha^{*(Y_1 Y_2)}}  
  C_{\alpha^{\prime*(Y_1 Y_2)}} \,    \langle \alpha^{*(Y_1 Y_2)} |\,  H^{S=0}_{Y_1Y_2} |\,
 \alpha^{\prime*(Y_1 Y_2)} \rangle\nonumber\\[2pt]
  &  & +\sum_{\alpha^{*(\Xi)}, \alpha^{\prime*(\Xi)}}  C_{\alpha^{*(\Xi)}}
  C_{\alpha^{\prime*(\Xi)}}    \langle \alpha^{*(\Xi)} \, | H^{S=0}_{\Xi} |
 \alpha^{\prime*(\Xi)} \rangle\,, 
\end{eqnarray}
does not require any new transition coefficients,  and  can  be performed analogously as  done for the   $S=-1$ systems \cite{Le:2020zdu}.  
Furthermore, the combinatorial factors  that relate the $A$-body matrix elements \break $ \langle \alpha^{*(Y_1 Y_2)} |\,  H^{S=0}_{Y_1Y_2} |\,
 \alpha^{\prime*(Y_1 Y_2)} \rangle$  and    $  \langle \alpha^{*(\Xi)} \, | H^{S=0}_{\Xi} |\,
 \alpha^{\prime*(\Xi)} \rangle$  to  the two-nucleon matrix elements in the two-body sector
  are  given 
 by the binomial coefficients of $\binom{A_{\text{nucl}}}{2} = 
A_{\text{nucl}}(A_{\text{nucl}}-1)/2 $  with   $A_{\text{nucl}} =A-2$ and $A_{\text{nucl}} = A-1$, respectively, being the number of nucleons in the system (see~\ref{appendix:appendixA}
for the definition of the combinatorial factors).

The matrix elements of the double-strange part \\ $H^{S=-2}$ of the Hamiltonian, 
\begin{eqnarray}\label{eq:S=-2-strangepart}
 &  & \langle  \Psi(\pi J T) \,| \,H^{S=-2} \, |\,   \Psi(\pi J T)  \rangle  = \nonumber\\[3pt]
 & &\! \sum_{\substack{\alpha^{*(Y_1 Y_2)} \\[3pt]  \alpha^{\prime*(Y_1 Y_2)}}} \!\! C_{\alpha^{*(Y_1 Y_2)}}
  C_{\alpha^{\prime*(Y_1 Y_2)}} \,    \langle \alpha^{*(Y_1 Y_2)} |\,  H^{S=-2}_{Y_1Y_2} |\,
 \alpha^{\prime*(Y_1 Y_2)} \rangle\nonumber\\[2pt]
&  & \!+ \sum_{\substack{\alpha^{*(Y_1 Y_2)} \\[3pt]  \alpha^{\prime*(\Xi)}}} \!2 C_{\alpha^{*(Y_1 Y_2)}}
  C_{\alpha^{\prime*(\Xi)}} \,    \langle \alpha^{*(Y_1 Y_2)} |\,  H^{S=-2}_{Y_1Y_2 ,\Xi N} |\,
 \alpha^{\prime*(\Xi)} \rangle\nonumber\\[2pt]
  &  & \!+\sum_{\alpha^{*(\Xi)}, \alpha^{\prime*(\Xi)}} \! C_{\alpha^{*(\Xi)}}
  C_{\alpha^{\prime*(\Xi)}}    \langle \alpha^{*(\Xi)} \, | H^{S=-2}_{\Xi} |\,
 \alpha^{\prime*(\Xi)} \rangle\,,   
\end{eqnarray}
are evaluated analogously. Indeed,   in order  to calculate the last two terms in Eq.~(\ref{eq:S=-2-strangepart}),  one simply  needs to expand the 
states $|\alpha^{*(\Xi)}\rangle$ in the complete  set of  intermediate states $|\alpha^{*(\Xi N)}\rangle$ that explicitly single out a $\Xi N$ pair, 
\begin{align}\label{eq:expansionXiN}
\begin{split}
 |\alpha^{*(\Xi)} \rangle = \sum_{\alpha^{*(\Xi N)}  }   \langle \alpha^{*(\Xi)}  |  \alpha^{*(\Xi N)} \rangle
   \,|\alpha^{*(\Xi N)} \rangle\,.  
\end{split}
\end{align}
 Here the transition coefficients $ \langle \alpha^{*(\Xi)}  |  \alpha^{*(\Xi N)} \rangle$ can be computed  using  the  expression   Eq.~(A.6) in Ref.~\cite{Le:2020zdu}. 
 It is easy to see that 
  the  last term  in Eq.~(\ref{eq:S=-2-strangepart}), \break 
 $  \langle \alpha^{*(\Xi)} \, | H^{S=-2}_{\Xi} |\,
 \alpha^{\prime*(\Xi)} \rangle,$  differs from   the matrix element  of the   two-body $\Xi N$ Hamiltonian in the  $|\Xi N \rangle$ basis 
 by a   combinatorial factor of $A-1$.  The  factor that relates $ \langle \alpha^{*(Y_1 Y_2)} |\,  H^{S=-2}_{Y_1Y_2 -\Xi N} |\,
 \alpha^{\prime*(\Xi)}\rangle $  to  the two-body transition potential  $V_{Y_1 Y_2 - \Xi N}$ is, however,  not obvious because of   possible couplings between identical and non-identical two-body states,  for instance,   $\Sigma \Sigma - \Xi N$ or $\Lambda \Lambda -\Xi N$. 
 In  \ref{appendix:appendixA},  we have  shown that,  in this case, the  corresponding   combinatorial factor is  $\sqrt{A-1}$ (see Table~\ref{tab:appendt2}).
\subsection{Separation of  a YN   pair}\label{sec:separation_NN_YN_YY}
Let us now  discuss the evaluation of the second term in Eq.~(\ref{eq:matrixE_S=-2}) that involves the singly-strange  Hamiltonian
$H^{S=-1}$ of Eq.~(\ref{eq:hamiltonian2Y}),
\begin{eqnarray}\label{eq:single-strangepart}
 &  & \langle  \Psi(\pi J T) \,|\, H^{S=-1}\,| \,  \Psi(\pi J T)  \rangle= \nonumber\\[4pt] 
&  & \!  \! \sum_{\substack{\alpha^{*(Y_1 Y_2)} \\[3pt]  \alpha^{\prime*(Y_1 Y_2)}}} \! \!\!C_{\alpha^{*(Y_1 Y_2)}}  
 C_{\alpha^{\prime*(Y_1 Y_2)}} \,    \langle \alpha^{*(Y_1 Y_2)} |\,  H^{S=-1}_{Y_1Y_2} |\,
 \alpha^{\prime*(Y_1 Y_2)} \rangle,\nonumber\\[-7pt]
 \end{eqnarray}
 in some details  since it requires 
 new sets of transition coefficients.  
Here, in order to  compute   the matrix elements $\langle \alpha^{*(Y_1 Y_2)} |\,  H^{S=-1}_{Y_1Y_2} |\,
 \alpha^{\prime*(Y_1 Y_2)} \rangle$,
one needs to employ other  sets  of intermediate  states  that explicitly separate out a YN  pair .  Obviously,  each of the  hyperons,   $Y_1$ and   $Y_2$,  can  interact with a nucleon  independently (as it is clearly seen from the expression for $H^{S=-1}_{Y_1 Y_2}$ in Eq.~(\ref{eq:hamiltonian2Y})).   It is then  instructive   to exploit  two separate intermediate sets, namely 
$|\big( \alpha^{*{(Y_1 N)}}  \big)^{*(Y_2)}\rangle$ and  $|\big( \alpha^{*{(Y_2 N)}}  \big)^{*(Y_1)}\rangle$. The first set, $|\big( \alpha^{*{(Y_1 N)}}  \big)^{*(Y_2)}\rangle$,  is needed when computing the matrix elements of the first two terms of  $H^{S=-1}_{Y_1 Y_2}$ where   $Y_1$  is the active hyperon while $Y_2$
plays the  role of a spectator. Similarly,  the second set, $|\big( \alpha^{*{(Y_2 N)}}  \big)^{*(Y_1)}\rangle$,  is useful  for
evaluating the two remaining terms in Eq.~(\ref{eq:hamiltonian2Y}) where the roles of $Y_1$ and $Y_2$ hyperons have been interchanged (i.e.,$Y_2$ is now the active particle).   The construction of these bases is straightforward. For example, 
the  first set     can be formed  by combining the hyperon  states $|Y_{2}\rangle$, depending on the Jacobi coordinate of 
the $Y_2$ hyperon  relative to  the  C.M.~of the     $ ((A-3)N + Y_1 N)$ subcluster, with  the  $|\alpha^{*(Y_1 N)}\rangle$ states  constructed  in Eq.~(9) in \cite{Le:2020zdu}.  Thus,
\begin{eqnarray}\label{eq:seprateYNpare_S=-2}
&  &    |\big( \alpha^{*{(Y_1 N)}}  \big)^{*(Y_2)}\rangle   = |\alpha^{*{(Y_1 N)}}\rangle \otimes |Y_2\rangle\nonumber\\[3pt]
&   &   = | \mathcal{N} J T, \alpha^{*(Y_1 N)}_{A-1} \tilde{n}_{Y_2}  \tilde{I}_{Y_2} \tilde{t}_{Y_2}; (J^{*(Y_1 N)}_{A-1} (\tilde{l}_{Y_2} s_{Y_2})\tilde{I}_{Y_2})J,\nonumber \\[4pt]
&  &  \qquad\qquad (T^{*(Y_1 N)}_{A-1} \tilde{t}_{Y_2})T \rangle\nonumber\\[-7pt]
&  &     \equiv \Big| \begin{tikzpicture}[baseline={([yshift=-0.5ex]current bounding box.center)},scale=0.65]
                \filldraw[color=black, ultra thick, fill=gray ]  (0.,0.) circle(0.20cm);
                 \filldraw[red] (-0.8,0.24)   circle(0.7mm) node[anchor=base, above=0.07em, text=black] {\scriptsize{$Y_1$}} ;
                \filldraw[black] (-0.8,-0.24)  circle(0.55mm); 
                  \filldraw[red] (-0.37,-0.5)   circle(0.7mm) node[anchor=base, below=0.07em, text=black] {\scriptsize{$Y_2$}} ;
                   \filldraw[white] (-0.37,0.5)   circle(0.7mm) node[anchor=base, above=0.07em, text=white] {\scriptsize{$Y_2$}} ;
                \draw[baseline,thick, -]  (-0.8,-0.24) -- (-0.8,0.17);
                \draw[baseline,thick, -]  (-0.8,0.) -- (-0.2,0.0);
                  \draw[baseline,thick, -]  (-0.37,-0.45) -- (-0.37,0.);
                \end{tikzpicture}\, \Big\rangle,
\end{eqnarray}
and,  similarly 
\begin{eqnarray}\label{eq:2seprateYNpare_S=-2}
&  & |\big( \alpha^{*{(Y_2 N)}}  \big)^{*(Y_1)}\rangle   = |\alpha^{*{(Y_2 N)}}\rangle \otimes |Y_1\rangle\nonumber\\[4pt]
&  &  = | \mathcal{N} J T, \alpha^{*(Y_2 N)}_{A-1} \tilde{n}_{Y_1}  \tilde{I}_{Y_1} \tilde{t}_{Y_1}; (J^{*(Y_2 N)}_{A-1} (\tilde{l}_{Y_1} s_{Y_1})\tilde{I}_{Y_1})J, \nonumber\\[3pt] 
&  &\qquad\qquad  (T^{*(Y_2 N)}_{A-1} \tilde{t}_{Y_1})T \rangle\nonumber\\[-7pt]
 & & \equiv \Big| \begin{tikzpicture}[baseline={([yshift=-0.5ex]current bounding box.center)},scale=0.65]
                \filldraw[color=black, ultra thick, fill=gray ]  (0.,0.) circle(0.20cm);
                 \filldraw[red] (-0.8,0.24)   circle(0.7mm) node[anchor=base, above=0.07em, text=black] {\scriptsize{$Y_2$}} ;
                \filldraw[black] (-0.8,-0.24)  circle(0.55mm); 
                  \filldraw[red] (-0.37,-0.5)   circle(0.7mm) node[anchor=base, below=0.07em, text=black] {\scriptsize{$Y_1$}} ;
                      \filldraw[white] (-0.37,0.5)   circle(0.7mm) node[anchor=base, above=0.07em, text=white] {\scriptsize{$Y_1$}} ;
                \draw[baseline,thick, -]  (-0.8,-0.24) -- (-0.8,0.17);
                \draw[baseline,thick, -]  (-0.8,0.) -- (-0.2,0.0);
                  \draw[baseline,thick, -]  (-0.37,-0.45) -- (-0.37,0.);
                \end{tikzpicture}\, \Big\rangle.
\end{eqnarray}
In both of these basis states, we have one momentum/position of the spectator pointing towards the spectator, the one of the pair pointing towards the hyperon and the third momentum/position pointing towards the $A-3$ cluster. 

Clearly, each of the above two auxiliary sets  is  complete with respect to the basis states  $|\alpha^{*(Y_1 Y_2)}\rangle$ in Eq.~(\ref{eq:basisS=Y1Y2}). This in turn  allows for  the following expansions
\begin{eqnarray}  \label{eq:expansionS=-2_1}
&  &   |\alpha^{*(Y_1 Y_2)}\rangle=\nonumber\\[3pt]
 &  &   \!\!\!\sum_{\substack{ { } \\[4pt]  (\alpha^{*{(Y_1 N)}} )^{*(Y_2)}}} \!\!\!\!\! \!\!\big\langle \big( \alpha^{*{(Y_1 N)}}  \big)^{*(Y_2)} \big|  \alpha^{*(Y_1 Y_2)}\big\rangle \,  \big|\big( \alpha^{*{(Y_1 N)}}  \big)^{*(Y_2)}\big\rangle,\nonumber\\[-8pt]
\end{eqnarray}
or,
\begin{eqnarray}\label{eq:expansionS=-2}
&  &  |\alpha^{*(Y_1 Y_2)}\rangle  = \nonumber\\
&  &     \!\!\!\sum_{\substack{ { } \\[4pt]  (\alpha^{*{(Y_2 N)}} )^{*(Y_1)}}}  \!\!\! \!\!\!\! \big\langle \big( \alpha^{*{(Y_2 N)}}  \big)^{*(Y_1)} \big|  \alpha^{*(Y_1 Y_2)}\big\rangle \,  \big|\big( \alpha^{*{(Y_2 N)}}  \big)^{*(Y_1)}\big\rangle.\nonumber\\[-8pt] 
\end{eqnarray}
 Obviously, when $Y_1$  and $Y_2$  are identical,   the two auxiliary sets Eqs.~(\ref{eq:seprateYNpare_S=-2},\ref{eq:2seprateYNpare_S=-2})  are  the same,  and there is no need to distinguish between the two expansions. In any case, 
the expansion  coefficients in Eqs.~(\ref{eq:expansionS=-2_1},\ref{eq:expansionS=-2})  are very similar to each other and can be  computed analogously. In the following, we  focus on   the transition coefficients of  the first expansion. 
 For computing the  overlap,\break 
 $  \langle ( \alpha^{*{(Y_1 N)}})^{*(Y_2)}  |  \alpha^{*(Y_1 Y_2)} \rangle$,  we make use of  another set of auxiliary  states, $|(\alpha^{*(Y_1)})^{*{( Y_2)}}\rangle $ that explicitly single out the $Y_1$
 and $Y_2$ hyperons. These states are 
 obtained by coupling the hyperon states $|Y_{2}\rangle$  to the basis states   of the $((A-2)N + Y_1)$  system, $|\alpha^{*(Y_1)}\rangle_{A-1}$,  defined in   
 Eq.~(4) in \cite{Le:2020zdu},
\begin{eqnarray}\label{eq:auxiliary_S=-2}
&  & |\big( \alpha^{*{(Y_1)}}  \big)^{*(Y_2)}\rangle   = |\alpha^{*{(Y_1)}}\rangle_{A-1} \otimes |Y_2\rangle \nonumber\\[3pt]
&   &  = | \mathcal{N} J T, \alpha^{*(Y_1)}_{A-1} {n}_{Y_2} {I}_{Y_2} {t}_{Y_2}; (J^{*(Y_1 )}_{A-1} ({l}_{Y_2} s_{Y_2}){I}_{Y_2})J,\nonumber\\
& & \qquad\qquad  (T^{*(Y_1 )}_{A-1}\, t_{Y_2})T \rangle\nonumber\\[3pt]
&   &  = | \mathcal{N} J T, \mathcal{N}^{\prime}_{A-2} n_{Y_1}  I_{Y_1}  t_{Y_1} {n}_{Y_2}  {I}_{Y_2} {t}_{Y_2}; ( (J^{\prime}_{A-2} (l_{Y_1} s_{Y_1}) I_{Y_1})\nonumber\\[3pt]
&  & \qquad  J^{*(Y_1 )}_{A-1} ({l}_{Y_2} s_{Y_2}){I}_{Y_2})J,
 (( \,T^{\prime}_{A-2} \,  t_{Y_1}) T^{*(Y_{1})}_{A-1}\, t_{Y_2})T \,\rangle
\nonumber\\[-7pt]
&  & \equiv
  \Big| \begin{tikzpicture}[baseline={([yshift=-0.5ex]current bounding box.center)},scale=0.65]
                \filldraw[color=black, ultra thick, fill=gray ]  (0.,0.) circle(0.20cm);
                 \filldraw[red] (-0.8,0.)   circle(0.7mm) node[anchor=base, above=0.07em, text=black] {\scriptsize{$Y_1$}} ;
                  \filldraw[red] (-0.37,-0.5)   circle(0.7mm) node[anchor=base, below=0.07em, text=black] {\scriptsize{$Y_2$}} ;
                    \filldraw[white] (-0.37,0.5)   circle(0.7mm) node[anchor=base, above=-0.10em, text=white] {\scriptsize{$Y_2$}} ;
                \draw[baseline,thick, -]  (-0.75,0.) -- (-0.2,0.0);
                  \draw[baseline,thick, -]  (-0.37,-0.45) -- (-0.37,0.);
                \end{tikzpicture}\, \Big\rangle.
\end{eqnarray}
The third line in Eq.~(\ref{eq:auxiliary_S=-2}) is to illustrate how the quantum number of  the three subclusters:  (A-2) nucleons, $Y_1$  and $Y_2$  hyperons, are combined to form
the intermediate states with the definite  quantum numbers $\mathcal{N}, J$  and $T$.
 Exploiting the completeness of the auxiliary states $|\big( \alpha^{*{(Y_1)}}  \big)^{*(Y_2)}\big\rangle$,  the transition coefficient in Eq.~(\ref{eq:expansionS=-2_1}) then becomes
\begin{eqnarray}\label{eq:transitionS=-2}
&  & \big\langle \big( \alpha^{*{(Y_1 N)}}  \big)^{*(Y_2)}  |  \alpha^{*(Y_1 Y_2)}\big\rangle   = \nonumber\\[5pt]
 &  &  \big\langle \big( \alpha^{*{(Y_1 N)}}  \big)^{*(Y_2)}  |  \big( \alpha^{*{(Y_1)}}  \big)^{*(Y_2)}\big\rangle\,
\big \langle \big( \alpha^{*{(Y_1)}}  \big)^{*(Y_2)} | \alpha^{*(Y_1 Y_2)} \rangle\nonumber\\[-2pt]
& & \equiv  \Big\langle \begin{tikzpicture}[baseline={([yshift=-0.5ex]current bounding box.center)},scale=0.65]
                \filldraw[color=black, ultra thick, fill=gray ]  (0.,0.) circle(0.20cm);
                 \filldraw[red] (0.8,0.24)   circle(0.7mm) node[anchor=base, above=-0.1em, text=black] {\scriptsize{$Y_1$}} ;
                \filldraw[black] (0.8,-0.24)  circle(0.55mm); 
                  \filldraw[red] (0.37,-0.5)   circle(0.7mm) node[anchor=base, below=-0.1em, text=black] {\scriptsize{$Y_2$}} ;
                    \filldraw[white] (0.37,0.5)   circle(0.7mm) node[anchor=base, above=-0.1em, text=white] {\scriptsize{$Y_2$}} ;
                \draw[baseline,thick, -]  (0.8,-0.24) -- (0.8,0.17);
                \draw[baseline,thick, -]  (0.8,0.) -- (0.2,0.0);
                  \draw[baseline,thick, -]  (0.37,-0.45) -- (0.37,0.);
                \end{tikzpicture}        
                 \Big| \begin{tikzpicture}[baseline={([yshift=-0.5ex]current bounding box.center)},scale=0.65]
                \filldraw[color=black, ultra thick, fill=gray ]  (0.,0.) circle(0.20cm);
                 \filldraw[red] (-0.8,0.)   circle(0.7mm) node[anchor=base, above=0.07em, text=black] {\scriptsize{$Y_1$}} ;
                  \filldraw[red] (-0.37,-0.5)   circle(0.7mm) node[anchor=base, below=0.07em, text=black] {\scriptsize{$Y_2$}} ;
                    \filldraw[white] (-0.37,0.5)   circle(0.7mm) node[anchor=base, above=-0.10em, text=white] {\scriptsize{$Y_2$}} ;
                \draw[baseline,thick, -]  (-0.75,0.) -- (-0.2,0.0);
                  \draw[baseline,thick, -]  (-0.37,-0.45) -- (-0.37,0.);
                \end{tikzpicture} \, \Big\rangle
                          \, \Big\langle \begin{tikzpicture}[baseline={([yshift=-0.5ex]current bounding box.center)},scale=0.65]
                \filldraw[color=black, ultra thick, fill=gray ]  (0.,0.) circle(0.20cm);
                 \filldraw[red] (0.8,0.)   circle(0.7mm) node[anchor=base, above=-0.10em, text=black] {\scriptsize{$Y_1$}} ;
                  \filldraw[red] (0.37,-0.5)   circle(0.7mm) node[anchor=base, below=-0.10em, text=black] {\scriptsize{$Y_2$}} ;
                    \filldraw[white] (0.37,0.5)   circle(0.7mm) node[anchor=base, above=-0.10em, text=white] {\scriptsize{$Y_2$}} ;
                \draw[baseline,thick, -]  (0.75,0.) -- (0.2,0.0);
                  \draw[baseline,thick, -]  (0.37,-0.45) -- (0.37,0.);
                \end{tikzpicture}\, \Big | 
                 \begin{tikzpicture}[baseline={([yshift=-0.5ex]current bounding box.center)},scale=0.65]
                \filldraw[color=black, ultra thick, fill=gray ]  (0.,0.) circle(0.20cm);
                \filldraw[red] (-0.7,0.24)   circle(0.7mm) node[anchor=base, above=-0.1em, text=black] {\scriptsize{$Y_1$}} ;
                \filldraw[red] (-0.7,-0.24)  circle(0.7mm) node[anchor=base, below=-0.01em, text=black] {\scriptsize{$Y_2$}} ;
                \draw[baseline,thick, -]  (-0.7,-0.19) -- (-0.7,0.17);
                \draw[baseline,thick, -]  (-0.7,0.) -- (-0.2,0.0);
                \end{tikzpicture} \Big\rangle\nonumber\\[-5pt]
                &  &=\delta_{Y_2' Y_2} \ \delta_{A-2',A-2} 
                \nonumber \\ & & \qquad \times  \begin{tikzpicture}[baseline={([yshift=-0.5ex]current bounding box.center)},scale=0.65]
                    \draw[baseline,thick, -]  (-0.5,-0.) -- (-0.33,0.4);
                   \draw[baseline,thick, -]  (-0.5,0.) -- (-0.33,-0.5);
                \filldraw[color=black, ultra thick, fill=gray ]  (0.,0.) circle(0.20cm);
                 \filldraw[red] (0.8,0.24)   circle(0.7mm) node[anchor=base, above=-0.02em, text=black] {\scriptsize{$Y_1$}} ;
                \filldraw[black] (0.8,-0.24)  circle(0.55mm); 
                \draw[baseline,thick, -]  (0.8,-0.24) -- (0.8,0.17);
                \draw[baseline,thick, -]  (0.8,0.) -- (0.2,0.0);
                \end{tikzpicture}
                 \Big| \begin{tikzpicture}[baseline={([yshift=-.5ex]current bounding box.center)},scale=0.65]
                \filldraw[color=black, ultra thick, fill=gray ]  (0.,0.) circle(0.22cm);
                \filldraw[red]  (-0.6,0.) circle (0.7mm)  node[anchor=base, above=0.02em, text=black] {\scriptsize{$Y_1$}} ;
                \filldraw[red]  (-0.6,0.) circle (0.7mm)  node[anchor=base, below=0.03em, text=white] {\scriptsize{$Y_1$}} ;
                           \filldraw[white] (-0.6,0.24)   circle(0.7mm) node[anchor=base, above=-0.02em, text=white] {\scriptsize{$Y_1$}} ;
                              \filldraw[white] (-0.6,-0.24)   circle(0.7mm) node[anchor=base, above=-0.02em, text=white] {\scriptsize{$Y_1$}} ;
                \draw[baseline,thick, -]  (-0.22,0.) -- (-0.53,0.);
                \end{tikzpicture}  \Big\rangle_{\!(A-1)}   \Big\langle
                 \begin{tikzpicture}[baseline={([yshift=-0.5ex]current bounding box.center)},scale=0.65]
                \filldraw[color=black, ultra thick, fill=gray ]  (0.,0.) circle(0.20cm);
                 \filldraw[red] (0.8,0.)   circle(0.7mm) node[anchor=base, above=-0.10em, text=black] {\scriptsize{$Y_1$}} ;
                  \filldraw[red] (0.37,-0.5)   circle(0.7mm) node[anchor=base, below=-0.10em, text=black] {\scriptsize{$Y_2$}} ;
                     \filldraw[white] (0.37,0.5)   circle(0.7mm) node[anchor=base, above=-0.10em, text=white] {\scriptsize{$Y_2$}} ;
                \draw[baseline,thick, -]  (0.75,0.) -- (0.2,0.0);
                  \draw[baseline,thick, -]  (0.37,-0.45) -- (0.37,0.);
                \end{tikzpicture}\, \Big | 
                 \begin{tikzpicture}[baseline={([yshift=-0.5ex]current bounding box.center)},scale=0.65]
                \filldraw[color=black, ultra thick, fill=gray ]  (0.,0.) circle(0.20cm);
                \filldraw[red] (-0.7,0.24)   circle(0.7mm) node[anchor=base, above=-0.1em, text=black] {\scriptsize{$Y_1$}} ;
                \filldraw[red] (-0.7,-0.24)  circle(0.7mm) node[anchor=base, below=-0.1em, text=black] {\scriptsize{$Y_2$}} ;
                \draw[baseline,thick, -]  (-0.7,-0.19) -- (-0.7,0.17);
                \draw[baseline,thick, -]  (-0.7,0.) -- (-0.2,0.0);
                \end{tikzpicture} \Big\rangle\,, 
\end{eqnarray}
where a summation over the states $|\big( \alpha^{*{(Y_1)}}  \big)^{*(Y_2)}\big\rangle$  is implied.  The first overlap  $ \big\langle \big( \alpha^{*{(Y_1 N)}}  \big)^{*(Y_2)}  |  \big( \alpha^{*{(Y_1)}}  \big)^{*(Y_2)}\big\rangle
$ in   Eq.~(\ref{eq:transitionS=-2}) is essentially given by   the transition coefficients  of a system consisting of  $(A-2)$ nucleons and the $Y_1$ hyperon
(see Eq.~(11) in \cite{Le:2020zdu}),  whereas the second
term $\big \langle \big( \alpha^{*{(Y_1)}}  \big)^{*(Y_2)} | \alpha^{*(Y_1 Y_2)} \rangle$ can quickly be   deduced from 
 Eq.~(11) in \cite{Liebig:2015kwa},
  \begin{eqnarray} \label{eq:BtransitionS=-2}
&  & \! \big\langle \big(\alpha^{*(Y_1)}\big)^{*(Y_2)}| \alpha^{*(Y_1 Y_2)} \big\rangle =  \nonumber\\[4pt]
&  &  \! \delta_{{T}^{\prime}_{A-2} T_{A-2}} \delta_{{J}^{\prime}_{A-2} J_{A-2}} 
 \delta_{\mathcal{{N}^{\prime}}_{A-2} \mathcal{N}_{A-2}}\delta_{{\zeta}^{\prime}_{A-2} \zeta_{A-2}}\nonumber\\[8pt]
 &  &  \times  \hat{I}_{Y_{1}} \hat{I}_{Y_{2}}  \, \hat{J}_{Y_1 Y_2}  \hat{S}_{Y_1 Y_2}\, \hat{T}_{Y_1 Y_2} \,\hat{I}_{\lambda} 
 \hat{J}^{*(Y_1)}_{A-1}\, \hat{T}^{*(Y_1)}_{A-1}\nonumber\\[8pt]
&  &  \! \times (-1)^{ 3J_{A-2} + 2T_{A-2} + T_{Y_1 Y_2} + S_{Y_1Y_2} +\lambda + t_{Y_1} + l_{Y_1}  + t_{Y_2} + l_{Y_2} + I_{Y_1}}\nonumber\\[5pt]
&  & \!  \times \sum_{S_{A-1}=J_{A-2} + s_{Y_1}} \!\! \!\!\!  (-1)^{S_{A-1} +1} \, \hat{S}^{2}_{A-1} 
                   \left\{ \begin{array}{ccc}
                  J_{A-2} & s_{Y_1} &  S_{A-1}\\
                  l_{Y_1} & J^{*(Y_1)}_{A-1} & I_{Y_1}  
                  \end{array} \right\} \nonumber\\[8pt]
&   &\!  \times\!\!\! \!\!\!\! \sum_{\substack{{L = l_{Y_1} + l_{Y_2}}\\[3pt]{S= S_{A-1} +s_{Y_2}}}} \!\!\!\!\! \!\!\!\!\hat{L}^2 \hat{S}^2
           \left\{\begin{array}{ccc}
                   l_{Y_1}  & S_{A-1} & J^{*(Y_{1})}_{A-1}\\
                   l_{Y_2} & s_{Y_2} & I_{Y_2}\\
                   L & S  &{J}
                  \end{array}\right\}  \!  
                  \left\{\begin{array}{ccc}
                   l_{Y_1 Y_2}  & S_{Y_1 Y_2} & J_{Y_1 Y_2}\\
                   \lambda & J_{A-2} & I_{\lambda}\\
                   L & S  & {J}
                  \end{array}\!\right\}\nonumber \\[6pt]
&  & \! \times   \left\{ \begin{array}{ccc}
                  s_{Y_2} & s_{Y_1} &  S_{Y_1 Y_2}\\
                  J_{A-2} & S & S_{A-1}  
                  \end{array} \right\}
                   \left\{ \begin{array}{ccc}
                  t_{Y_2} & t_{Y_1} &  T_{Y_1Y_2}\\
                  T_{A-2} & {T} & T^{*(Y_1)}_{A-1}  
                  \end{array} \right\} \nonumber\\[4pt]
  &  &\!  \times   \langle n_{Y_1} \,l_{Y_1}\, n_{Y_2}\, l_{Y_2} : L \,|\,  n_{Y_1 Y_2} \,l_{Y_1Y_2} \, n_{\lambda} \, \lambda: L \rangle_{d}\,,
\end{eqnarray}
with, 
 $$ d=\frac{(A-2)  m_N \,  m(t_{Y_2})} { m(t_{Y_1})  \big ((A-2)\,m_N + m(t_{Y_1})  + m(t_{Y_2})\big)}\,.$$
Here, we use the notation $\hat j = \sqrt{2j+1}$ 
and abbreviate the summations running from $|J_1-J_2|$ to $J_1+J_2$ simply by $J_1+J_2$.

The transition coefficients for  the second expansion in  Eq.~(\ref{eq:expansionS=-2}) are computed analogously. 
Taking into account   the expansions   Eqs.~(\ref{eq:expansionS=-2_1},\ref{eq:expansionS=-2}), the matrix element  $ \langle \alpha^{*(Y_1 Y_2)} |\,  H^{S=-1}_{Y_1Y_2} |\,
 \alpha^{\prime*(Y_1 Y_2)} \rangle $ in Eq.~(\ref{eq:single-strangepart}) is then decomposed into,
 \begin{eqnarray}\label{eq:decompose}
&    &    \langle \alpha^{*(Y_1 Y_2)} |\,  H^{S=-1}_{Y_1Y_2} |\,
 \alpha^{\prime*(Y_1 Y_2)} \rangle   \nonumber\\[5pt]
   & & \qquad\qquad   =  \langle \alpha^{*(Y_1 Y_2)} |\,  H^{S=-1}_{Y_1Y_2} |\,
 \alpha^{\prime*(Y_1 Y_2)} \rangle_{Y_2}  \nonumber\\[5pt]
 &  & \qquad\qquad  \quad +\,\,  \langle \alpha^{*(Y_1 Y_2)} |\,  H^{S=-1}_{Y_1Y_2} |\,
 \alpha^{\prime*(Y_1 Y_2)} \rangle_{Y_1 }\,.
 \end{eqnarray}
 The subscript  in each term  on the right-hand side of Eq.~(\ref{eq:decompose}) specifies  the hyperon spectator. 
 The first contribution is further given by
\begin{eqnarray}\label{eq:EvalS=-1part}
& &\!\!  \langle \alpha^{*(Y_1 Y_2)} |\,  H^{S=-1}_{Y_1Y_2} |\,
 \alpha^{\prime*(Y_1 Y_2)} \rangle_{Y_2}     \nonumber\\[2pt]
 & & \!\!=  \big\langle   \begin{tikzpicture}[baseline={([yshift=-0.5ex]current bounding box.center)},scale=0.55]
                \filldraw[color=black, ultra thick, fill=gray ]  (0.,0.) circle(0.20cm);
                \filldraw[red] (0.7,0.24)   circle(0.7mm) node[anchor=base, above=-0.1em, text=black] {\scriptsize{$Y_1$}} ;
                \filldraw[red] (0.7,-0.24)  circle(0.7mm) node[anchor=base, below=-0.1em, text=black] {\scriptsize{$Y_2$}} ;
                \draw[baseline,thick, -]  (0.7,-0.19) -- (0.7,0.17);
                \draw[baseline,thick, -]  (0.7,0.) -- (0.2,0.0);
                \end{tikzpicture} \! \big | \! \!
                \begin{tikzpicture}[baseline={([yshift=-0.5ex]current bounding box.center)},scale=0.55]
                \filldraw[color=black, ultra thick, fill=gray ]  (0.,0.) circle(0.20cm);
                 \filldraw[red] (-0.8,0.24)   circle(0.7mm) node[anchor=base, above=-0.1em, text=black] {\scriptsize{$Y_1$}} ;
                \filldraw[black] (-0.8,-0.24)  circle(0.55mm); 
                  \filldraw[red] (-0.37,-0.5)   circle(0.7mm) node[anchor=base, below=-0.1em, text=black] {\scriptsize{$Y_2$}} ;
                \draw[baseline,thick, -]  (-0.8,-0.24) -- (-0.8,0.17);
                \draw[baseline,thick, -]  (-0.8,0.) -- (-0.2,0.0);
                  \draw[baseline,thick, -]  (-0.37,-0.45) -- (-0.37,0.);
                \end{tikzpicture}\big\rangle 
                  \big\langle  \begin{tikzpicture}[baseline={([yshift=-0.5ex]current bounding box.center)},scale=0.55]
                \filldraw[color=black, ultra thick, fill=gray ]  (0.,0.) circle(0.20cm);
                 \filldraw[red] (0.8,0.24)   circle(0.7mm) node[anchor=base, above=-0.1em, text=black] {\scriptsize{$Y_1$}} ;
                \filldraw[black] (0.8,-0.24)  circle(0.55mm); 
                  \filldraw[red] (0.37,-0.5)   circle(0.7mm) node[anchor=base, below=-0.1em, text=black] {\scriptsize{$Y_2$}} ;
                \draw[baseline,thick, -]  (0.8,-0.24) -- (0.8,0.17);
                \draw[baseline,thick, -]  (0.8,0.) -- (0.2,0.0);
                  \draw[baseline,thick, -]  (0.37,-0.45) -- (0.37,0.);
                \end{tikzpicture}\! \big |  H^{S=-1}_{Y_1 Y_2} \big |  \!
                   \begin{tikzpicture}[baseline={([yshift=-0.5ex]current bounding box.center)},scale=0.55]
                \filldraw[color=black, ultra thick, fill=gray ]  (0.,0.) circle(0.20cm);
                 \filldraw[red] (-0.8,0.24)   circle(0.7mm) node[anchor=base, above=-0.1em, text=black] {\scriptsize{$Y_1$}} ;
                \filldraw[black] (-0.8,-0.24)  circle(0.55mm); 
                  \filldraw[red] (-0.37,-0.5)   circle(0.7mm) node[anchor=base, below=-0.1em, text=black] {\scriptsize{$Y_2$}} ;
                \draw[baseline,thick, -]  (-0.8,-0.24) -- (-0.8,0.17);
                \draw[baseline,thick, -]  (-0.8,0.) -- (-0.2,0.0);
                  \draw[baseline,thick, -]  (-0.37,-0.45) -- (-0.37,0.);
                \end{tikzpicture}\big\rangle 
                \big \langle   
                 \begin{tikzpicture}[baseline={([yshift=-0.5ex]current bounding box.center)},scale=0.55]
                \filldraw[color=black, ultra thick, fill=gray ]  (0.,0.) circle(0.20cm);
                 \filldraw[red] (0.8,0.24)   circle(0.7mm) node[anchor=base, above=-0.1em, text=black] {\scriptsize{$Y_1$}} ;
                \filldraw[black] (0.8,-0.24)  circle(0.55mm); 
                  \filldraw[red] (0.37,-0.5)   circle(0.7mm) node[anchor=base, below=-0.1em, text=black] {\scriptsize{$Y_2$}} ;
                \draw[baseline,thick, -]  (0.8,-0.24) -- (0.8,0.17);
                \draw[baseline,thick, -]  (0.8,0.) -- (0.2,0.0);
                  \draw[baseline,thick, -]  (0.37,-0.45) -- (0.37,0.);
                \end{tikzpicture} \! \!\big | \!\!
                  \begin{tikzpicture}[baseline={([yshift=-0.5ex]current bounding box.center)},scale=0.55]
                \filldraw[color=black, ultra thick, fill=gray ]  (0.,0.) circle(0.20cm);
                \filldraw[red] (-0.7,0.24)   circle(0.7mm) node[anchor=base, above=-0.1em, text=black] {\scriptsize{$Y_1$}} ;
                \filldraw[red] (-0.7,-0.24)  circle(0.7mm) node[anchor=base, below=-0.1em, text=black] {\scriptsize{$Y_2$}} ;
                \draw[baseline,thick, -]  (-0.7,-0.19) -- (-0.7,0.17);
                \draw[baseline,thick, -]  (-0.7,0.) -- (-0.2,0.0);
                \end{tikzpicture} \big\rangle \nonumber\\[1pt]
&  & \!\! =\big\langle   \begin{tikzpicture}[baseline={([yshift=-0.5ex]current bounding box.center)},scale=0.55]
                \filldraw[color=black, ultra thick, fill=gray ]  (0.,0.) circle(0.20cm);
                \filldraw[red] (0.7,0.24)   circle(0.7mm) node[anchor=base, above=-0.1em, text=black] {\scriptsize{$Y_1$}} ;
                \filldraw[red] (0.7,-0.24)  circle(0.7mm) node[anchor=base, below=-0.1em, text=black] {\scriptsize{$Y_2$}} ;
                \draw[baseline,thick, -]  (0.7,-0.19) -- (0.7,0.17);
                \draw[baseline,thick, -]  (0.7,0.) -- (0.2,0.0);
                \end{tikzpicture} \!\big | \!\!
                \begin{tikzpicture}[baseline={([yshift=-0.5ex]current bounding box.center)},scale=0.55]
                \filldraw[color=black, ultra thick, fill=gray ]  (0.,0.) circle(0.20cm);
                 \filldraw[red] (-0.8,0.24)   circle(0.7mm) node[anchor=base, above=-0.1em, text=black] {\scriptsize{$Y_1$}} ;
                \filldraw[black] (-0.8,-0.24)  circle(0.55mm); 
                  \filldraw[red] (-0.37,-0.5)   circle(0.7mm) node[anchor=base, below=-0.1em, text=black] {\scriptsize{$Y_2$}} ;
                \draw[baseline,thick, -]  (-0.8,-0.24) -- (-0.8,0.17);
                \draw[baseline,thick, -]  (-0.8,0.) -- (-0.2,0.0);
                  \draw[baseline,thick, -]  (-0.37,-0.45) -- (-0.37,0.);
                \end{tikzpicture}\big\rangle 
                \delta_{Y_2 Y^{\prime}_2}
                  \big\langle 
                   \begin{tikzpicture}[baseline={([yshift=-0.5ex]current bounding box.center)},scale=0.55]
                \filldraw[color=black, ultra thick, fill=gray ]  (0.,0.0) circle(0.20cm);
                 \filldraw[red] (0.8,0.25)   circle(0.7mm) node[anchor=base, above=-0.1em, text=black] {\scriptsize{$Y_1$}} ;
                \filldraw[black] (0.8,-0.24)  circle(0.55mm) node[anchor=base, below=-0.1em, text=white] {\scriptsize{$Y_1$}} ;
                \draw[baseline,thick, -]  (0.8,-0.24) -- (0.8,0.17);
                \draw[baseline,thick, -]  (0.8,0.0) -- (0.2,0.0);
                \end{tikzpicture} \! \big |  H^{S=-1}_{Y_1 Y_2} \big |  \!
                   \begin{tikzpicture}[baseline={([yshift=-0.5ex]current bounding box.center)},scale=0.55]
                \filldraw[color=black, ultra thick, fill=gray ]  (0.,0.) circle(0.20cm);
                 \filldraw[red] (-0.8,0.24)   circle(0.7mm) node[anchor=base, above=-0.1em, text=black] {\scriptsize{$Y_1$}} ;
                \filldraw[black] (-0.8,-0.24)  circle(0.55mm) node[anchor=base, below=-0.1em, text=white] {\scriptsize{$Y_1$}} ;
                \draw[baseline,thick, -]  (-0.8,-0.24) -- (-0.8,0.17);
                \draw[baseline,thick, -]  (-0.8,0.) -- (-0.2,0.0);
                \end{tikzpicture}\big\rangle 
                \big \langle   
                 \begin{tikzpicture}[baseline={([yshift=-0.5ex]current bounding box.center)},scale=0.55]
                \filldraw[color=black, ultra thick, fill=gray ]  (0.,0.) circle(0.20cm);
                 \filldraw[red] (0.8,0.24)   circle(0.7mm) node[anchor=base, above=-0.1em, text=black] {\scriptsize{$Y_1$}} ;
                \filldraw[black] (0.8,-0.24)  circle(0.55mm); 
                  \filldraw[red] (0.37,-0.5)   circle(0.7mm) node[anchor=base, below=-0.1em, text=black] {\scriptsize{$Y_2$}} ;
                \draw[baseline,thick, -]  (0.8,-0.24) -- (0.8,0.17);
                \draw[baseline,thick, -]  (0.8,0.) -- (0.2,0.0);
                  \draw[baseline,thick, -]  (0.37,-0.45) -- (0.37,0.);
                \end{tikzpicture} \!\! \big  | \!\!
                  \begin{tikzpicture}[baseline={([yshift=-0.5ex]current bounding box.center)},scale=0.55]
                \filldraw[color=black, ultra thick, fill=gray ]  (0.,0.) circle(0.20cm);
                \filldraw[red] (-0.7,0.24)   circle(0.7mm) node[anchor=base, above=-0.1em, text=black] {\scriptsize{$Y_1$}} ;
                \filldraw[red] (-0.7,-0.24)  circle(0.7mm) node[anchor=base, below=-0.1em, text=black] {\scriptsize{$Y_2$}} ;
                \draw[baseline,thick, -]  (-0.7,-0.19) -- (-0.7,0.17);
                \draw[baseline,thick, -]  (-0.7,0.) -- (-0.2,0.0);
                \end{tikzpicture} \big\rangle\nonumber  \\
\end{eqnarray}
The expression for the second  term 
in Eq.~(\ref{eq:decompose})  is  obtained from  Eq.~(\ref{eq:EvalS=-1part})  by  interchanging the roles of the 
$Y_1$  and $Y_2$ hyperons in the intermediate states,
 \begin{eqnarray}\label{eq:EvalS=-1part_1}
& &\!\!  \langle \alpha^{*(Y_1 Y_2)} |\,  H^{S=-1}_{Y_1Y_2} |\,
 \alpha^{\prime*(Y_1 Y_2)} \rangle_{Y_1}     \nonumber\\[2pt]
 &  & \!\! =\big\langle   \begin{tikzpicture}[baseline={([yshift=-0.5ex]current bounding box.center)},scale=0.55]
                \filldraw[color=black, ultra thick, fill=gray ]  (0.,0.) circle(0.20cm);
                \filldraw[red] (0.7,0.24)   circle(0.7mm) node[anchor=base, above=-0.1em, text=black] {\scriptsize{$Y_1$}} ;
                \filldraw[red] (0.7,-0.24)  circle(0.7mm) node[anchor=base, below=-0.1em, text=black] {\scriptsize{$Y_2$}} ;
                \draw[baseline,thick, -]  (0.7,-0.19) -- (0.7,0.17);
                \draw[baseline,thick, -]  (0.7,0.) -- (0.2,0.0);
                \end{tikzpicture} \!\big | \!\!
                \begin{tikzpicture}[baseline={([yshift=-0.5ex]current bounding box.center)},scale=0.55]
                \filldraw[color=black, ultra thick, fill=gray ]  (0.,0.) circle(0.20cm);
                 \filldraw[red] (-0.8,0.24)   circle(0.7mm) node[anchor=base, above=-0.1em, text=black] {\scriptsize{$Y_2$}} ;
                \filldraw[black] (-0.8,-0.24)  circle(0.55mm); 
                  \filldraw[red] (-0.37,-0.5)   circle(0.7mm) node[anchor=base, below=-0.1em, text=black] {\scriptsize{$Y_1$}} ;
                \draw[baseline,thick, -]  (-0.8,-0.24) -- (-0.8,0.17);
                \draw[baseline,thick, -]  (-0.8,0.) -- (-0.2,0.0);
                  \draw[baseline,thick, -]  (-0.37,-0.45) -- (-0.37,0.);
                \end{tikzpicture}\big\rangle 
                \delta_{Y_1 Y^{\prime}_1}
                  \big\langle 
                   \begin{tikzpicture}[baseline={([yshift=-0.5ex]current bounding box.center)},scale=0.55]
                \filldraw[color=black, ultra thick, fill=gray ]  (0.,0.0) circle(0.20cm);
                 \filldraw[red] (0.8,0.25)   circle(0.7mm) node[anchor=base, above=-0.1em, text=black] {\scriptsize{$Y_2$}} ;
                \filldraw[black] (0.8,-0.24)  circle(0.55mm) node[anchor=base, below=-0.1em, text=white] {\scriptsize{$Y_2$}} ;
                \draw[baseline,thick, -]  (0.8,-0.24) -- (0.8,0.17);
                \draw[baseline,thick, -]  (0.8,0.0) -- (0.2,0.0);
                \end{tikzpicture} \! \big |  H^{S=-1}_{Y_1 Y_2} \big |  \!
                   \begin{tikzpicture}[baseline={([yshift=-0.5ex]current bounding box.center)},scale=0.55]
                \filldraw[color=black, ultra thick, fill=gray ]  (0.,0.) circle(0.20cm);
                 \filldraw[red] (-0.8,0.24)   circle(0.7mm) node[anchor=base, above=-0.1em, text=black] {\scriptsize{$Y_2$}} ;
                \filldraw[black] (-0.8,-0.24)  circle(0.55mm) node[anchor=base, below=-0.1em, text=white] {\scriptsize{$Y_2$}} ;
                \draw[baseline,thick, -]  (-0.8,-0.24) -- (-0.8,0.17);
                \draw[baseline,thick, -]  (-0.8,0.) -- (-0.2,0.0);
                \end{tikzpicture}\big\rangle 
                \big \langle   
                 \begin{tikzpicture}[baseline={([yshift=-0.5ex]current bounding box.center)},scale=0.55]
                \filldraw[color=black, ultra thick, fill=gray ]  (0.,0.) circle(0.20cm);
                 \filldraw[red] (0.8,0.24)   circle(0.7mm) node[anchor=base, above=-0.1em, text=black] {\scriptsize{$Y_2$}} ;
                \filldraw[black] (0.8,-0.24)  circle(0.55mm); 
                  \filldraw[red] (0.37,-0.5)   circle(0.7mm) node[anchor=base, below=-0.1em, text=black] {\scriptsize{$Y_1$}} ;
                \draw[baseline,thick, -]  (0.8,-0.24) -- (0.8,0.17);
                \draw[baseline,thick, -]  (0.8,0.) -- (0.2,0.0);
                  \draw[baseline,thick, -]  (0.37,-0.45) -- (0.37,0.);
                \end{tikzpicture} \!\! \big  | \!\!
                  \begin{tikzpicture}[baseline={([yshift=-0.5ex]current bounding box.center)},scale=0.55]
                \filldraw[color=black, ultra thick, fill=gray ]  (0.,0.) circle(0.20cm);
                \filldraw[red] (-0.7,0.24)   circle(0.7mm) node[anchor=base, above=-0.1em, text=black] {\scriptsize{$Y_1$}} ;
                \filldraw[red] (-0.7,-0.24)  circle(0.7mm) node[anchor=base, below=-0.1em, text=black] {\scriptsize{$Y_2$}} ;
                \draw[baseline,thick, -]  (-0.7,-0.19) -- (-0.7,0.17);
                \draw[baseline,thick, -]  (-0.7,0.) -- (-0.2,0.0);
                \end{tikzpicture} \big\rangle\nonumber\\
\end{eqnarray}
 Although Eqs.(\ref{eq:EvalS=-1part},\ref{eq:EvalS=-1part_1}) are very similar to the expression  for  computing the Hamiltonian matrix elements in $S=-1$ systems,  the presence of a hyperon spectator  $Y_2 (Y_1)$ makes it rather difficult
  to determine  the proper   combinatorial factors that
 relate the many-body matrix elements 
$       \delta_{Y_2 Y^{\prime}_2}
                  \big\langle 
                   \begin{tikzpicture}[baseline={([yshift=-0.5ex]current bounding box.center)},scale=0.5]
                \filldraw[color=black, ultra thick, fill=gray ]  (0.,0.0) circle(0.20cm);
                 \filldraw[red] (0.8,0.25)   circle(0.7mm) node[anchor=base, above=-0.1em, text=black] {\tiny{$Y_1$}} ;
                \filldraw[black] (0.8,-0.24)  circle(0.55mm) node[anchor=base, below=-0.1em, text=white] {\tiny{$Y_1$}} ;
                \draw[baseline,thick, -]  (0.8,-0.24) -- (0.8,0.17);
                \draw[baseline,thick, -]  (0.8,0.0) -- (0.2,0.0);
                \end{tikzpicture}\big |  H^{S=-1}_{Y_1 Y_2} \big |  
                   \begin{tikzpicture}[baseline={([yshift=-0.5ex]current bounding box.center)},scale=0.5]
                \filldraw[color=black, ultra thick, fill=gray ]  (0.,0.) circle(0.20cm);
                 \filldraw[red] (-0.8,0.24)   circle(0.7mm) node[anchor=base, above=-0.1em, text=black] {\tiny{$Y_1$}} ;
                \filldraw[black] (-0.8,-0.24)  circle(0.55mm) node[anchor=base, below=-0.1em, text=white] {\tiny{$Y_1$}} ;
                \draw[baseline,thick, -]  (-0.8,-0.24) -- (-0.8,0.17);
                \draw[baseline,thick, -]  (-0.8,0.) -- (-0.2,0.0);
                \end{tikzpicture}\big\rangle$    and  \break $ \delta_{Y_1 Y^{\prime}_1}
                  \big\langle 
                   \begin{tikzpicture}[baseline={([yshift=-0.5ex]current bounding box.center)},scale=0.5]
                \filldraw[color=black, ultra thick, fill=gray ]  (0.,0.0) circle(0.20cm);
                 \filldraw[red] (0.8,0.25)   circle(0.7mm) node[anchor=base, above=-0.1em, text=black] {\tiny{$Y_2$}} ;
                \filldraw[black] (0.8,-0.24)  circle(0.55mm) node[anchor=base, below=-0.1em, text=white] {\tiny{$Y_2$}} ;
                \draw[baseline,thick, -]  (0.8,-0.24) -- (0.8,0.17);
                \draw[baseline,thick, -]  (0.8,0.0) -- (0.2,0.0);
                \end{tikzpicture}\big |  H^{S=-1}_{Y_1 Y_2} \big |  
                   \begin{tikzpicture}[baseline={([yshift=-0.5ex]current bounding box.center)},scale=0.5]
                \filldraw[color=black, ultra thick, fill=gray ]  (0.,0.) circle(0.20cm);
                 \filldraw[red] (-0.8,0.24)   circle(0.7mm) node[anchor=base, above=-0.1em, text=black] {\tiny{$Y_2$}} ;
                \filldraw[black] (-0.8,-0.24)  circle(0.55mm) node[anchor=base, below=-0.1em, text=white] {\tiny{$Y_2$}} ;
                \draw[baseline,thick, -]  (-0.8,-0.24) -- (-0.8,0.17);
                \draw[baseline,thick, -]  (-0.8,0.) -- (-0.2,0.0);
                \end{tikzpicture}\big\rangle $
                 to the YN Hamiltonian matrix elements in the two-body sector. These factors are also provided  in 
                  Table \ref{tab:appendt1} in  \ref{appendix:appendixA}. From   Table \ref{tab:appendt1}, one
                 can clearly see that the corresponding factors  depend not only on the total number of nucleons  but also on the two  hyperons $Y_1$  and  $ Y_2$
                in the intermediate  states.   
\section{Results}\label{sec:result}
In this section, as a first application, we report results for the 
$\Lambda \Lambda$  s-shell   hypernuclei $^{\text{ }\text{ }\text{ } \text{}4}_{\Lambda \Lambda}\text{H}(1^+, 0)$,
\break 
$^{\text{ }\text{ }\text{ } \text{}5}_{\Lambda \Lambda}\text{He}(\frac{1}{2}^+, \frac{1}{2})$, and  $^{\text{ }\text{ }\text{ } \text{}6}_{\Lambda \Lambda}\text{He}(0^+, 0)$.  To zeroth approximation,
 these  systems can be   regarded as a $\Lambda \Lambda$ pair  in  the $^1S_0$ state  being  attached to the corresponding  core-nuclei   predominantly in their ground states. 
 While the quantum numbers of  $^{\text{ }\text{ }\text{ } \text{}5}_{\Lambda \Lambda}\text{He}$,  $(J^+, T) = (\frac{1}{2}^+, \frac{1}{2}),$ are obvious,  those  for 
 the   $^{\text{ }\text{ }\text{ } \text{}4}_{\Lambda \Lambda}\text{H}$  hypernucleus are chosen according to our observations that the state  with  $(J^+, T) = (1^+,0) $ is  
 the lowest-lying level  and in many calculations 
 the one closest to binding of all $A=4$ $S=-2$ hypernuclei. Therefore, we will report our results 
 for this state below. 
 
For all calculations presented here, we employ BB interactions that 
are derived within chiral EFT \cite{Epelbaum:2008ga}. 
The high-order semilocal momentum-space regularized potential with a regulator of $\Lambda_{N}= 450$ MeV \break (N\textsuperscript{4}LO{+}(450)) \cite{Reinert:2017usi},
SRG-evolved to  $\lambda_{NN} =1.6$  fm\textsuperscript{-1}, is adopted 
for describing the NN interaction. The next-to-leading order potential
NLO19 \cite{Haidenbauer:2019boi} with a chiral cutoff of  
$\Lambda_{Y} = 650$~MeV and an SRG parameter of  
 $\lambda_{YN} =0.868$~fm\textsuperscript{-1} is used for the YN interaction. We remark that the chosen  NN and YN potentials 
 successfully  predict  the empirical 
 $\Lambda$-separation energies for $^3_{\Lambda}\text{H}$, $^4_{\Lambda}\text{He}(1^+)$  and $^5_{\Lambda}\text{He}$  but 
 slightly underbind  
 $^4_{\Lambda}\text{He}(0^+)$  \cite{Le:2020zdu}. To describe the two-body  interactions in the $S=-2$ sector, we utilize the chiral YY interactions at LO \cite{Polinder:2007mp} and up to NLO   \cite{Haidenbauer:2015zqb,Haidenbauer:2018gvg},  
with a chiral cutoff of $\Lambda_{YY} =600 $ MeV. 

One of our primary aims here is to establish the
predictions of these chiral YY potentials  for double-$\Lambda $ s-shell hypernuclei.
 Ultimately,
 it is expected that results from such a study may provide useful additional constraints for constructing realistic $S=-2$ BB interaction potentials,
given the scarcity of direct empirical information on the underlying
two-body systems ($\Lambda\Lambda$, $\Xi N$, ...). 
Due to the latter circumstance, 
 in the chiral approach (as well as in 
 meson-exchange and/or constituent 
 quark models) the assumption of SU(3)\textsubscript{f} symmetry is
 an essential prerequisite for deriving pertinent
 potentials.
For example, in chiral 
 EFT the short-distance dynamics is represented 
 by contact terms which 
 involve low-energy constants (LECs) that 
 need to be determined from a fit to data \cite{Epelbaum:2008ga}. SU(3) symmetry strongly
 limits the number of independent LECs \cite{Haidenbauer:2013oca}. 
However, at NLO, there are two LECs which are only present in the $S=-2$ sector, and which 
 contribute to the interaction in the spin- and isospin zero channel, specifically to the $^1S_0$
 partial wave of $\Lambda\Lambda$. 
 They correspond to the SU(3) singlet irreducible 
 representation, see Ref.~\cite{Haidenbauer:2015zqb}, and are 
 denoted by $\tilde C^{1}$ and $C^{1}$, respectively, in that work.  
 These have been fixed by considering the extremely sparse and 
 uncertain YY data (i.e.,~a total cross section for 
$\Xi^{-} p - \Lambda \Lambda$ \cite{Kim:2015} and 
the 
upper limits of elastic and inelastic $\Xi^{-} p$ cross 
sections \cite{AHN2006214}). Clearly, 
such poor empirical data do not allow for a reliable quantitative 
determination of the unknown strength of the two contact terms in question.   
Nevertheless, it turned out that reasonable choices for the $C^{1}$'s can be made
\cite{Haidenbauer:2015zqb,Haidenbauer:2018gvg}
and the YY cross sections predicted by the two
  NLO potentials are fairly consistent with the experiments. Furthermore, the $\Lambda \Lambda$ $^1S_0$ scattering lengths predicted by these
  interactions are compatible with values 
  inferred from empirical information \cite{Gasparyan:2011kg,Ohnishi:2016elb}. 
  The LO interaction yields a somewhat large
  scattering length in comparison to those values
  and it also exhibits a rather 
  strong regulator dependence \cite{Polinder:2007mp}. 

  It should be pointed out that our initial NLO interaction 
  for $S=-2$ \cite{Haidenbauer:2015zqb}  
  and the updated version \cite{Haidenbauer:2018gvg} 
  differ only in the antisymmetric SU(3)\textsubscript{f} component 
 which means essentially only in the strength of 
  the $\Xi N$ interaction in the $^3S_1$ partial
  wave. This has an impact on the corresponding
  in-medium properties of the $\Xi$.
Specifically, the updated version from 2019 \cite{Haidenbauer:2018gvg} 
yields a moderately attractive $\Xi$ single-particle potential that   
  is roughly in line \cite{Kohno:2019oyw} with recent 
  experimental evidence that the existence of bound $\Xi$-hypernuclei 
  is very likely \cite{Nakazawa:2015joa}. 
 With regard to  $\Lambda\Lambda$ systems, we observe that the two realizations yield very similar binding energies for the double-$\Lambda$ s-shell hypernuclei.
   This indicates that, in general, the actual
  strength  of the spin-triplet $\Xi N$ interaction has little
  influence on few-body observables related to 
  $\Lambda \Lambda$. 
  In the following, we therefore present results  for  the LO   and the updated NLO interactions for a chiral cutoff of $\Lambda_{YY} =600$ MeV. In order to speed up the convergence,   both YY potentials are also SRG-evolved. We use a wide range of the SRG  flow parameters, namely  $1.4 \leq \lambda_{YY}  \leq 3.0$~fm\textsuperscript{-1}, to
  quantify the contribution of possible SRG-induced YYN three-body forces. 
\subsection{$^{\text{ }\text{ }\text{ } \text{}6}_{\Lambda \Lambda}\text{He}(0^+, 0)$}\label{sec:6He2lambda}
The $^{\text{ }\text{ }\text{ } \text{}6}_{\Lambda \Lambda}\text{He}$ hypernucleus is so far  the lightest   double-$\Lambda$ system being unambiguously established. Since the observation of the Nagara event
 \cite{PhysRevLett.87.212502}, its $\Lambda\Lambda$ separation energy, 
 defined as $B_{\Lambda \Lambda}( ^{\text{ }\text{ }\text{ } \text{}6}_{\Lambda \Lambda}\text{He}) =  E(^4\text{He}) - E( ^{\text{ }\text{ }\text{ } \text{}6}_{\Lambda \Lambda}\text{He})$,  has been  exploited as a crucial constraint for constructing  effective   potentials  that  are then employed in many-body calculations  like the Gaussian expansion method  \cite{Nemura:2004xb,Hiyama:2018lgs} or the cluster Faddeev-Yakubovsky approach \cite{PhysRevLett.89.172502,Filikhin:2002wm}.  
 The re-analysis of the Nagara event using the updated $\Xi$ mass   yielded a slightly smaller $\Lambda \Lambda$ separation energy, $B_{\Lambda \Lambda}(^{\text{ }\text{ }\text{ } \text{}6}_{\Lambda \Lambda}\text{He}) =6.91 \pm 0.16 $ MeV 
 \cite{Nakazawa:2010zza,PhysRevC.88.014003}, 
 as compared to the initially estimated 
value of \break $B_{\Lambda \Lambda}(^{\text{ }\text{ }\text{ } \text{}6}_{\Lambda \Lambda}\text{He}) = 7.25 \pm 0.19 $ \cite{PhysRevLett.87.212502}. 
This, in turn,   may  have direct consequences for theoretical predictions for  potentially   observable   bound states  of  other s-shell  $\Lambda\Lambda$  hypernuclei,   particularly the $A=4$ double-$\Lambda$ system \cite{NakaichiMaeda:1990kr,CONTESSI2019134893}, see also the discussion in  \ref{sec:4H2lambda}.  We note that  
  the information   about  $B_{\Lambda \Lambda}(^{\text{ }\text{ }\text{ } \text{}6}_{\Lambda \Lambda}\text{He})$ has not been directly utilized in order to constrain  the LECs  appearing in the chiral LO and NLO potentials.  It  is therefore  of enormous interest to 
explore 
this double-$\Lambda$ system  using the two chiral  interactions to scrutinize their consistency with the measured $\Lambda \Lambda$ separation energy.

\begin{figure*}[htbp] 
      \begin{center}
      \hspace{0.3cm}{
      \subfigure[$E_{\mathcal{N}}(^{\text{ }\text{ }6}_{\Lambda \Lambda}\text{He})$ as a function of $\omega$.]{\includegraphics[width=0.45\textwidth,trim={0.0cm 0.00cm 0.0cm 0 cm},clip]{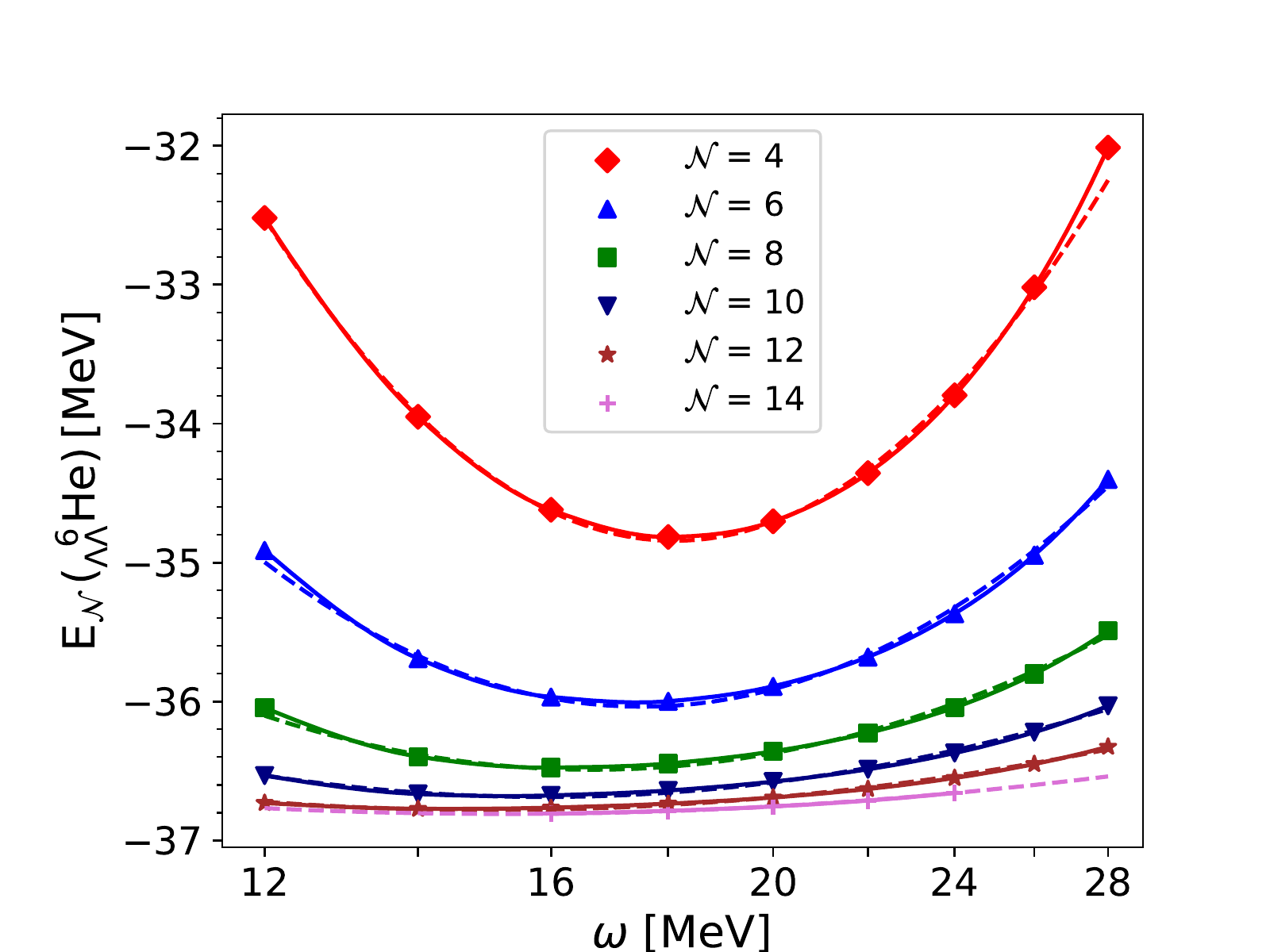}}
      \subfigure[$E(^{\text{ }\text{ } 6}_{\Lambda \Lambda}\text{He})$ as a function of $\mathcal{N}$.]{ \includegraphics[width=0.45\textwidth,trim={0.0cm 0.00cm 0.0cm 0 cm},clip]{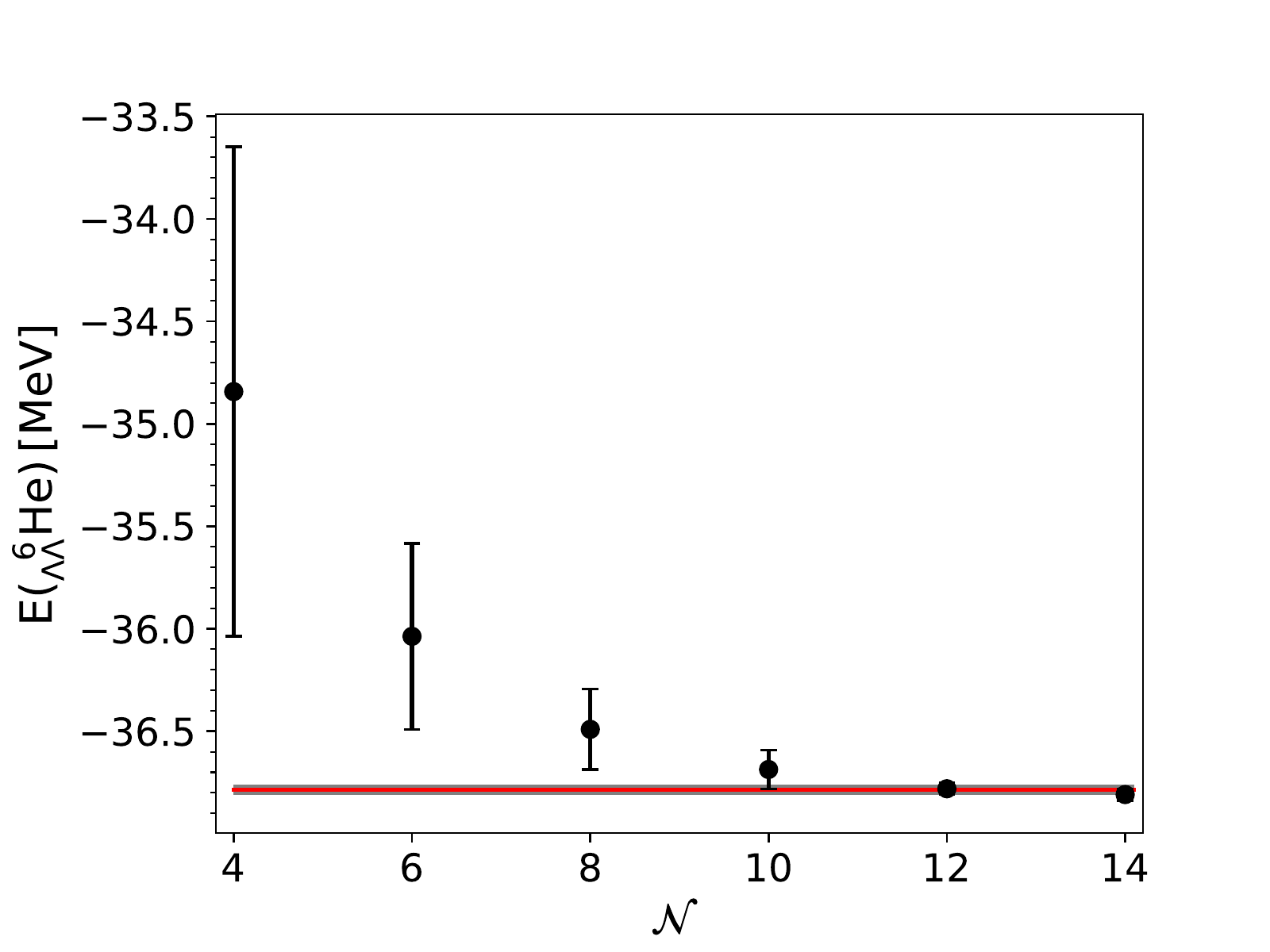}}}\\ 
         \vskip 0.15cm
      \hspace{0.3cm}{\subfigure[$B_{\Lambda \Lambda}(^{\text{ }\text{ } 6}_{\Lambda \Lambda}\text{He})$ as a function of $\mathcal{N}$.]{\includegraphics[width=0.45\textwidth,trim={0.0cm 0.00cm 0.0cm 0.0cm},clip] {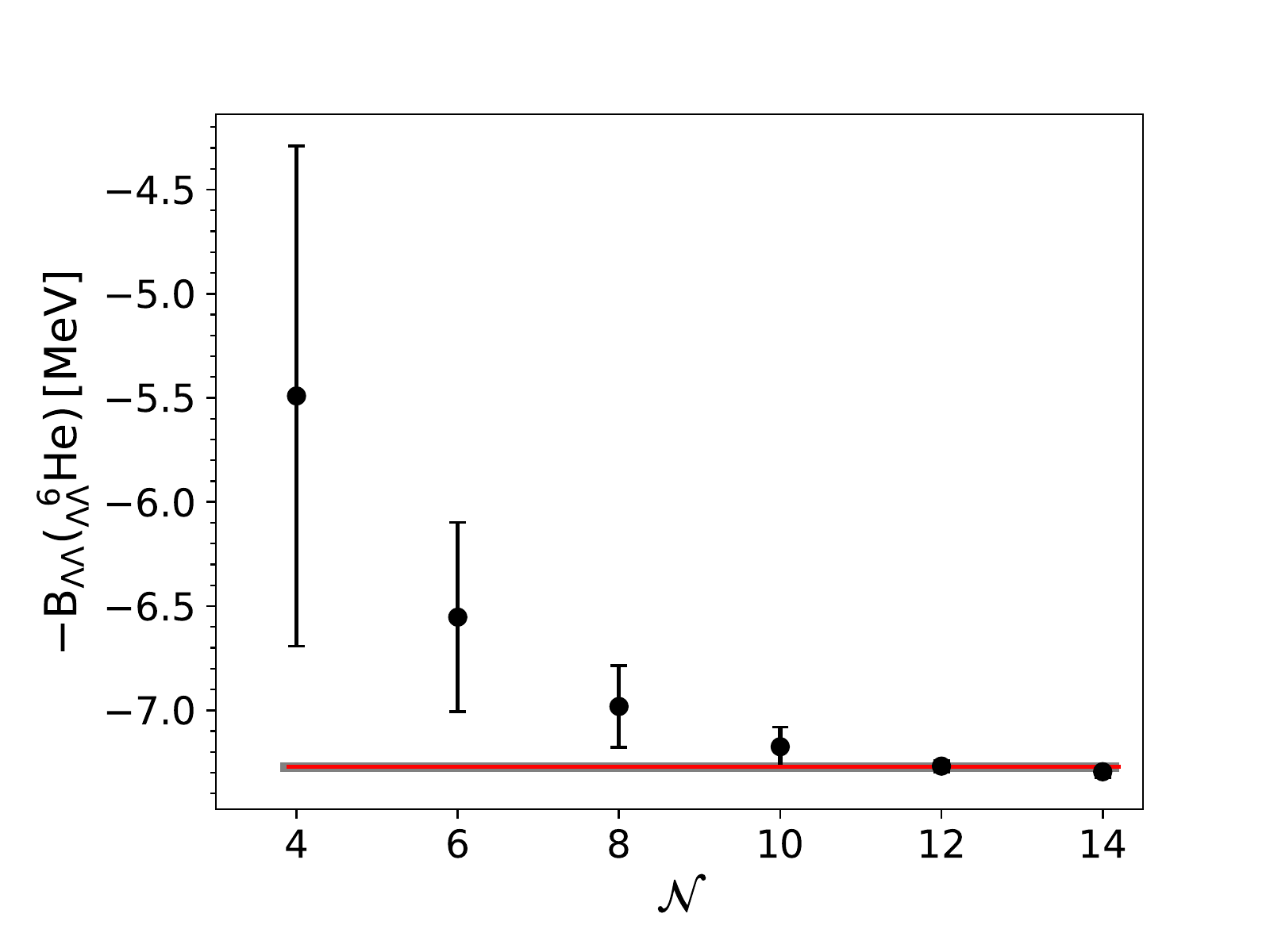}}
      \subfigure[$\Delta B_{\Lambda \Lambda}(^{\text{ }^\text{ } 6}_{\Lambda \Lambda}\text{He})$ as a function of $\mathcal{N}$.]{ \includegraphics[width=0.45\textwidth,trim={0.0cm 0.00cm 0.0cm 0.0cm},clip]{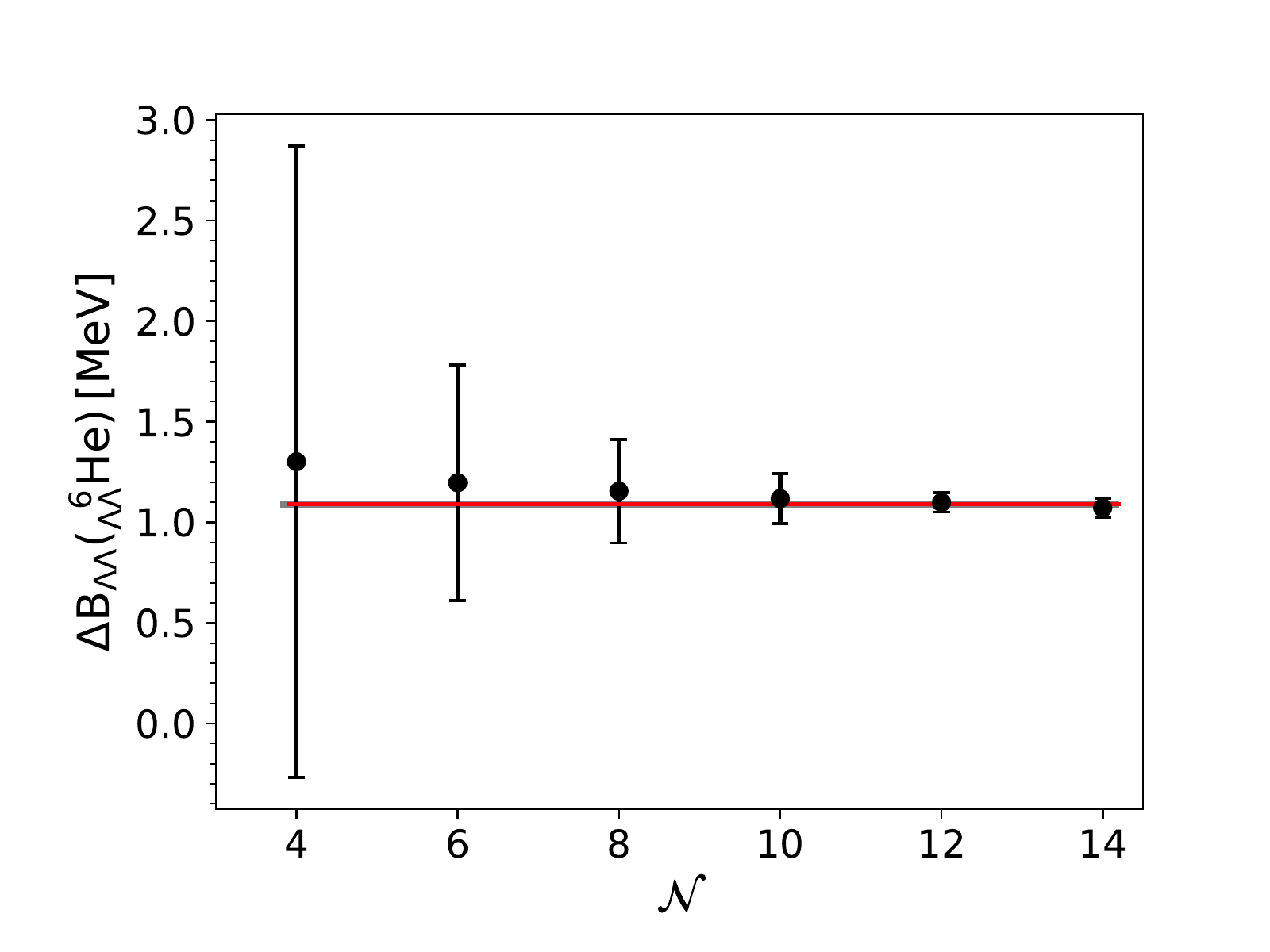}}}
      \end{center}
          \caption{  Binding energy $E$,  $\Lambda \Lambda$-separation energy $B_{\Lambda \Lambda}$  and  separation-energy difference  $\Delta B_{\Lambda \Lambda}$ for    $^{\text{ }\text{ }6}_{\Lambda \Lambda}$He computed using   the YY NLO(600) interaction \cite{Haidenbauer:2018gvg}  that is SRG evolved to a flow parameter of $\lambda_{YY} =1.8 $ fm\textsuperscript{-1}. The SMS N\textsuperscript{4}LO+(450) potential \cite{Reinert:2017usi} with $\lambda_{NN}=1.6$ fm\textsuperscript{-1}  and the NLO19(650) potential 
          \cite{Haidenbauer:2019boi} with $\lambda_{YN} =0.868$ fm\textsuperscript{-1} are employed for the NN and YN interactions, respectively.}
    \label{fig:Convergence_6He2Lanbda}
         \end{figure*}

As mentioned earlier, in order to eliminate the effect of the finite-basis truncation on the binding energies, we follow the
two-step extrapolation procedure   as explained in  \cite{Le:2020zdu}.    The $\omega$- and $\mathcal{N}$-space extrapolations for  $E(^{\text{ }\text{ }\text{ } \text{}6}_{\Lambda \Lambda}\text{He})$ are illustrated in   
 panels (a) and (b) of Fig.~\ref{fig:Convergence_6He2Lanbda}, respectively. Here, for illustration purposes, we  present 
 results for  the NLO potential with  $\lambda_{YY}=1.8$~fm\textsuperscript{-1} but stress that  the  convergence trend  is  similar for all other  values of  $\lambda_{YY}$, and for the LO interaction.   Also,  the  behavior
of  $E(^{\text{ }\text{ }\text{ } \text{}6}_{\Lambda \Lambda}\text{He})$ with respect to $\omega$ and $\mathcal{N}$  resembles that of  the binding  energy of the parent hypernucleus
$^5_{\Lambda}\text{He}$ \cite{Le:2020zdu}.
Furthermore,  panel (b)  clearly  demonstrates a nice convergence pattern  of the binding energy  $E(^{\text{ }\text{ }\text{ } \text{}6}_{\Lambda \Lambda}\text{He})$  computed  for   model spaces up to $\mathcal{N}_{max}=14$.   Likewise,  the  $ \Lambda \Lambda$-separation energy
$B_{\Lambda \Lambda}(^{\text{ }\text{ }\text{ } \text{}6}_{\Lambda \Lambda}\text{He})$, displayed in panel (c),   is also well-converged 
for $\mathcal{N}_{max}=14$ (practically with the same speed    as that of  $E(^{\text{ }\text{ }\text{ } \text{}6}_{\Lambda \Lambda}\text{He})$).  Note that, for   single-$\Lambda$ hypernuclei,
 the  separation energy 
 $B_{\Lambda}$ converges somewhat faster than the individual binding energies. 
 For $S=-2$ systems, we are also interested in  the so-called $\Lambda \Lambda$ excess binding energy
  \begin{align}\label{eq:deltaB2lambda}
  \begin{split}
 \Delta B_{\Lambda \Lambda} (^{\text{ }\text{ } A}_{\Lambda \Lambda} \text{X} ) & =  B_{\Lambda \Lambda}(^{\text{ }\text{ } A}_{\Lambda \Lambda} \text{X} ) - 2 \bar B_{\Lambda}(^{A-1}_{\Lambda}\text{X})\\
 & = 2 \bar E(^{A-1}_{\Lambda}\text{X})  - E(^{\text{ }\text{ }A}_{\Lambda \Lambda} \text{X})  - E(^{A-2}\text{X})
 \end{split}
 \end{align}
which  
 provides   information about the strength of the $\Lambda\Lambda$ interaction. 
 $\bar B$ and $\bar E$  are spin averaged $\Lambda$-separation 
 and binding energies of the hypernuclear core if the core 
 supports several spin states. Cleary, this difference is also 
 affected by the spin-dependent  part of the $\Lambda$-core interaction,  dynamical changes in the core-nucleus  structure as well as   the mass-polarization effect  \cite{Danysz:1963zza,PhysRevC.66.024007}.       For    $^{\text{ }\text{ }\text{ } \text{}6}_{\Lambda \Lambda}$He,  the spin-dependent 
 part of the $\Lambda$-core interaction vanishes because of the spin zero  
 the parent nucleus $^4$He, hence the difference 
  $$\Delta B_{\Lambda \Lambda} (^{\text{ }\text{ }\text{ } \text{}6}_{\Lambda \Lambda} \text{He} ) = B_{\Lambda \Lambda}(^{\text{ }\text{ }\text{ } \text{}6}_{\Lambda \Lambda} \text{He} ) - 2B_{\Lambda}(^{5}_{\Lambda}\text{He}),$$
 will  reflect  the net contributions of the $\Lambda\Lambda$ interactions and the $^4\text{He}$ core-distortion\footnote{Our preliminary results  for the RMS distances of an  NN pair and point-nucleon  radii  in $^{\text{ }\text{ }\text{ } \text{}6}_{\Lambda \Lambda}\text{He}$, $^5_{\Lambda}\text{He}$ and $^4\text{He}$ are very similar to each other which  implies that the distortions of the $^4\text{He}$ core  are rather small. However,  we also note that  Hiyama \etal   in their study for $A=7-10$ double-strangeness systems  using the Gaussian-basis coupled cluster method found that the dynamical changes in the nuclear core structures are quite visible \cite{PhysRevC.66.024007}. Further studies are  necessary in order to clarify the discrepancy.} (polarization) effects. 
  In panel (d),   we exemplify the model-space extrapolation for   $\Delta B_{\Lambda \Lambda}(^{\text{ }\text{ }\text{ } \text{}6}_{\Lambda \Lambda} \text{He} )$.  Interestingly, $\Delta B_{\Lambda \Lambda}$  converges with respect to $\mathcal{N}$  visibly  faster than    both the  $\Lambda\Lambda$-separation    and    the binding energies.
  
\begin{figure*}[tbp]
\begin{subfigure}
  \centering
     \hspace{0.5cm}
  \includegraphics[width=.45\linewidth]{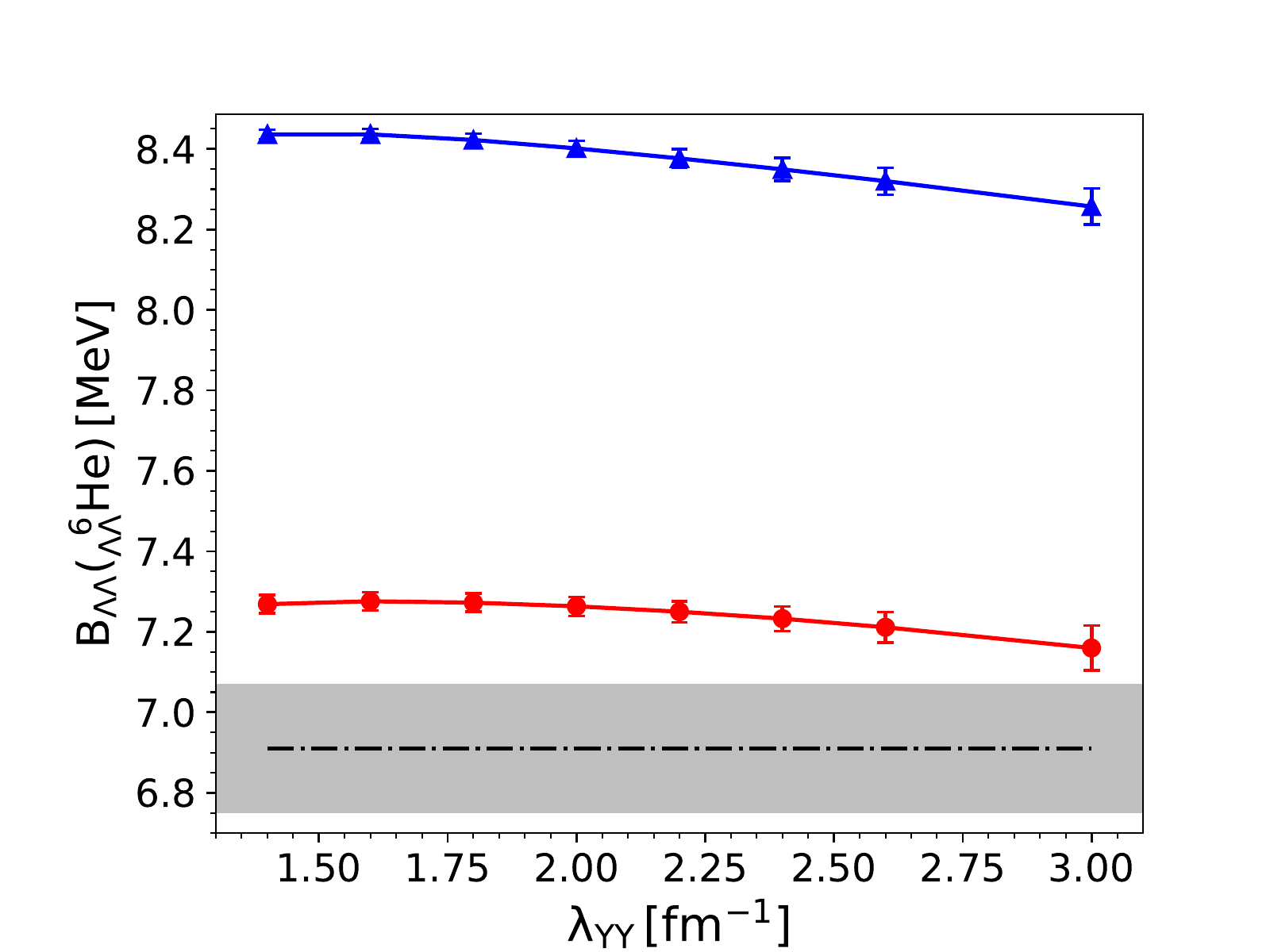}  
\end{subfigure}
\begin{subfigure}
\centering
  \includegraphics[width=.45\linewidth]{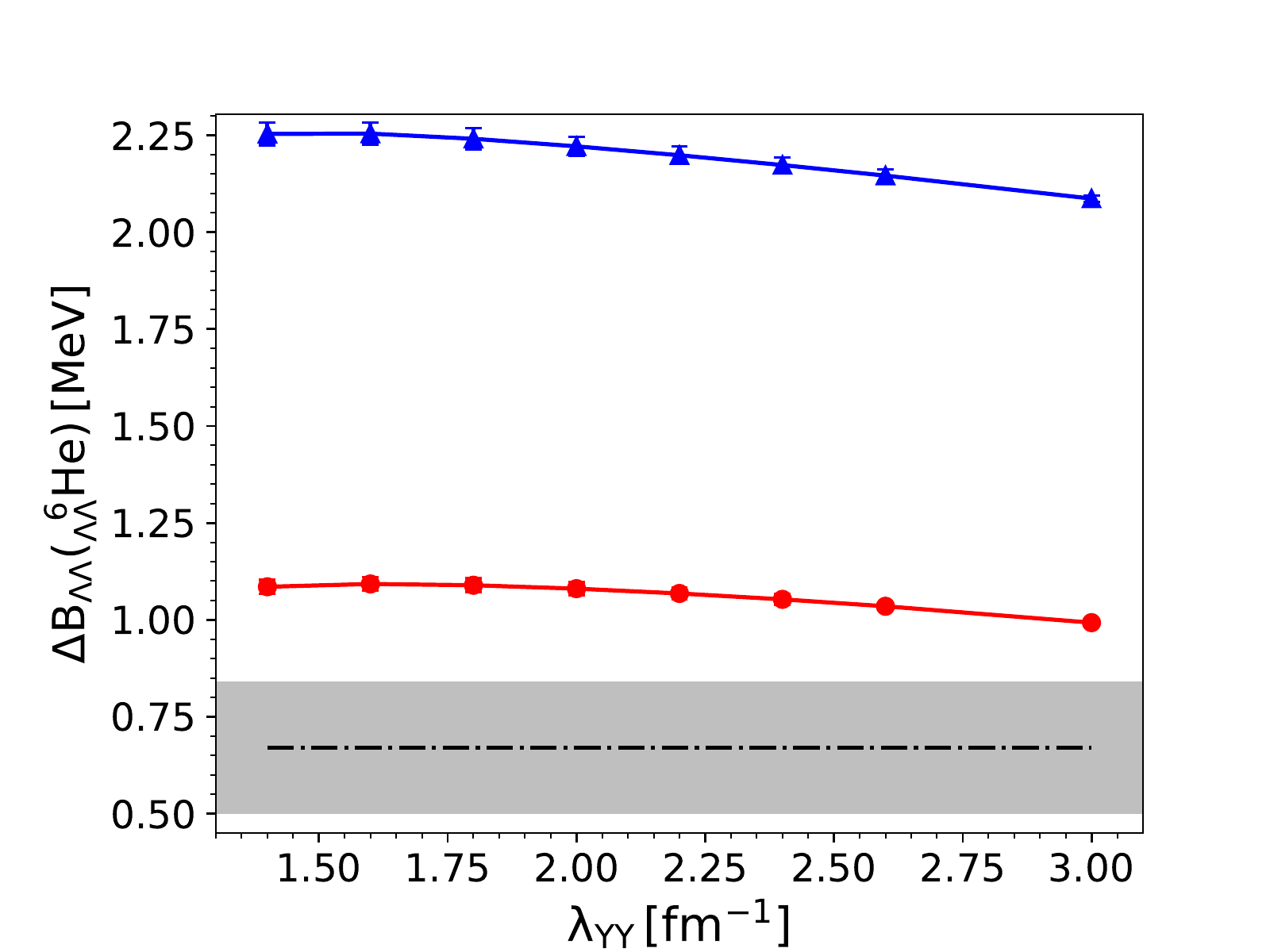}  
\end{subfigure}
  \caption{ $B_{\Lambda \Lambda}(^{\text{ }\text{ }6}_{\Lambda \Lambda} \text{He} )$  (left) and $\Delta B_{\Lambda \Lambda }{(^{\text{ }\text{ } 6}_{\Lambda \Lambda} \text{He} )}$   (right) as functions of the flow parameter $\lambda_{YY}$. Calculations are  based on the YY LO(600)  (blue triangles) and NLO(600) (red circles) potentials. Dash-dotted line with grey band  represents the experimental value and the  uncertainty  of the Nagara event \cite{Nakazawa:2010zza}. Same NN and YN interactions as in Fig.~\ref{fig:Convergence_6He2Lanbda}.}
    \label{fig:compare_6He2Lanbda}
\end{figure*}

 Being able to accurately extract $B_{\Lambda \Lambda}(^{\text{ }\text{ }\text{ } \text{}6}_{\Lambda \Lambda} \text{He} )$  and $\Delta B_{\Lambda \Lambda }{(^{\text{ }\text{ }\text{ } \text{}6}_{\Lambda \Lambda} \text{He} )}$,  we are in a position to study the impact  of the two chiral interactions on these quantities.  The converged results for $B_{\Lambda \Lambda}$  and $\Delta B_{\Lambda \Lambda}$, calculated for a wide range of the SRG flow parameter $\lambda_{YY}$,   are presented in the left and right plots of  Fig.~\ref{fig:compare_6He2Lanbda}, respectively.
  Evidently, the LO YY potential (blue triangles) produces too much attraction (more than 2 MeV as can be seen in 
  the right panel), which, as a  consequence, 
 leads to overbinding  by about $1.5$ MeV   in     $^{\text{ }\text{ }\text{ } \text{}6}_{\Lambda \Lambda} \text{He}$ 
as can be seen in the left panel. On the other hand, the moderately attractive NLO interaction
 predicts a $\Lambda \Lambda$ excess energy of   $\Delta B_{\Lambda \Lambda} \approx 1.1$~MeV, that  is only slightly larger than the empirical  value of  $\Delta B^{exp}_{\Lambda \Lambda}  = 0.67 \pm 0.17 $ MeV \cite{Nakazawa:2010zza,PhysRevC.88.014003}.
For completeness, let us mention that the pertinent
$\Lambda \Lambda$ $^1S_0$ scattering lengths are 
$a=-1.52$~fm (LO \cite{Polinder:2007mp}) and
$a=-0.66$~fm (NLO \cite{Haidenbauer:2015zqb}), respectively. 
 
 It is rather remarkable that both, $B_{\Lambda \Lambda}(^{\text{ }\text{ }\text{ } \text{}6}_{\Lambda \Lambda} \text{He} )$  and $\Delta B_{\Lambda \Lambda }{(^{\text{ }\text{ }\text{ } \text{}6}_{\Lambda \Lambda} \text{He} )}$, exhibit a rather weak dependence on
 the SRG YY parameter $\lambda_{YY}$. With an order of 100 keV, it is at least  one  order of magnitude smaller than the variation of, say,   $B_{\Lambda}(^5_{\Lambda}\text{He})$  with respect to the SRG YN flow parameter $\lambda_{YN}$ \cite{Le:2020zdu}.   The insensitivity of the $\Lambda\Lambda$-separation energy to the SRG evolution  indicates that the SRG-induced  YYN forces are negligibly small.  
 This is probably the result of a rather weak $\Lambda \Lambda$ interaction.

Finally, we  benchmark  the probabilities of finding one $\Sigma$ $(P_{\Lambda \Sigma})$ or two $\Sigma$ $(P_{\Sigma\Sigma})$, or the $\Xi$ hyperon          
 $(P_{\Xi})$ in the ground-state wave function  of $^{\text{ }\text{ }\text{ } \text{}6}_{\Lambda \Lambda} \text{He}$  obtained  for  the two chiral potentials.  Such probabilities   are summarized  in Table \ref{tab:Probability_6He2lambda} for several values of  $\lambda_{YY}$.  Overall,  the $P_{\Lambda \Sigma}$  and $P_{\Sigma  \Sigma}$  probabilities
 are fairly  small, but almost stable with respect to the SRG evolution of the YY interaction. Also, their dependence on the two considered potentials is practically negligible. We remark that  the probability of finding a  $\Sigma$ in $^5_{\Lambda}\text{He}$
 for the employed NN and  YN interactions is also very small, $P_{\Sigma}(^5_{\Lambda}\text{He}) = 0.07 \%$.   In contrast,  $P_{\Xi}$ is more  sensitive to the evolution 
 and also strongly influenced by the interactions. Surprisingly, the updated NLO potential, that yields a  more attractive $\Xi$-nuclear interaction \cite{Haidenbauer:2018gvg}, predicts a considerably 
 smaller $\Xi$ probability (less than 0.2  $\%$ for $\lambda_{YY} =3.0$  fm\textsuperscript{-1}) as compared to the value of $P_{\Xi} =1.1 \%$ obtained for  the LO  at the same  $\lambda_{YY}$.  This  reflects our observation in the $S=-1$ sector that there is no simple one-to-one connection between the probabilities of finding a hyperon particle 
 $(\Sigma, \Xi)$ and the interaction strength. 
 \renewcommand{\arraystretch}{1.7}
 \begin{table}
 \vskip 0.3cm
\begin{center}
   \setlength{\tabcolsep}{0.25cm}
\begin{tabular}{|c|ccc|ccc|}
\cline{1-7}
 $\lambda_{YY} $ & \multicolumn{3}{c|}{NLO(600)}      &\multicolumn{3}{c|}{LO(600)} \\
  fm\textsuperscript{-1}  & $P_{\Lambda\Sigma}$  & $P_{
  \Sigma\Sigma}$  & $P_{\Xi}$ & \multicolumn{1}{l}{$P_{\Lambda \Sigma}$}   & $P_{\Sigma\Sigma}$  & $ P_{\Xi}  $\\
\cline{1-7}
1.4  &  0.13   &  0.11  & {0.02}  &       0.17   & 0.04   & {0.5} \\
2.0  &  0.13   & 0.11   & {0.07}  &      0.17  & 0.05   & {0.84}\\
3.0  & 0.12   & 0.13   & {0.12 }  &    0.18  & 0.08   & { 1.08}\\
\cline{1-7}
\end{tabular}
\end{center}
\caption{Probabilities $(\%)$ of finding a single and  double  $\Sigma$,   and a $\Xi$ hyperons in the ground-state wavefunction of  $^{\text{ }\text{ }\text{ } \text{}6}_{\Lambda \Lambda}\text{He}$.  Note that $P_{\Sigma}(^{5}_{\Lambda}\text{He})=0.07 \%$. }
\label{tab:Probability_6He2lambda}
\end{table}
 
      \begin{figure*}[htbp] 
      \begin{center}
      \hspace{0.3cm}{
      \subfigure[$E_{\mathcal{N}}(^{\text{ }\text{ }5}_{\Lambda \Lambda}\text{He})$ as a function of $\omega$.]{\includegraphics[width=0.45\textwidth,trim={0.0cm 0.00cm 0.0cm 0 cm},clip]{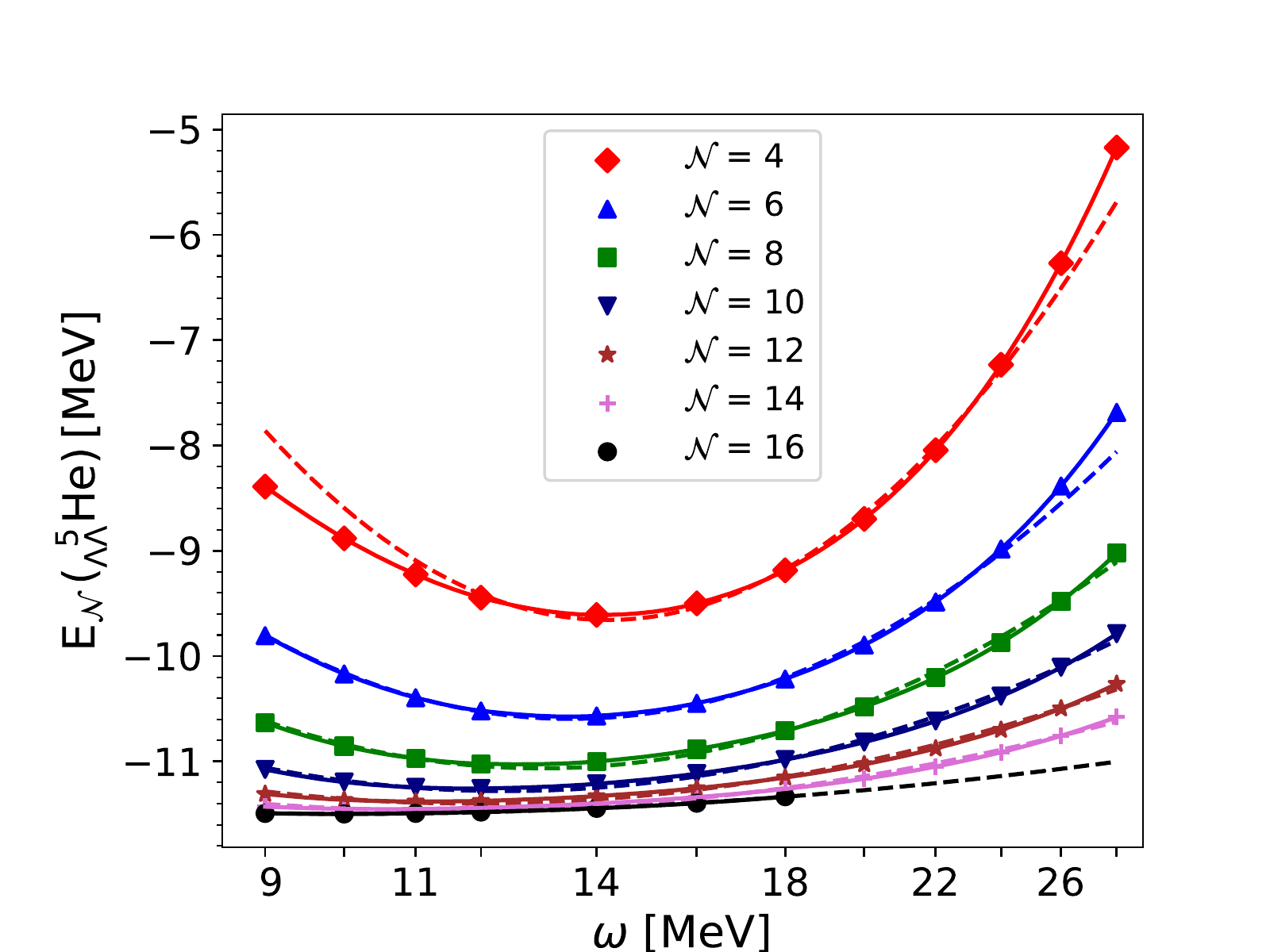}}
      \subfigure[$E(^{\text{ }\text{ } 5}_{\Lambda \Lambda}\text{He})$ as a function of $\mathcal{N}$.]{ \includegraphics[width=0.45\textwidth,trim={0.0cm 0.00cm 0.0cm 0 cm},clip]{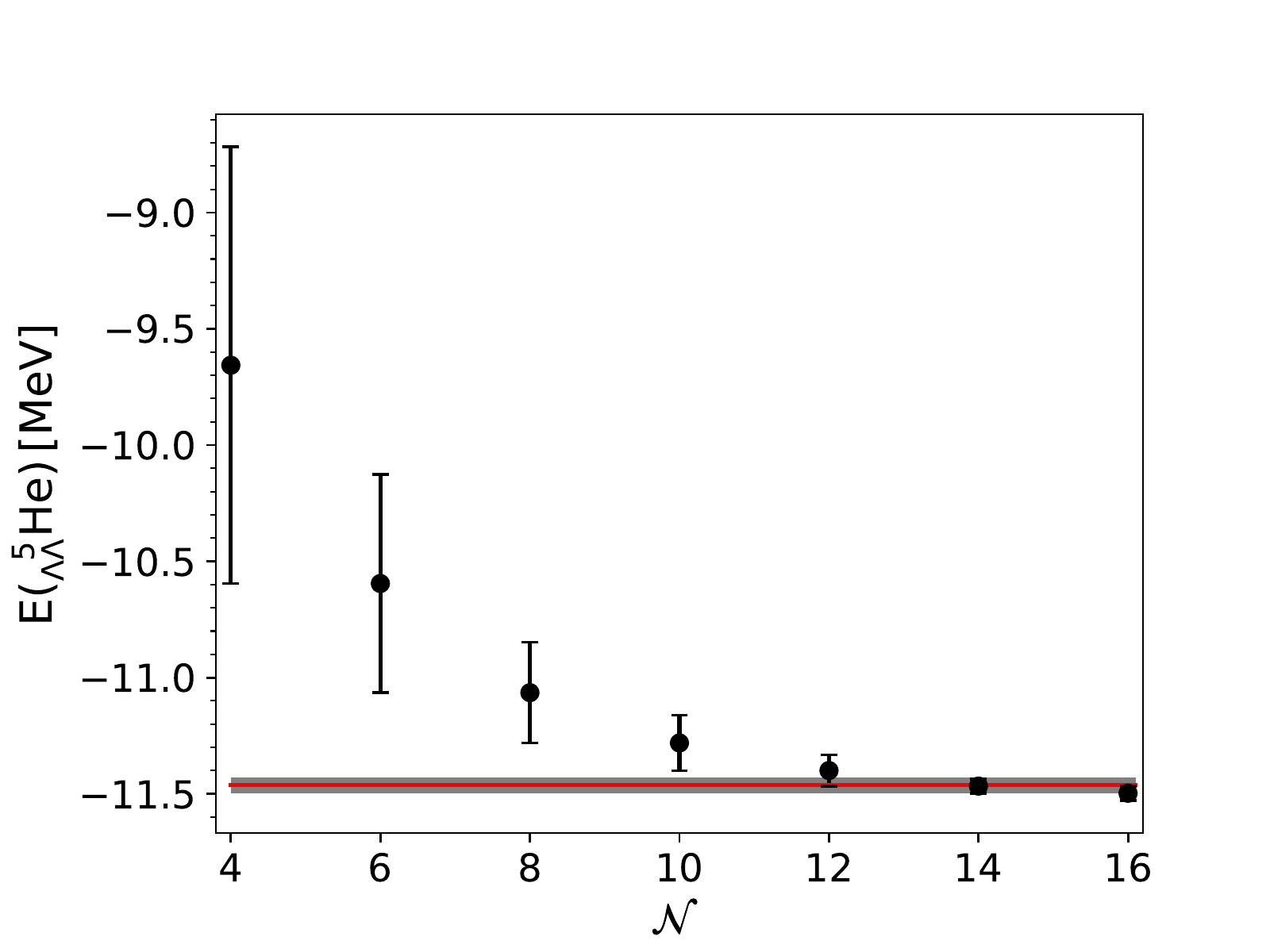}}}\\ 
         \vskip 0.15cm
      \hspace{0.3cm}{\subfigure[$B_{\Lambda \Lambda}(^{\text{ }\text{ }5}_{\Lambda \Lambda}\text{He})$ as a function of $\mathcal{N}$.]{\includegraphics[width=0.45\textwidth,trim={0.0cm 0.00cm 0.0cm 0.0cm},clip] {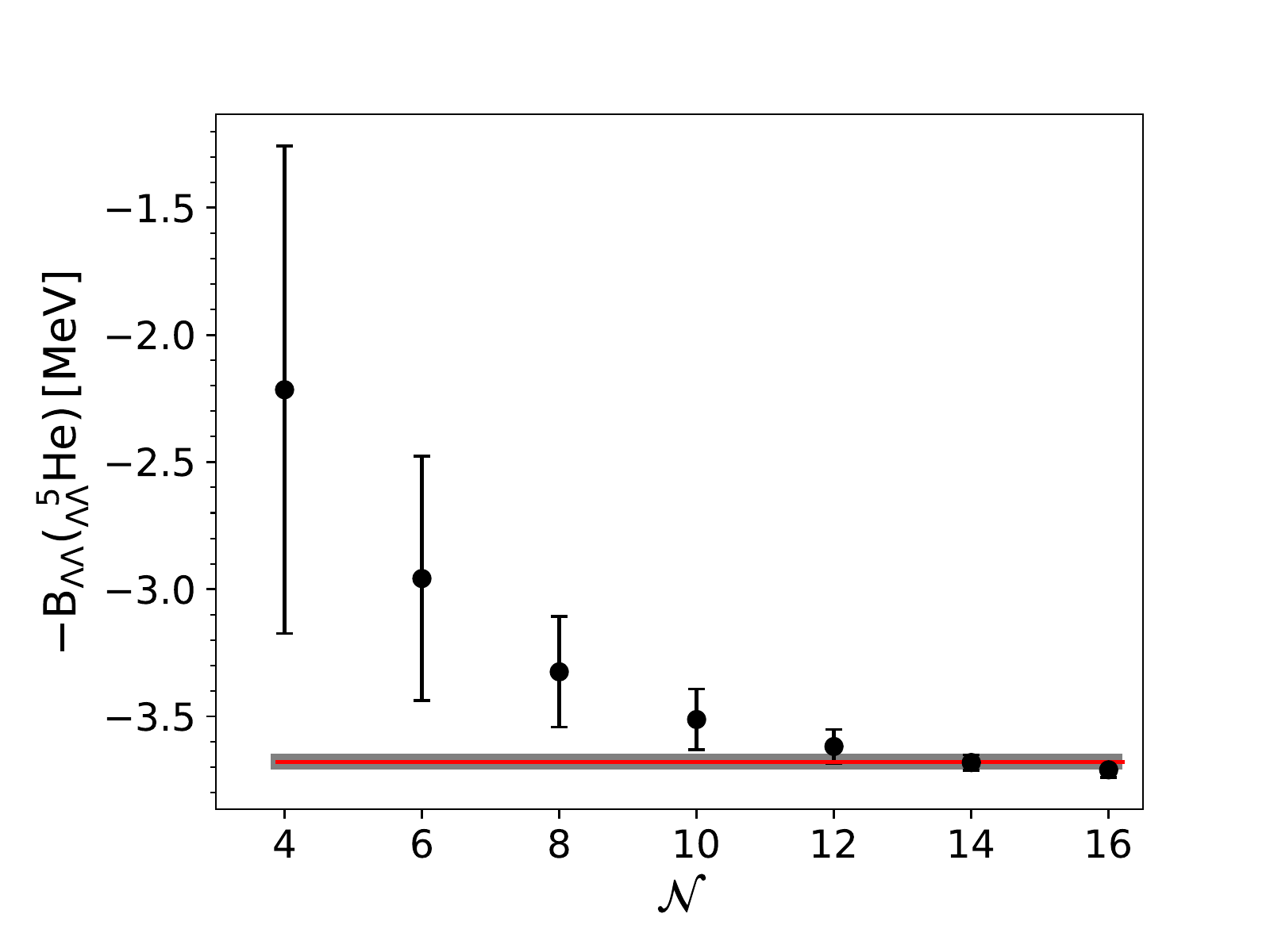}}
      \subfigure[$\Delta B_{\Lambda \Lambda}(^{\text{ }\text{ }\text{ } \text{}5}_{\Lambda \Lambda}\text{He})$ as a function of $\mathcal{N}$.]{ \includegraphics[width=0.45\textwidth,trim={0.0cm 0.00cm 0.0cm 0.0cm},clip]{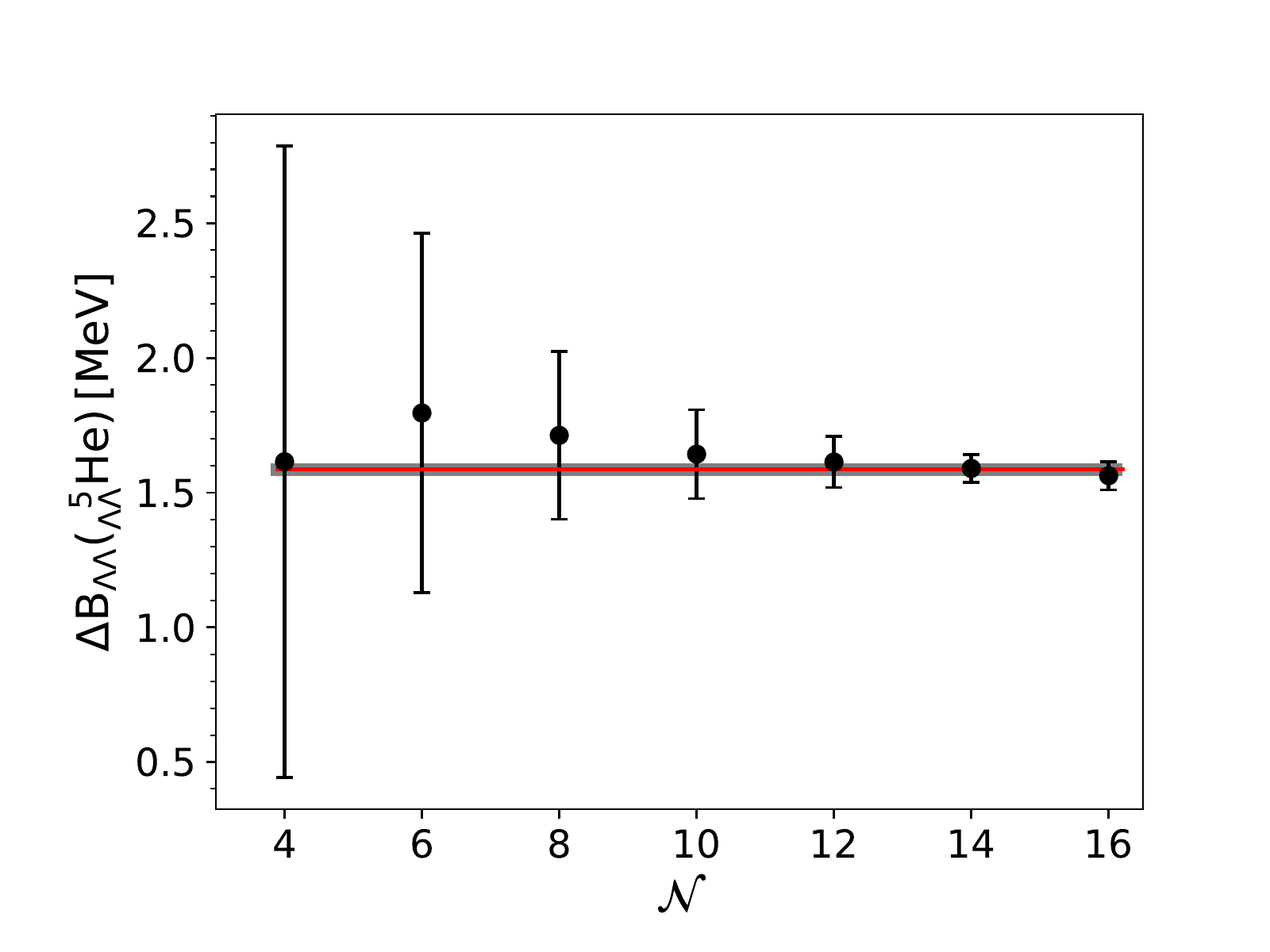}}}
      \end{center}
          \caption{  Binding energy $E$,  $\Lambda \Lambda$-separation energy $B_{\Lambda \Lambda}$  and  separation-energy difference  $\Delta B_{\Lambda \Lambda}$ for    $^{\text{ }\text{ }5}_{\Lambda \Lambda}$He computed using   the YY NLO(600) interaction   that is SRG evolved to a flow parameter of $\lambda_{YY} =1.8 $ fm\textsuperscript{-1}. Same NN and YN interactions as in Fig.~\ref{fig:Convergence_6He2Lanbda}.}
    \label{fig:Convergence_5He2Lanbda}
         \end{figure*}
         
\subsection{ $^{\text{ }\text{ }\text{ } \text{}5}_{\Lambda \Lambda}\text{He}(\frac{1}{2}^+, \frac{1}{2})$}
\label{sec:5He2lambda}
 The next system  that we  investigate is   $^{\text{ }\text{ }\text{ } \text{}5}_{\Lambda \Lambda}\text{He}$. Although the existence of   $^{\text{ }\text{ }\text{ } \text{}5}_{\Lambda \Lambda}\text{He}$  has not been experimentally confirmed yet, most of the  many-body calculations 
 employing    effective  potentials that reproduce the separation energy $B_{\Lambda \Lambda}(^{\text{ }\text{ }6}_{\Lambda \Lambda} \text{He})$  predict a particle-stable bound state of  $^{\text{ }\text{ }\text{ } \text{}5}_{\Lambda \Lambda}\text{He}$   \cite{NakaichiMaeda:1990kr,Nemura:2004xb,Filikhin:2002wm}. However,  there are visible discrepancies among the  values of 
$B_{\Lambda\Lambda}(^{\text{ }\text{ }\text{ } \text{}5}_{\Lambda \Lambda}\text{He})$ predicted by different numerical approaches or different interaction models.  Additionally,  it has been observed in   Faddeev cluster calculations that there is an almost linear correlation between the calculated values of $B_{\Lambda \Lambda}$ for   
the  $^{\text{ }\text{ }\text{ } \text{}5}_{\Lambda \Lambda}\text{He}$ ($^{\text{ }\text{ }\text{ } \text{}5}_{\Lambda \Lambda}\text{H}$) and    $^{\text{ }\text{ }\text{ } \text{}6}_{\Lambda \Lambda}\text{He}$ hypernuclei \cite{Filikhin:2002wm}.  
Such a behavior was also seen in the
study based on pionless EFT \cite{CONTESSI2019134893}. 
It will be of interest  to see whether one observes a similar correlation using other realizations of the chiral interactions. 
However, at this stage, we postpone that question to a future investigation
and focus on the different 
 effects of the LO and NLO potentials  on $B_{\Lambda\Lambda}(^{\text{ }\text{ }\text{ } \text{}5}_{\Lambda \Lambda}\text{He})$ instead.
 
\begin{figure*}[tbp]
\begin{subfigure}
  \centering
     \hspace{0.5cm}
  \includegraphics[width=.45\linewidth]{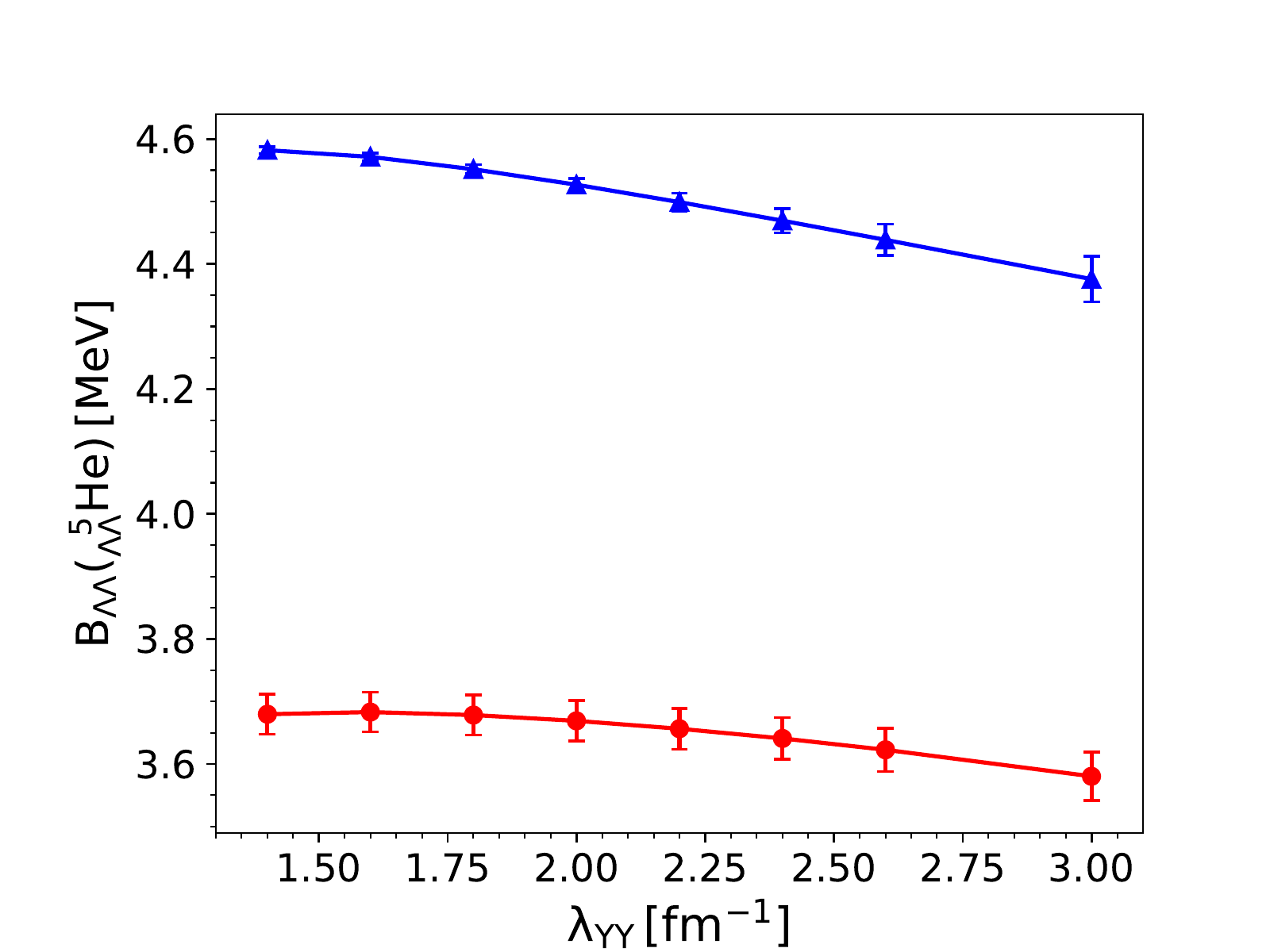}  
\end{subfigure}
\begin{subfigure}
\centering
  \includegraphics[width=.45\linewidth]{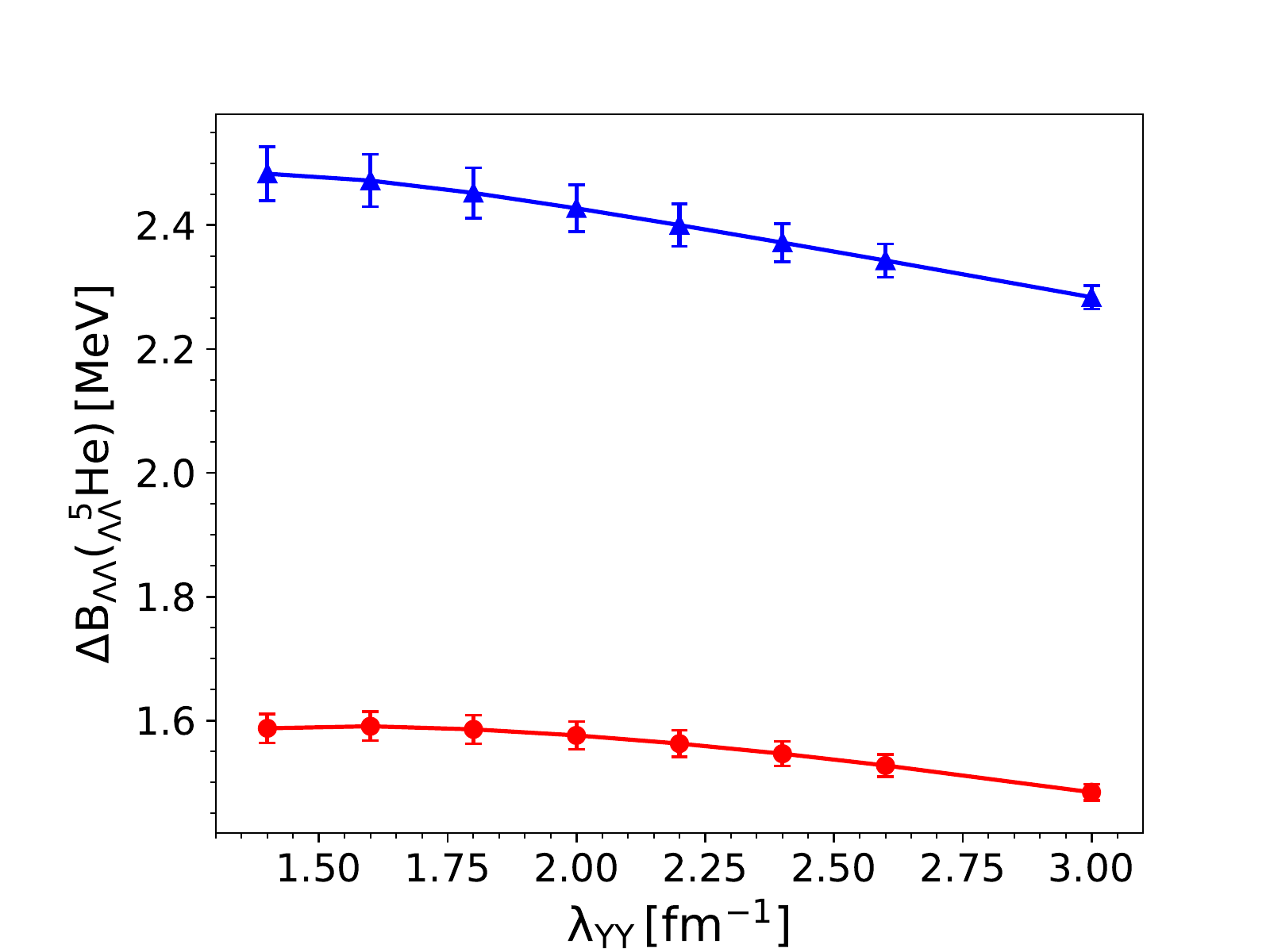}  
\end{subfigure}
   \caption{ $B_{\Lambda \Lambda}(^{\text{ }\text{ } 5}_{\Lambda \Lambda} \text{He} )$  (left) and $\Delta B_{\Lambda \Lambda }{(^{\text{ }\text{ } 5}_{\Lambda \Lambda} \text{He} )}$   (right) as functions of the flow parameter $\lambda_{YY}$. Calculations are  based on the YY LO(600) (blue triangles) and NLO(600) (red circles)
   potentials.  Same NN and YN interactions as in  Fig.~\ref{fig:Convergence_6He2Lanbda}.}
    \label{fig:Compare_5He2Lanbda}
\end{figure*}
 The $\omega$- and $\mathcal{N}$-extrapolation of the binding energy $E$, $\Lambda\Lambda$-separation energy $B_{\Lambda\Lambda}$ and the separation-energy difference  $\Delta B_{\Lambda \Lambda}$ of  
  $^{\text{ }\text{ }\text{ } \text{}5}_{\Lambda \Lambda}\text{He}$   are illustrated in  Fig.~\ref{fig:Convergence_5He2Lanbda}. 
           Here, the results are shown   for the   NLO potential with a flow parameter of $\lambda_{YY}=1.8$
  fm\textsuperscript{-1}  and for  model spaces up to $\mathcal{N}_{max} =16$.  Note that in the case of $^4_{\Lambda}\text{He}$, the  energy calculations were performed for model spaces up to $\mathcal{N}_{max}(^4_{\Lambda}\text{He})=22$ in order to achieve a good convergence.  Calculations with such large model spaces are currently  not feasible for  $^{\text{ }\text{ }\text{ } \text{}5}_{\Lambda \Lambda}\text{He}$  because of computer-memory constraints.   Nonetheless,  the illustrative results in Fig.~\ref{fig:Convergence_5He2Lanbda}  clearly indicate that  well-converged results  are achieved for 
  this double-$\Lambda$ hypernucleus already for   model spaces  up to  $\mathcal{N}_{max}=16$. Moreover, the employed  two-step  extrapolation procedure 
   also allows
  for a reliable estimate  of the truncation  uncertainty.  Let us  further remark that,  when calculating the  excess energy   
  \begin{align}\label{eq:spinave}
  \Delta B_{\Lambda\Lambda}(^{\text{ }\text{ }\text{ } \text{}5}_{\Lambda \Lambda}\text{He}) = B_{\Lambda\Lambda}(^{\text{ }\text{ }\text{ } \text{}5}_{\Lambda \Lambda}\text{He}) -2 {B}_{\Lambda}(^4_{\Lambda}\text{He}) \ ,
  \end{align}
   we do not simply assign  the  ground-state $\Lambda$-separation energy $B_{\Lambda}(^4_{\Lambda}\text{He}, 0^+)$  to  $B_{\Lambda}(^4_{\Lambda}\text{He})$  but  rather use a  \break  spin-averaged value      $\overline{B}_{\Lambda}(^4_{\Lambda}\text{He}) $ of  the  ground-state doublet 
 \cite{PhysRevC.66.024007}
 \begin{align}
  \overline{B}_{\Lambda}(^4_{\Lambda}\text{He}) = \frac{1}{4}  B_{\Lambda}(^4_{\Lambda}\text{He}, 0^+)  + \frac{3}{4}  B_{\Lambda}(^4_{\Lambda}\text{He}, 1^+).
    \end{align} 
   By doing so, the computed quantity $ \Delta B_{\Lambda\Lambda}(^{\text{ }\text{ }\text{ } \text{}5}_{\Lambda \Lambda}\text{He}) $  will be less dependent on the  spin-dependence effect of the $\Lambda$-core interactions, and, therefore,    can be used as a measure of the $\Lambda\Lambda$ interaction strength, provided that the nuclear contraction  effects are small.
  The results for 
         $ B_{\Lambda\Lambda}(^{\text{ }\text{ }\text{ } \text{}5}_{\Lambda \Lambda}\text{He}) $  and   $ \Delta B_{\Lambda\Lambda}(^{\text{ }\text{ }\text{ } \text{}5}_{\Lambda \Lambda}\text{He}) $ calculated for the two interactions and a wide 
         range of flow parameter, $1.4 \leq \lambda_{YY} \leq 3.0$ fm\textsuperscript{-1}  are shown in  Fig.~\ref{fig:Compare_5He2Lanbda}.
Overall, we observe  a very weak dependence of these two quantities on the SRG flow parameter, like for 
$ ^{\text{ }\text{ } 6}_{\Lambda \Lambda}\text{He}$, reinforcing the insignificance of SRG-induced YYN forces. Again, the LO interaction predicts a much larger $\Lambda\Lambda$-separation energy 
and a more significant $\Lambda \Lambda$ interaction strength than 
the one at NLO. In either case, the $\Lambda \Lambda$ excess energy $\Delta B_{\Lambda \Lambda}$ computed for $^{\text{ }\text{ }\text{ } \text{}5}_{\Lambda \Lambda}\text{He}$,   slightly exceeds  the corresponding  one for $ ^{\text{ }\text{ }\text{ } \text{}6}_{\Lambda \Lambda}\text{He}$, by about 0.23 and 0.5~MeV for the LO and NLO interactions, respectively. The main deviations should  come from the nuclear-core distortion and the suppression of the $\Lambda \Lambda -\Xi N$ coupling in $ ^{\text{ }\text{ }\text{ } \text{}6}_{\Lambda \Lambda}\text{He}$
as discussed in \cite{PhysRevC.49.R1768,Myint:2002dp,PhysRevC.68.024002}. However, it is necessary to carefully study the impact of the employed interactions on the results  before a final conclusion can be drawn.
We further note that Filikhin and Gal \cite{Filikhin:2002wm} in their Faddeev cluster calculations, based on potentials that
simulate the low-energy $s$-wave scattering parameters of 
some Nijmegen interaction models, obtained 
  an opposite relation, namely  $\Delta B_{\Lambda\Lambda}(^{\text{ }\text{ }\text{ } \text{}5}_{\Lambda \Lambda}\text{He} ) <  \Delta B_{\Lambda\Lambda}(^{\text{ }\text{ }\text{ } \text{}6}_{\Lambda \Lambda}\text{He})$. As a consequence, our results 
  do  also not fit into the correlation of $\Delta B_{\Lambda\Lambda}(^{\text{ }\text{ }\text{ } \text{}5}_{\Lambda \Lambda}\text{He} )$  and  \break $\Delta B_{\Lambda\Lambda}(^{\text{ }\text{ }\text{ } \text{}6}_{\Lambda \Lambda}\text{He})$ shown in the same work. We will need to study more interactions in future to 
  understand whether such a correlation can also be established using chiral 
  interactions. 

  It is also very interesting to point out that the $\Lambda\Lambda$-separation energies $B_{\Lambda\Lambda}$ for both $^{\text{ }\text{ }\text{ } \text{}5}_{\Lambda \Lambda}\text{He}$ and $^{\text{ }\text{ }\text{ } \text{}6}_{\Lambda \Lambda}\text{He}$ predicted by the NLO potential are surprisingly close to the results obtained  by Nemura et al.,  $B_{\Lambda\Lambda}(^{\text{ }\text{ }\text{ } \text{}5}_{\Lambda \Lambda}\text{He}) = 3.66$~MeV, $B_{\Lambda\Lambda}(^{\text{ }\text{ }\text{ } \text{}6}_{\Lambda \Lambda}\text{He}) =  7.54 $ MeV, using the modified Nijmegen YY potential (mND\textsubscript{s}) \cite{Nemura:2004xb}.  
 \renewcommand{\arraystretch}{1.7}
 \begin{table}
 \vskip 0.5cm
\begin{center}
   \setlength{\tabcolsep}{0.25cm}
\begin{tabular}{|c|ccc|ccc|}
\cline{1-7}
 $\lambda_{YY} $ & \multicolumn{3}{c|}{YY-NLO(600)}      &\multicolumn{3}{c|}{YY-LO(600)} \\
  fm\textsuperscript{-1}  & $P_{\Lambda \Sigma}$  & $P_{
  \Sigma\Sigma}$  & $P_{\Xi}$ & \multicolumn{1}{l}{$P_{\Lambda \Sigma}$}   & $P_{\Sigma\Sigma}$  & $ P_{\Xi}  $\\
  \cline{1-7}
1.4  &  0.61   &  0.07   & 0.4  &       0.53   & 0.02   & 1.25 \\
2.0  &  0.6     & 0.08   & 0.38  &      0.51  & 0.03   & 1.36\\
3.0  & 0.57    & 0.08   & 0.23   &    0.51   & 0.05   & 1.35\\
\cline{1-7}
\end{tabular}
\end{center}
\caption{Probabilities (in percentage) of finding  a  $\Sigma$ $(P_{\Lambda \Sigma})$, double $\Sigma$  $(P_{\Sigma \Sigma})$and a $\Xi$ $(P_{\Xi})$ hyperons in $^{\text{ }\text{ }\text{ } \text{}5}_{\Lambda \Lambda}\text{He}$.  $P_{\Sigma}(^4_{\Lambda}\text{He}) = 0.43\,\%. $ }
\label{tab:Probability_5He2lambda}
\end{table}
Finally,  we  provide in  Table~\ref{tab:Probability_5He2lambda}  the probabilities of finding a $\Sigma$  ($P_{\Lambda \Sigma}$), double $\Sigma $ $  (P_{\Sigma \Sigma})$, or a $\Xi$ $(P_{\Xi})$ in the  $^{\text{ }\text{ }\text{ } \text{}5}_{\Lambda \Lambda} \text{He}$ ground-state wave function, computed with  the two   potentials    and    several SRG values, $\lambda_{YY}=1.4, 2.0 $ and 3.0 fm\textsuperscript{-1}. Apparently, all the probabilities including also  $P_{\Xi}$ exhibit a rather weak  sensitivity to the flow parameter $\lambda_{YY}$. The two interactions seem to  have little impact  on the $\Sigma$-probabilities ($P_{ \Lambda\Sigma}$ and $P_{\Sigma \Sigma}$) but   strongly influence $P_{ \Xi}$. Like in
 the  $^{\text{ }\text{ }\text{ } \text{}6}_{\Lambda \Lambda} \text{He}$  system, here,   the LO potential  yields considerably larger $\Xi$-probabilities  as compared to the values  predicted by the NLO
 interaction.  It also clearly sticks out  from 
 Tables~\ref{tab:Probability_6He2lambda} and \ref{tab:Probability_5He2lambda}    that the probabilities of finding a  $\Sigma$ or $\Xi$  hyperon in 
  $^{\text{ }\text{ }\text{ } \text{}5}_{\Lambda \Lambda} \text{He}$  are visibly  larger than the corresponding ones  in  $^{\text{ }\text{ }6}_{\Lambda \Lambda} \text{He}$. This is indeed consistent with the 
  $ \Sigma$-probabilities in the ground-state wave functions of  their parent   hypernuclei   (e.g., $P_{\Sigma}(^4_{\Lambda}\text{He} ) =0.43\,\%$ and $P_{\Sigma}(^5_{\Lambda}\text{He} ) =0.07\, \% $), and more importantly, is consistent with the 
  suppression of particle conversions such as $\Lambda\Lambda - \Xi N$ in p-shell hypernuclei \cite{PhysRevC.49.R1768}.  
   
          \begin{figure*}[htbp] 
      \begin{center}
      \hspace{0.3cm}{
      \subfigure[$E_{\mathcal{N}}(^{\text{ }\text{ }4}_{\Lambda \Lambda}\text{H})$ as a function of $\omega$.]{\includegraphics[width=0.45\textwidth,trim={0.0cm 0.00cm 0.0cm 0 cm},clip]{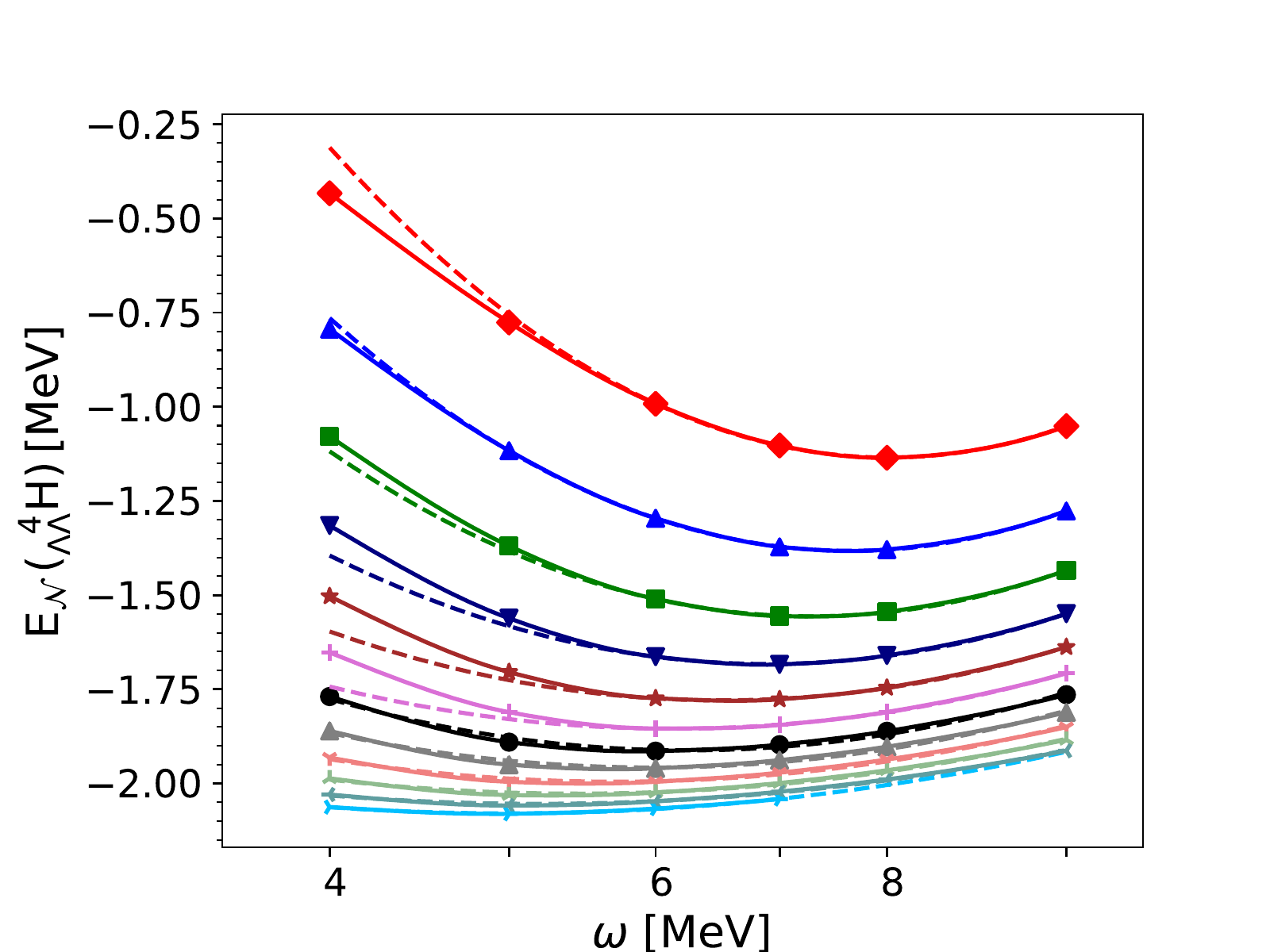}}
      \subfigure[$E(^{\text{ }\text{ } 4}_{\Lambda \Lambda}\text{H})$ as a  function of  $\mathcal{N}$.]{ \includegraphics[width=0.45\textwidth,trim={0.0cm 0.00cm 0.0cm 0 cm},clip]{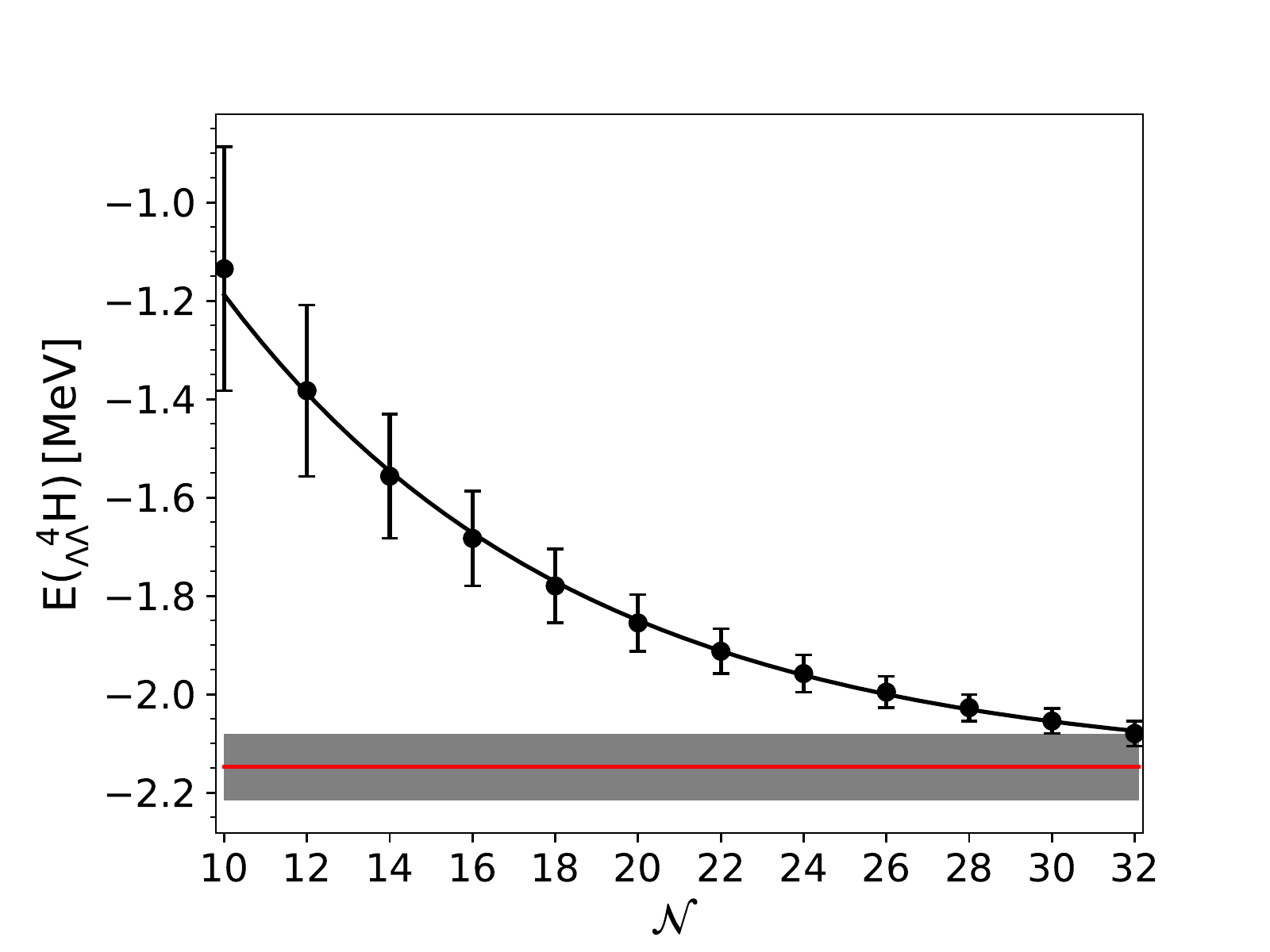}}}\\ 
         \vskip 0.15cm
      \hspace{0.3cm}{\subfigure[$E(^3_{\Lambda }\text{H})$ as a function of $\mathcal{N}$.]{\includegraphics[width=0.45\textwidth,trim={0.0cm 0.00cm 0.0cm 0.0cm},clip] {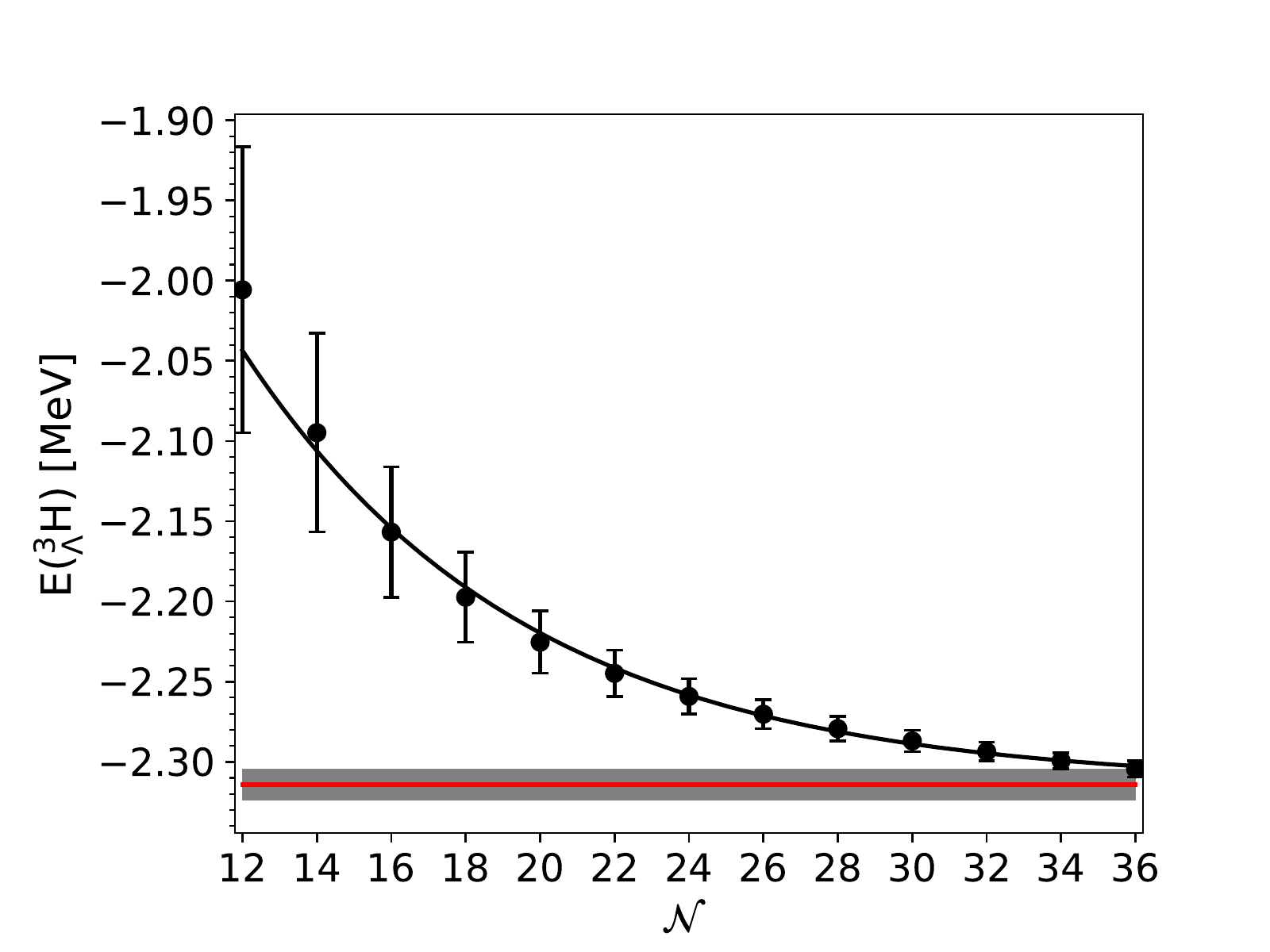}}
      \subfigure[$E(^{\text{ }\text{ } 4}_{\Lambda \Lambda}\text{H})$ as a  function SRG flow parameter $\lambda_{YY}$.]{ \includegraphics[width=0.45\textwidth,trim={0.0cm 0.00cm 0.0cm 0.0cm},clip]{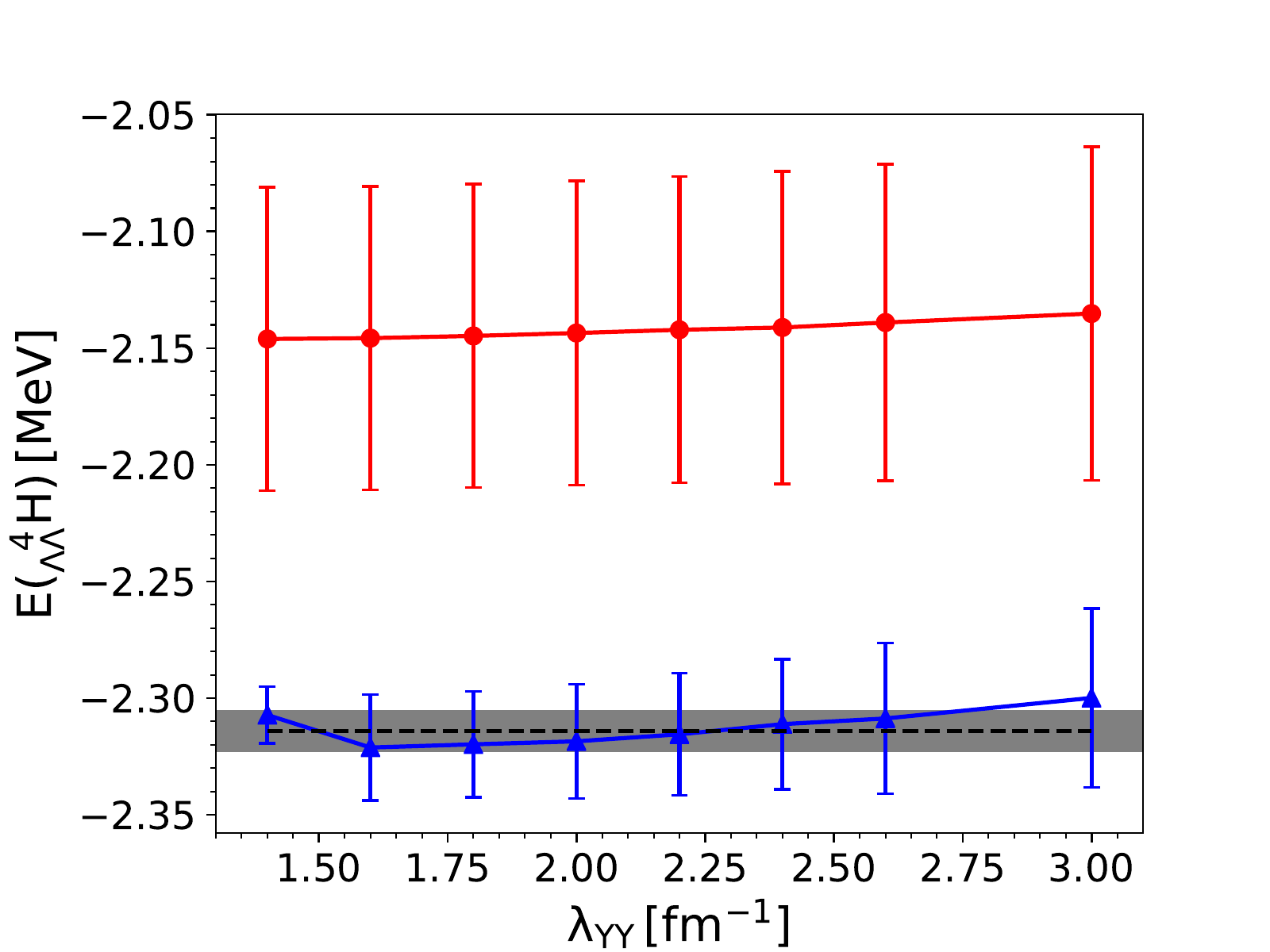}}}
      \end{center}
                 \caption{ (a):  Ground-state energies of  $^{\text{ }\text{ } 4}_{\Lambda \Lambda}\text{He}$ as functions of $\omega$ for model space $\mathcal{N}=10 - 32$. Calculations are performed with the YY NLO(600) potential  evolved to a flow parameter of $\lambda_{YY} =1.8$ fm\textsuperscript{-1}. (b): model space extrapolation of $E(^{\text{ }\text{ }4}_{\Lambda \Lambda}\text{H})$  with the same YY interaction as in (a).  (c): model space extrapolation of $E(^3_{\Lambda}\text{H})$.
                 (d): Converged  $E(^{\text{ }\text{ } 4}_{\Lambda \Lambda}\text{H})$ as functions of flow parameter  for  the  LO(600) (blue triangles)  and  NLO(600)
                 (red circles) potentials.  The dashed line with grey band represents the computed $E(^3_{\Lambda}\text{H})$
                  and  the theoretical  uncertainty, respectively.   Same NN and YN interactions as in  Fig.~\ref{fig:Convergence_6He2Lanbda} }
    \label{fig:Convergence_4H2Lanbda}
         \end{figure*}
         
 \subsection{$^{\text{ }\text{ }\text{ }\text{}4}_{\Lambda \Lambda}\text{H}(1^+, 0)$}
\label{sec:4H2lambda}
Our final exploratory  s-shell hypernucleus is 
$^{\text{ }\text{ }\text{ } \text{}4}_{\Lambda \Lambda}\text{H}$.
This system has been the subject of many theoretical and experimental studies. It turned out
that  theoretical predictions of the  stability of   $^{\text{ }\text{ }\text{ } \text{}4}_{\Lambda \Lambda}\text{H}$   against   the $^3_{\Lambda}\text{H} + \Lambda $ breakup  are very sensitive to the   interpretations  of \\ double-strangeness   hypernuclear  data,  in particular,  the $^{\text{ }\text{ }\text{ } \text{}6}_{\Lambda \Lambda}\text{He}$ hypernucleus   \cite{NakaichiMaeda:1990kr}.  Indeed,
 Nemura \etal  \cite{Nemura:2004xb}  
 observed     a 
 particle-stable  but  loosely  bound   state  of  $^{\text{ }\text{ }\text{ } \text{}4}_{\Lambda \Lambda}\text{H}$ (just only about 2 keV below the $^3_{\Lambda}\text{H} + \Lambda$
 threshold for the mND\textsubscript{s} potential)
 using the 
fully coupled-channel  stochastic variational method in combination with  effective YY potentials that are  fitted to reproduce 
the initially extracted   value  of  $B_{\Lambda \Lambda}(^{\text{ }\text{ }\text{ } \text{}6}_{\Lambda \Lambda}\text{He}) = 7.25 \pm 0.19 $ MeV  \cite{PhysRevLett.87.212502}. 
 The study by Filikhin and Gal \cite{PhysRevLett.89.172502} indicated, however, that there is a sizable model dependence. The authors found 
 no bound state within an exact four-body (Faddeev-Yakubovsky)
 calculation for the $\Lambda  \Lambda p n $ system, but a 
 particle-stable  $^{\text{ }\text{ }\text{ } \text{}4}_{\Lambda \Lambda}\text{H}$ hypernucleus  
 when solving the (three-body) Faddeev equation for the 
 $\Lambda \Lambda d$ cluster system.
 A more recent calculation by Contessi \etal \cite{CONTESSI2019134893}, based on the pionless EFT interaction at LO, showed that the existence of a bound state in $^{\text{ }\text{ }\text{ } \text{}4}_{\Lambda \Lambda}\text{H}$ is not compatible with the corrected value of 
 $B_{\Lambda \Lambda}(^{\text{ }\text{ }\text{ } \text{}6}_{\Lambda \Lambda}\text{He}) = 6.91 \pm 0.16 $ MeV.
  Although  the  observation of $^{\text{ }\text{ }\text{ } \text{}4}_{\Lambda \Lambda}\text{H}$ was reported in an  experiment  at 
   BNL \cite{PhysRevLett.87.132504}, it     has been recently invalidated by  a thorough re-examination of the recorded  events \cite{PhysRevC.76.064308}.  
   Nevertheless, the existence of a stable  $^{\text{ }\text{ }\text{ } \text{}4}_{\Lambda \Lambda}\text{H}$ hypernucleus cannot be completely ruled out  and the search for its experimental 
   confirmation or exclusion is still ongoing. 
   
In view of the previous calculations, it is interesting to see whether the chiral YY potential at NLO, that predicts similar results for $A=5-6$ $\Lambda\Lambda$ hypernuclei as the mND\textsubscript{s} interaction
\cite{Nemura:2004xb}, also results in a loosely bound state for $^{\text{ }\text{ }\text{ } \text{}4}_{\Lambda \Lambda}\text{H}$. 
  It is well-known that NCSM calculations for very loosely bound systems like the hypertriton converge very slowly. Hence, in 
  order to unambiguously answer that question,
  converged results for the binding energy of the parent $^3_\Lambda$H and the ground-state energy of $^{\text{ }\text{ }\text{ } \text{}4}_{\Lambda \Lambda}\text{H}$ are crucial. In panels (a) and (b) of Fig.~\ref{fig:Convergence_4H2Lanbda},   we examine  the convergence of   $E(^{\text{ }\text{ }\text{ } \text{}4}_{\Lambda \Lambda}\text{H})$ in 
$\omega$- and $\mathcal{N}$-space, respectively, using   model spaces  to $\mathcal{N}_{max} =32$. 
          The results are shown 
 for the  NLO(600) potential  with   a flow
parameter of $\lambda_{YY}=2.4 $ fm\textsuperscript{-1}.  For a better comparison,  the  $\mathcal{N}$-space extrapolation  of
$E(^3_{\Lambda}\text{H})$  computed with  model spaces up to $\mathcal{N}=36$ is also  presented in panel~(c).   As  expected, due to the weak binding of the hypertriton, the binding energy calculations for both hypernuclei,  $^{\text{ }\text{ }\text{ } \text{}4}_{\Lambda \Lambda}\text{H}$ and $^3_{\Lambda}\text{H}$,   converge very slowly when using HO  bases.  
 It also clearly sticks  out that the optimal HO frequencies  $\omega$  for large model space sizes  are  around $\omega_{opt} \approx 6$ MeV  which  is much smaller than the value   of  $\omega_{opt} \approx 16$ MeV for the  $A=4,5$  systems. This again reflects the large spatial extension of the wave functions of 
 $^{\text{ }\text{ }\text{ } \text{}4}_{\Lambda \Lambda}\text{H}$  and $^3_{\Lambda}\text{H}$.  Nevertheless, one can still   observe a slightly faster convergence speed for 
   $E(^{\text{ }\text{ }\text{ } \text{}4}_{\Lambda \Lambda}\text{H})$ (especially with the LO potential) as for  $E(^3_{\Lambda}\text{H})$.   
  Moreover,  our extrapolated value of $E(^3_{\Lambda}\text{H}) = -2.314 \pm 0.009 $ MeV  (for model space  up to $\mathcal{N}=36$) agrees within 10 keV  with the exact 
 Faddeev result $ E_{\rm Fa}(^3_{\Lambda}\text{H}) = -2.333 \pm 0.002 $ MeV  \cite{LE2020135189}.  We  conclude that a model space truncation of   $\mathcal{N}_{max} =32$  for 
the energy calculations in  $^{\text{ }\text{ }\text{ } \text{}4}_{\Lambda \Lambda}\text{H}$   should  be sufficient in order to draw  conclusions  about the 
 stability   of the system against  $\Lambda$ emission.    

 The extrapolated ground-state
  energies $E(^{\text{ }\text{ }\text{ } \text{}4}_{\Lambda \Lambda}\text{H})$ for the NLO (red circles) and LO (blue triangles) potentials evolved to a wide range of flow parameters are displayed in panel (c) of Fig.~\ref{fig:Convergence_4H2Lanbda}.  Here, the dashed black line together with the grey band represent the computed  $E(^3_{\Lambda}\text{H})$ and the 
  estimated uncertainty. Calculations with the NLO  potential seem to converge  more slowly than the ones for the LO interaction.  
  The NLO potential  clearly leads to an unbound    $^{\text{ }\text{ }\text{ } \text{}4}_{\Lambda \Lambda}\text{H}$ hypernucleus. Although our results for 
  $A=5$ and 6 are similar to the ones of Ref.~\cite{Nemura:2004xb}, our results 
  for $A=4$ do not support the existence of a bound  $^{\text{ }\text{ }\text{ } \text{}4}_{\Lambda \Lambda}\text{H}$ state.
 The LO results for  $^{\text{ }\text{ }\text{ } \text{}4}_{\Lambda \Lambda}\text{H}$ likely hint at a particle-unstable
  system with respect to the hypertriton $^3_{\Lambda}$H. 
Admittedly, in order to draw a definite conclusion on the actual
situation, the uncertainties of the calculation would have to be reduced. 
However, since the LO interaction considerably overbinds $^{\text{ }\text{ }\text{ } \text{}6}_{\Lambda \Lambda}\text{He}$,
very likely it overpredicts the actual attraction in 
the $A=4$ system, too. Interestingly, 
in pionless EFT~\cite{CONTESSI2019134893}
a $\Lambda\Lambda$ scattering length practically identical to that 
of our LO interaction was found as limit for 
which the $^{\text{ }\text{ }\text{ } \text{}4}_{\Lambda \Lambda}\text{H}$ 
system becomes bound.

  \section{Conclusions and outlook}\label{sec:concl}
 In this work, we have generalized the J-NCSM formalism in order to include  strangeness $S=-2$ hyperons. Using the second quantization approach, we 
  systematically derived the necessary combinatorial factors that relate the
 Hamiltonian matrix elements in a many-body basis  to the corresponding ones in  a two-body basis for the $S=0,-1$ and $-2$ sectors. 
 A generalization to higher-strangeness sectors will be straightforward. 
 
 We then applied the J-NCSM approach to compute predictions of the chiral YY interactions at LO and NLO for $\Lambda \Lambda$ s-shell hypernuclei.   To speed up the convergence, the two interactions are also evolved via SRG. Unlike for the 
 $S=-1$ systems, here,  we  observed a very small  effect of the  SRG  YY evolution  on the $\Lambda \Lambda$-separation energies, implying  negligible  contributions  of  SRG-induced YYN forces.  Furthermore, 
 we found that the binding energy for $^{\text{ } \text{ }\text{ }\text{} 6}_{\Lambda \Lambda}$He predicted by the YY NLO potential
 is close to the
 empirical value while the LO interaction overbinds the system. 
 Both interactions also yield  a particle-stable   
 $^{\text{ } \text{ }\text{ }\text{} 5}_{\Lambda \Lambda}$He hypernucleus, whereas  
 $^{\text{ } \text{ }\text{ }\text{} 4}_{\Lambda \Lambda}$H 
 is found to be unstable against
 a breakup to $^3_{\Lambda}\text{H} + \Lambda$.  
 However, for a final conclusion,  a more  elaborate study that involves a more careful estimate of uncertainties stemming from various NN, YN and YY interactions is definitely necessary. 
 It will be also very interesting to study the predictions of the chiral YY interactions for other s-shell $\Lambda\Lambda$ systems such as     $^{\text{ } \text{ }\text{ }\text{} 4}_{\Lambda \Lambda}$n or   $^{\text{ } \text{ }\text{ }\text{} 4}_{\Lambda \Lambda}\text{He}$,  
 as well as for p-shell hypernuclei. 
  Finally, investigating possible Tjon-line like correlations for $B_{\Lambda \Lambda}$ of different systems is also of importance. 

\vskip 0.3cm
{\bf Acknowledgements:} This work is supported in part by the NSFC and the Deutsche Forschungsgemeinschaft (DFG, German Research
Foundation) through the funds provided to the Sino-German Collaborative
Research Center TRR110 ``Symmetries and the Emergence of Structure in QCD''
(NSFC Grant \break No. 12070131001, DFG Project-ID 196253076 - TRR 110). We
also acknowledge support of the THEIA net-working
activity of the Strong 2020 Project. The numerical calculations
have been performed on JURECA and the
JURECA booster of the JSC, J\"ulich, Germany. The
work of UGM was supported in part by the Chinese
Academy of Sciences (CAS) President's International
Fellowship Initiative (PIFI) (Grant No. 2018DM0034)
and by VolkswagenStiftung (Grant No. 93562).
 
\appendix

\section{Many-body Schr\"odinger equation  in second quantization}\label{appendix:appendixA}

Generally,  baryon-baryon (BB) interactions in the  $S=-2$ sector 
can lead to
 couplings  between states with identical particles and with 
 non-identical particles,  
for example $\Sigma \Sigma \rightarrow N \Xi$. Such   transitions  make it not \\ straightforward to 
 properly determine the  combinatorial factors of  free-space  two-body potentials that are embedded in the $A$-body Hamiltonian matrix elements.  In this appendix,  we  demonstrate that 
 these  factors can  systematically  be deduced  by comparing 
 the  Schr\"odinger equation for  A-body systems  with  the free-space two-body Schr\"odinger equation, provided that
  these    equations  are derived in a consistent way.  
We   show explicit examples for systems of  two and three particles, 
  and then generalize to the $A$-baryon problems. We note that Gl\"ockle and Miyagawa  \cite{Gloeckle:2000bt} have also derived a system of coupled-Faddeev equations for three-baryon systems taking into account full particle conversions. However it is not clear to us how to read off the involved combinatorial factors based on their equations. The authors of Ref.~\cite{Shevchenko:2007zz} have formulated the problem taking all permutations of particles explictly into account. This is however not consistent with the 
  approach of BB interactions used in \cite{Haidenbauer:2015zqb,Polinder:2007mp,Nagels:2020oqo}. For directly taking these interactions into account, we therefore require to derive the combinatorial factors consistent with these interactions. 
  
 To  derive the general Schr\"odinger equation,  we will  work with  second quantization.  
 The  many-body  Hamiltonian then has the form,
\begin{eqnarray} \label{eq:appendHam}
 &  & {H} = \sum_{k_1 k_1^{\prime}} {T}_{ k_1^{\prime} k_1}
a^{\dagger}_{k_1^{\prime}} a_{k_{1}}  + \frac{1}{2} \sum_{\substack{ {k_1 k_2}
\\ { k_1^{\prime} k_2^{\prime}}}} 
 V_{k_1^{\prime} k_2^{\prime}, k_1 k_2}
a^{\dagger}_{k_1^{\prime}} a^{\dagger}_{k_2^{\prime}} a_{k_2} a_{k_{1}},\nonumber\\[-7pt]
\end{eqnarray}
where $k_i$ stands for  a set of quantum numbers characterizing the particle state, i.e.,
momentum, spin, isospin as well as particle species $\lambda_{i} \,(\text{N}, \Lambda, \Sigma \,\text{or}\, \Xi)$. When it is
necessary  to separate the particle species  $\lambda_{i}$ from other quantum numbers, we use $k_i = \lambda_i \, \tilde{k}_i $.  Let us further assume  that  the potential matrix elements $ V_{k_1^{\prime} k_2^{\prime}, k_1 k_2}$
 in Eq.~(\ref{eq:appendHam})
are antisymmetric under exchanges of two indices, i.e.,   $V_{k_1^{\prime}
k_2^{\prime}, k_1 k_2 } = -V_{k_1^{\prime} k_2^{\prime}, k_2 k_1} =
-V_{k_2^{\prime} k_1^{\prime}, k_1 k_2}=  V_{k_2^{\prime} k_1^{\prime}, k_2
k_1}$.  Note that,   there is no ordering imposed 
 for quantum numbers of the incoming particles $k_1 $ and $k_2$ or of the outgoing pair
$k^{\prime}_1$  and $k^{\prime}_2$ in Eq.~(\ref{eq:appendHam}).
\subsection{Two-body  Schr\"odinger equation }
We start with the derivation 
 of  the  Schr\"odinger equation in a 
two-particle basis. For that, we define  the ordered two-body  antisymmetrized   basis states as
\begin{align} \label{eq:append2basis}
|\{p_1 p_2  \} \rangle \equiv a^{\dagger}_{p_{1}}   a^{\dagger}_{p_2} | 0
\rangle = \frac{1}{\sqrt2} \big(|p_1\rangle | p_2\rangle  - |p_2\rangle | p_1 \rangle \big),
\end{align}
with the right-hand side being the states in first quantization. Here, $p_1$ and $p_2$  also stand for the sets of quantum numbers (momentum, spin, isospin  and
particle species) describing particles 1 and 2, respectively.
 The completeness relation  of the basis  Eq.~(\ref{eq:append2basis}) for bases with particle species $\lambda_1 \ne \lambda_2$ reads
\begin{eqnarray} \label{eq:appendComplet}
&  &\!\!\!\!  \sum_{p_1 < p_2} |\{p_1 p_2  \} \rangle  \langle \{p_1 p_2  \} | \equiv\nonumber\\
&  &\!\!\!\!  \sum_{\lambda_1 < \lambda_2} \int d^3{\tilde{p}_1} d^3 
\tilde{p}_2 |\{\lambda_1 \tilde{p}_1  \lambda_2 \tilde{p}_2  \} \rangle
\langle \{\lambda_1 \tilde{p}_1 \lambda_2\tilde{p}_2  \} |   = \mathbbm{1},
\end{eqnarray}
where the inequality $p_1 < p_2$ accounts for the ordering of the  states in Eq.~(\ref{eq:append2basis}) where the leading 
sorting key is assumed to be particle species. Note that by  exploiting the antisymmetry of the basis functions, the
left hand side of   Eq.~(\ref{eq:appendComplet})  is  equivalent to 
\begin{eqnarray} \label{eq:appendComplet_identical_P}
&  &\!\!  \sum_{p_1 < p_2} |\{p_1 p_2  \} \rangle  \langle \{p_1 p_2  \} |\nonumber \\
&  &\!\! =  \frac{1}{2}
 \Big \{    \sum_{p_1 < p_2} |\{p_1 p_2  \} \rangle  \langle \{p_1 p_2  \} |   + \sum_{p_1 < p_2} |\{p_1 p_2  \} \rangle  \langle \{p_1 p_2  \} |    \Big\}\nonumber\\[3pt]
&  & \!\! = \frac{1}{2}   \Big \{    \sum_{p_1 < p_2} |\{p_1 p_2  \} \rangle  \langle \{p_1 p_2  \} |   + \sum_{p_1 >  p_2} |\{p_2 p_1  \} \rangle  \langle \{p_2 p_1  \} |    \Big\}\nonumber \\[3pt]
 &  & \!\! = \frac{1}{2}      \sum_{p_1,  p_2} |\{p_1 p_2  \} \rangle  \langle \{p_1 p_2  \} |\,.
  \end{eqnarray}
Hence, the summation over the ordered particle species on the left hand side of  Eq.~(\ref{eq:appendComplet}) can be replaced by a  normal summation over all particle species but  with a  factor of $\frac{1}{2}$. For the case of two identical particles, i.e., $\lambda_{1} = \lambda_{2}$, the completeness relation   becomes
\begin{align} \label{eq:appendCompletIden}
\frac{1}{2} \int d^3{\tilde{p}_1} d^3 
\tilde{p}_2 |\{\lambda_1 \tilde{p}_1 \lambda_1 \tilde{p}_2  \} \rangle  \langle
\{\lambda_1 \tilde{p}_1 \lambda_1 \tilde{p}_2  \} |   =  \mathbbm{1}  
\end{align}
following similar lines.
  The factor $\frac{1}{2}$ can also be absorbed  into the definition of
the states when one rewrites  Eq.~(\ref{eq:appendCompletIden}) as follows
\begin{align} \label{eq:appendCompletIden2}
 \int d^3{\tilde{p}_1} d^3 
\tilde{p}_2 \frac{1}{\sqrt2}|\{\lambda_1 \tilde{p}_1 \lambda_1 \tilde{p}_2  \} \rangle  \langle
\{\lambda_1 \tilde{p}_1 \lambda_1 \tilde{p}_2  \} | \frac{1}{\sqrt2}  = \mathbbm{1}.
\end{align}
Now, exploiting the anticommutator relation  for  the creation and annihilation operators, the kinetic and potential matrix elements  in the basis  Eq.~(\ref{eq:append2basis}) are easily obtained
\begin{eqnarray}\label{eq:appendVT}
&   &  \!\!\langle \{p_1^{\prime}   p_2^{\prime}\} | {T} | \{ p_1 p_2 \} \rangle  \nonumber\\[3pt]
&   & =  \delta_{p_1^{\prime} p_1 } {T}_{p_2^{\prime} p_2} - \delta_{p^{\prime}_{1} p_2}
{T}_{p_2^{\prime}p_1 } + \delta_{p_2^{\prime} p_2 }{T}_{p_1^{\prime} p_1} -
\delta_{p^{\prime}_{2} p_1}{T}_{p_1^{\prime}p_2 } \nonumber\\[4pt]
&   & = \delta_{p^{\prime}_1 p_1}  \delta_{p^{\prime}_2 p_2} {t}_{p^{\prime}_2}  - 
\delta_{p^{\prime}_1 p_2}  \delta_{p^{\prime}_2 p_1} {t}_{p^{\prime}_2}  +
\delta_{p^{\prime}_2 p_2}  \delta_{p^{\prime}_1 p_1} {t}_{p^{\prime}_1} \nonumber\\[2pt]
&  & \qquad -  \delta_{p^{\prime}_2 p_1}  \delta_{p^{\prime}_1 p_2} {t}_{p^{\prime}_1} \nonumber\\[6pt]
&  &  \!\! \langle \{p_1^{\prime} p_2^{\prime}\} | {V} | \{ p_1 p_2\} \rangle \nonumber\\[2pt]
& &=  \frac{1}{2} \big ( {V}_{p^{\prime}_1 p_2^{\prime}, p_1 p_2}   -
{V}_{p^{\prime}_1 p_2^{\prime}, p_2 p_1}    - {V}_{p^{\prime}_2
p_1^{\prime}, p_1 p_2}
   + {V}_{p^{\prime}_2 p_1^{\prime}, p_2 p_1} \big)\nonumber\\[2pt]
&  &  = 2{V}_{p_1^{\prime}p^{\prime}_2, p_1 p_2 }.
\end{eqnarray}
In the second line of Eq.~(\ref{eq:appendVT}), we have exploited  the fact that the
kinetic operator is diagonal in the momentum basis. The Schr\"odinger equation,
\begin{align} \label{eq:appendSchro}
\begin{split}
{H} | \Psi \rangle = {E} | \Psi \rangle,
\end{split}
\end{align}
in the two-body basis  Eq.~(\ref{eq:append2basis})  then  reads
\begin{align} \label{eq:appendSchro1}
\begin{split}
\sum_{p_1 < p_2} \langle \{ p^{\prime}_1 p^{\prime}_2 \} | {H} | \{ p_1 p_2\}
\rangle  \langle \{p_1 p_2\} | \Psi \rangle = {E}  \underbrace{\langle\{p_1^{\prime} 
p_2^{\prime}\} |  \Psi \rangle}_{\equiv \Psi(p_1^{\prime}p_2^{\prime}) }.
\end{split}
\end{align}
Here,  it will be sufficient to consider  only  those components of $\Psi(p_1^{\prime} p_2^{\prime})$ with 
 $p_1^{\prime} < p_2^{\prime}$. Since the basis states are antisymmetric,  
the other components of $\Psi(p_1^{\prime} p_2^{\prime})$  with  $p_1^{\prime} > p_2^{\prime}$ will differ from
the ones  with   $p_1^{\prime} <  p_2^{\prime}$ by a simple  phase factor. 
Plugging Eq.~(\ref{eq:appendVT})  into Eq.~(\ref{eq:appendSchro1}) and using $p_1^{\prime} <  p_2^{\prime}$, one
arrives at a general two-body Schr\"odinger equation 
\begin{eqnarray}\label{eq:appendSchro2}
&  &   {t}_{p_1^{\prime}} \Psi(p_1^{\prime} p_2^{\prime}) + {t}_{p_2^{\prime}}
\Psi(p_1^{\prime} p_2^{\prime}) + \sum_{p_1 < p_2 } 2 {V}_{p_1^{\prime}
p_2^{\prime}, p_1 p_2} \Psi(p_1 p_2) \nonumber\\
&  &   ={E} \Psi(p_1^{\prime} p_2^{\prime}).
\end{eqnarray}
We note that there is a factor of $2$ in front of the potential
matrix elements, which  drops out for the case of the two-identical particle
basis,  i.e., $\lambda_1 = \lambda_2$.  In that case,  we use $\sum_{p_1<p_2} \rightarrow 1/2 \ \sum_{p_1,p_2}$ and equation Eq.~(\ref{eq:appendSchro2}) 
becomes
\begin{eqnarray}\label{eq:appendSchro2Identical}
&  &  {t}_{p_1^{\prime}}  \Psi(p_1^{\prime} p_2^{\prime}) + {t}_{p_2^{\prime}}
\Psi(p_1^{\prime} p_2^{\prime}) + \sum_{p_1,  p_2 }  {V}_{p_1^{\prime}
p_2^{\prime}, p_1 p_2} \Psi(p_1 p_2) \nonumber\\
&  &   ={E} \Psi(p_1^{\prime} p_2^{\prime}).
\end{eqnarray}
To better understand the prefactors of  the potential matrix elements  present  in
 Eqs.~(\ref{eq:appendSchro2},\ref{eq:appendSchro2Identical}), let us      consider  some explicit  bases. In the first example,   the basis consists of two two-particle states, 
 one with identical particles  and one with distinguishable particles, e.g., 
 $|\{\Lambda \Lambda \} \rangle$   and $|\{N \Xi  \}\rangle$.  Then,  the  
completeness relation is obtained by combining 
Eqs.~(\ref{eq:appendComplet}) and (\ref{eq:appendCompletIden2}) 
\begin{eqnarray} \label{eq:appendCompletLXN}
&  &\!\! \int d^3{\tilde{p}_1} d^3 
\tilde{p}_2 \,\Big \{  |\{\Xi  \tilde{p}_1  N \tilde{p}_2  \} \rangle
\langle \{\Xi \tilde{p}_1 N \tilde{p}_2  \} |  \nonumber\\[3pt]
&  &   \qquad  +  \frac{1}{\sqrt2}|\{\Lambda \tilde{p}_1 \Lambda \tilde{p}_2  \} \rangle  \langle
\{\Lambda \tilde{p}_1 \Lambda \tilde{p}_2  \} | \frac{1}{\sqrt2} \Big\} = \mathbbm{1},
\end{eqnarray}
leading to the following expression for  the norm of the wave function 
\begin{align} \label{eq:appendnormLXN}
 \! \langle \Psi |  \Psi \rangle = \!\int \!\!d^3{\tilde{p}_1} d^3 
\tilde{p}_2 \Big\{ |\Psi_{N \Xi }(\tilde{p}{_1} \tilde{p}_2)|^2  
   +  |\frac{1}{\sqrt2}\Psi_{\Lambda \Lambda}(\tilde{p}_1 \tilde{p}_2)|^2 \Big\}.
\end{align}
Therefore,  we absorb the   $\frac{1}{\sqrt2}$-factor   into the amplitude of  states  by
introducing a new set of the wave-function components, 
\begin{align}
 \Phi_{\Lambda \Lambda}(\tilde{p}_1 \tilde{p}_2)
= \frac{1}{\sqrt2} \Psi_{\Lambda \Lambda}(\tilde{p}_1 \tilde{p}_2);
 \,\,\,    \Phi_{N \Xi }(\tilde{p}_1 \tilde{p}_2)
=  \Psi_{N \Xi }(\tilde{p}_1 \tilde{p}_2),
\end{align}
 so that  the Schr\"odinger equation  Eqs.~(\ref{eq:appendSchro2},\ref{eq:appendSchro2Identical})
 for the two newly defined    components   possesses   a symmetric form
\begin{eqnarray} \label{eq:appendLLXiNbasis2}
& & \left( \begin{array}{c c}
2 {t}_{\Lambda} + {V}_{\Lambda\Lambda, \Lambda \Lambda}  & \sqrt2
{V}_{\Lambda \Lambda, N \Xi } \\[5pt]
\sqrt2 {V}_{N \Xi , \Lambda \Lambda} & {t}_{\Xi} + {t}_{N} + 2
{V}_{ N \Xi
, N \Xi } \end{array} \right) \left( \begin{array}{c}  \Phi_{\Lambda
\Lambda} \\[5pt]  \Phi_{N \Xi }  \end{array} \right)\nonumber\\[4pt]
&  & \qquad   = E \left( \begin{array}{c}  \Phi_{\Lambda
\Lambda} \\[5pt]  \Phi_{N \Xi }  \end{array} \right),
\end{eqnarray}
where,  for readability, we have omitted the dependence on $\tilde{p} $ and
$\tilde{p}^{\prime}$. Similarly, for the case where  the basis consists  of  four  states $\{|\Lambda \Lambda\} \rangle$,
$|\{\Sigma \Sigma \}\rangle$, $|\{ \Lambda \Sigma \}\rangle$  and  $|\{ \Xi N \}\rangle$, one analogously  defines  a new set of wave-function components
\begin{eqnarray}
&  &   \Phi_{\Lambda \Lambda} = \frac{1}{\sqrt2}
\Psi_{\Lambda \Lambda}\,; \quad \Phi_{\Sigma \Sigma} = \frac{1}{\sqrt2}  \Psi_{\Sigma \Sigma}\,;\nonumber\\[3pt]
 &  &  \Phi_{\Lambda \Sigma} = 
\Psi_{\Lambda \Sigma}\,;    \qquad   \Phi_{ N \Xi} = 
\Psi_{N \Xi}\,, 
\end{eqnarray}
  for which the 
 Schr\"odinger equation again possesses a symmetric form
%
\begin{eqnarray} \label{eq:appendLLSXiNbasis}
{H} \left( \begin{array}{c}  \Phi_{\Lambda
\Lambda} \\ \Phi_{\Sigma \Sigma} \\ \Phi_{\Lambda \Sigma} \\ 
\Phi_{N \Xi } \end{array} \right)
  = E \left( \begin{array}{c}  \Phi_{\Lambda \Lambda} \\ \Phi_{\Sigma \Sigma}
\\
\Phi_{\Lambda \Sigma} \\  \Phi_{N \Xi }  \end{array} \right),
\end{eqnarray}
\newline
with
\begin{align} \label{eq:appendHamilinLLSXiNbasis}
\resizebox{.47 \textwidth}{!}{ 
\!\!${H}= \left( \begin{array}{c c c c }
2 {t}_{\Lambda} + {V}_{\Lambda\Lambda, \Lambda \Lambda}  &
{V}_{\Lambda \Lambda, \Sigma \Sigma} & \sqrt2 {V}_{\Lambda
\Lambda, \Lambda \Sigma} & 
\sqrt2 {V}_{\Lambda \Lambda,N \Xi } \\
 {V}_{\Sigma\Sigma, \Lambda \Lambda}  & 2{t}_{\Sigma} +
{V}_{\Sigma\Sigma, \Sigma \Sigma} & 
\sqrt2{V}_{\Sigma \Sigma, \Lambda \Sigma} & \sqrt2 {V}_{\Sigma
\Sigma, N \Xi } \\
 \sqrt2{V}_{\Lambda\Sigma, \Lambda \Lambda}  & \sqrt2
{V}_{\Lambda\Sigma, \Sigma \Sigma} & 
{t}_{\Lambda} + {t}_{\Sigma} + 2{V}_{\Lambda \Sigma, \Lambda \Sigma}
& 2 {V}_{\Lambda\Sigma,N \Xi } \\
\sqrt2 {V}_{N \Xi , \Lambda \Lambda} & \sqrt2 {V}_{N \Xi , \Sigma \Sigma}
& 2 {V}_{N \Xi , \Lambda \Sigma} &  {t}_{\Xi} + {t}_{N} + 2
{V}_{N \Xi
, N\Xi } \end{array} \right)
$}\nonumber\\[3pt]
\end{align}
One sees that
  there is a  $\sqrt2$-factor for the transition between states of 
identical and  of distinguishable particles, and a factor of  $ 2 $ for the transition
between states of nonidentical particles.   It is important to mention that these factors are
 already included in the definition of the  two-body potentials 
derived from chiral EFT \cite{Polinder:2007mp,Haidenbauer:2015zqb} 
or phenomenological models
\cite{Nagels:2020oqo} (see, e.g., Eq.~(2) of \cite{Haidenbauer:2015zqb}). We therefore denote these initial two-body potentials
${V}_{\lambda_1 \lambda_2,\lambda^{\prime}_1 \lambda^{\prime}_2}$
 with an appropriate  factor of $\sqrt2$ or $2$ or $1$
to  be our new potential ${\tilde{V}}_{\lambda_1 \lambda_2,\lambda^{\prime}_1 \lambda^{\prime}_2}$. 
Expressing in terms of the new potentials $\tilde{V}$, the Hamiltonian Eq.~(\ref{eq:appendHamilinLLSXiNbasis})
now has a more intuitive form
\begin{align} \label{eq:appendHamilinLLSXiNbasis2}
\resizebox{.47 \textwidth}{!}{ 
${H}= \left( \begin{array}{c c c c }
2 {t}_{\Lambda} + {\tilde{V}}_{\Lambda\Lambda, \Lambda \Lambda}  &
{\tilde{V}}_{\Lambda \Lambda, \Sigma \Sigma} &  {\tilde{V}}_{\Lambda
\Lambda, \Lambda \Sigma} & 
 {\tilde{V}}_{\Lambda \Lambda,N \Xi } \\
 {\tilde{V}}_{\Sigma\Sigma, \Lambda \Lambda}  & 2{t}_{\Sigma} +
{\tilde{V}}_{\Sigma\Sigma, \Sigma \Sigma} & 
{\tilde{V}}_{\Sigma \Sigma, \Lambda \Sigma} & {\tilde{V}}_{\Sigma
\Sigma, N \Xi } \\
 {\tilde{V}}_{\Lambda\Sigma, \Lambda \Lambda}  & 
{\tilde{V}}_{\Lambda\Sigma, \Sigma \Sigma} & 
{t}_{\Lambda} + {t}_{\Sigma} + {\tilde{V}}_{\Lambda \Sigma, \Lambda \Sigma}
& {\tilde{V}}_{\Lambda\Sigma,N \Xi } \\
 {\tilde{V}}_{N \Xi , \Lambda \Lambda} &  {\tilde{V}}_{N \Xi , \Sigma \Sigma}
& {\tilde{V}}_{N \Xi , \Lambda \Sigma} &  {t}_{\Xi} + {t}_{N} + 
{\tilde{V}}_{N \Xi
, N\Xi } \end{array} \right). 
$}\nonumber\\[3pt]
\end{align}
In the next step, we are going to derive a similar \break Schr\"odinger equation in a
three-body basis. Then, by comparing the obtained equation with the one for  two-body
basis, we  will be  able to determine the corresponding combinatorial factors for the potentials in
each \break strangeness sector.

\subsection{Three-body Schr\"odinger equation}
  We define the ordered   three-body basis states in second quantization and its completeness relations as
\begin{eqnarray} \label{eq:append3basis}
&  &  |\{p_1 p_2 p_3 \} \rangle \equiv a^{\dagger}_{p_{1}}   a^{\dagger}_{p_2}
a^{\dagger}_{p_{3}} | 0
\rangle;\nonumber\\[4pt]
&  &   \sum_{p_1 < p_2 < p_3 } |\{p_1 p_2 p_3  \} \rangle  \langle
\{p_1 p_2 p_3 \} | = \mathbbm{1}.
\end{eqnarray}
The kinetic and potential matrix elements  in the basis Eq.~(\ref{eq:append3basis}) read
\begin{eqnarray}\label{eq:appendT2}
&  &  \,\langle \{p_1^{\prime} p_2^{\prime} p_3^{\prime}\}| {T} | \{ p_1 p_2 p_3
\} \rangle  =\nonumber\\[4pt]
  &   & \, {T}_{p_1^{\prime} p_1} \delta_{p_2^{\prime} p_2 } \delta_{p_3^{\prime} p_3 }-
  {T}_{p_1^{\prime} p_1} \delta_{p_2^{\prime} p_3 } \delta_{p_3^{\prime}
p_2}+ {T}_{p_1^{\prime} p_2} \delta_{p_2^{\prime} p_3 }
\delta_{p_3^{\prime} p_1 }\nonumber\\[4pt]
&  & - {T}_{p_1^{\prime} p_2} \delta_{p_2^{\prime} p_1 } \delta_{p_3^{\prime} p_3}+
 {T}_{p_1^{\prime} p_3} \delta_{p_2^{\prime} p_1 } \delta_{p_3^{\prime} p_2 }-
  {T}_{p_1^{\prime} p_3} \delta_{p_2^{\prime} p_2 } \delta_{p_3^{\prime} p_1 }\nonumber\\[4pt]
&   & + {T}_{p_2^{\prime} p_1} \delta_{p_1^{\prime} p_3 } \delta_{p_3^{\prime} p_2 }-
  {T}_{p_2^{\prime} p_1} \delta_{p_1^{\prime} p_2 } \delta_{p_3^{\prime}p_3}+ 
 {T}_{p_2^{\prime} p_2} \delta_{p_1^{\prime} p_1 } \delta_{p_3^{\prime} p_3 }\nonumber\\[4pt]
&  &  - {T}_{p_2^{\prime} p_2} \delta_{p_1^{\prime} p_3 } \delta_{p_3^{\prime} p_1}+
 {T}_{p_2^{\prime} p_3} \delta_{p_3^{\prime} p_1 } \delta_{p_1^{\prime} p_2 }-
  {T}_{p_2^{\prime} p_3} \delta_{p_1^{\prime} p_1 } \delta_{p_3^{\prime} p_2 }\nonumber\\[4pt]
& &  + {T}_{p_3^{\prime} p_1} \delta_{p_1^{\prime} p_2 } \delta_{p_2^{\prime} p_3 }-
  {T}_{p_3^{\prime} p_1} \delta_{p_1^{\prime} p_3 } \delta_{p_2^{\prime} p_2}+ 
 {T}_{p_3^{\prime} p_2} \delta_{p_1^{\prime} p_3 } \delta_{p_2^{\prime} p_1 }\nonumber\\[4pt]
&  & - {T}_{p_3^{\prime} p_2} \delta_{p_1^{\prime} p_1 } \delta_{p_2^{\prime} p_3}+
 {T}_{p_3^{\prime} p_3} \delta_{p_1^{\prime} p_1 } \delta_{p_2^{\prime} p_2 }-
  {T}_{p_3^{\prime} p_3} \delta_{p_1^{\prime} p_2 } \delta_{p_2^{\prime} p_1 },\nonumber\\[2pt]
\end{eqnarray}
and
\begin{eqnarray}\label{eq:appendV2}
&  & \!\!  \langle \{p_1^{\prime} p_2^{\prime} p_3^{\prime}\}| {V} | \{ p_1 p_2 p_3
\} \rangle = 2 \big\{\nonumber\\[4pt] 
&  &   {V}_{p^{\prime}_{2} p^{\prime}_3, p_2p_3}
\delta_{p^{\prime}_1 p_1}   +  {V}_{p^{\prime}_{2} p^{\prime}_3, p_3p_1}
\delta_{p^{\prime}_1 p_2} +{V}_{p^{\prime}_{2} p^{\prime}_3, p_1p_2}
\delta_{p^{\prime}_1 p_3}\nonumber\\[4pt]
&  & +{V}_{p^{\prime}_{3} p^{\prime}_1, p_2p_3}
\delta_{p^{\prime}_2 p_1}  +{V}_{p^{\prime}_{3} p^{\prime}_1, p_3p_1}
\delta_{p^{\prime}_2 p_2} + {V}_{p^{\prime}_{3} p^{\prime}_1, p_1p_2}
\delta_{p^{\prime}_2 p_3}\nonumber\\[5pt]
&  & +{V}_{p^{\prime}_{1} p^{\prime}_2, p_2p_3}
\delta_{p^{\prime}_3 p_1}  +{V}_{p^{\prime}_{1} p^{\prime}_2 ,p_3p_1}
\delta_{p^{\prime}_3 p_2} +  {V}_{p^{\prime}_{1} p^{\prime}_2, p_1p_2}
\delta_{p^{\prime}_3 p_3}\big\}\nonumber.\\[3pt]
\end{eqnarray}
Now, projecting the Schr\"odinger equation Eq.~(\ref{eq:appendSchro}) onto the state
$|\{p^{\prime}_1 p^{\prime}_2 p^{\prime}_3\} \rangle$   and 
then utilizing the
completeness relation in Eq.~(\ref{eq:append3basis}), one arrives at 
\begin{eqnarray} \label{eq:appendSchro31}
  &  & \!\! \! \sum_{p_1 < p_2 <p_3}\! \! \Big \{  \langle \{  p^{\prime}_1 p^{\prime}_2   p^{\prime}_3 \} |
{T} |  \{ p_1 p_2 p_3 \}
\rangle  \langle \{p_1 p_2 p_3 \} | \Psi \rangle + \nonumber\\[4pt]
 &  &\!\! \!\langle \{ p^{\prime}_1 p^{\prime}_2 p^{\prime}_3 \} |
{V} | \{ p_1 p_2 p_3 \}
\rangle  \langle \{p_1 p_2 p_3 \} | \Psi \rangle\!\Big\}  \!= {E}  \underbrace{\langle
\{p_1^{\prime} p_2^{\prime} p^{\prime}_3 \}| \Psi \rangle}_{\equiv \Psi(p_1^{\prime}
p_2^{\prime} p_3^{\prime}) } \nonumber\\[3pt]
\end{eqnarray}
Similar to the case of a  two-body basis, here it will be sufficient to consider 
only those components of $ \Psi(p_1^{\prime} p_2^{\prime} p_3^{\prime}) $
with $p^{\prime}_1 <
p^{\prime}_2 < p^{\prime}_3$. With  this   condition,  only three of the 18 kinetic terms in 
  Eq.~(\ref{eq:appendT2}) survive. Hence,
we have 
\begin{eqnarray} \label{eq:appendeqT}
  &  &  \sum_{p_1 < p_2 <p_3} \langle \{  p^{\prime}_1 p^{\prime}_2   p^{\prime}_3 \} |
{T} |  \{ p_1 p_2 p_3 \}
\rangle  \langle \{p_1 p_2 p_3 \} | \Psi \rangle  \nonumber\\[3pt] 
&  &  \qquad\qquad = ({t}_{p^{\prime}_1} +
{t}_{p^{\prime}_2} + {t}_{p^{\prime}_3})  \Psi(p^{\prime}_1 p^{\prime}_2 
p^{\prime}_3).
\end{eqnarray}
The contributions from the potential operator are a little bit more cumbersome, but can be reduced to a compact form by 
exploiting  the antisymmetry properties under the  exchange of two indices of the potential as well as of
the wave function. For example, 
the first three  terms in Eq.~(\ref{eq:appendV2}) give
\begin{eqnarray} \label{eq:appendeqV}
&  & \!\!\! \sum_{p_1 < p_2 <p_3} \!\!\Big\{  {V}_{p^{\prime}_{2}p^{\prime}_{3}, p_2 p_3}
\delta_{p^{\prime}_1 p_1} \Psi(p_1 p_2 p_3) +  \nonumber\\[3pt] 
&   &    \underset{p_2 \leftrightarrow p_1} {{V}_{p^{\prime}_{2}p^{\prime}_{3}, p_3 p_1}
\delta_{p^{\prime}_1 p_2} \Psi(p_1 p_2 p_3)}
+\underset{p_3 \leftrightarrow p_1} 
 {{V}_{p^{\prime}_{2}p^{\prime}_{3}, p_1 p_2}
\delta_{p^{\prime}_1 p_3} \Psi(p_1 p_2 p_3)}\Big\}\nonumber\\[5pt]
&  &  =  \sum_{p_1 < p_2 <p_3} {V}_{p^{\prime}_{2}p^{\prime}_{3}, p_2 p_3}
\delta_{p^{\prime}_1 p_1} \Psi(p_1 p_2 p_3)  \nonumber\\[3pt]
&  & \qquad  + \sum_{p_2 <p_1 <p_3} 
{{V}_{p^{\prime}_{2}p^{\prime}_{3}, p_3 p_2}
\delta_{p^{\prime}_1 p_1} \Psi(p_2 p_1 p_3)}\nonumber\\[4pt]
 &  &  \qquad   +\sum_{p_3 < p_2 < p_1} \underset{p_2 \leftrightarrow
p_3}{{{V}_{p^{\prime}_{2}p^{\prime}_{3}, p_3 p_2}
\delta_{p^{\prime}_1 p_1}}} \Psi(p_3 p_2 p_1)\nonumber\\[6pt]
&  &  =  \sum_{p_1 < p_2 <p_3} {V}_{p^{\prime}_{2}p^{\prime}_{3}, p_2 p_3}
\delta_{p^{\prime}_1 p_1} \Psi(p_1 p_2 p_3) \nonumber\\[3pt]
&  & \qquad +  \sum_{p_2 <p_1 <p_3} 
{{V}_{p^{\prime}_{2}p^{\prime}_{3}, p_2 p_3}
\delta_{p^{\prime}_1 p_1} \Psi(p_1 p_2 p_3)} \nonumber\\[4pt]
&  &  \qquad + \sum_{p_2 < p_3 < p_1} {{V}_{p^{\prime}_{2}p^{\prime}_{3}, p_2
p_3} \delta_{p^{\prime}_1 p_1} \Psi(p_1 p_2 p_3)}\nonumber\\[6pt]
&  & = \sum_{p_1} \sum_{p_2 < p_3 } {V}_{p^{\prime}_{2}p^{\prime}_{3}, p_2
p_3} \delta_{p^{\prime}_1 p_1} \Psi(p_1 p_2 p_3)\nonumber\\[6pt]
&  & =  \sum_{p_2 < p_3 } {V}_{p^{\prime}_{2}p^{\prime}_{3}, p_2
p_3}  \Psi(p^{\prime}_1 p_2 p_3).
\end{eqnarray}
 Analogously, the  next three terms in Eq.~(\ref{eq:appendV2}) yield 
\begin{eqnarray} \label{eq:appendeqV2}
 &  & \!\!\!  \sum_{p_1 < p_2 <p_3} \!\! \Big\{ \underset{p_1 \leftrightarrow p_2}
{{V}_{p^{\prime}_{3}p^{\prime}_{1}, p_2 p_3}
\delta_{p^{\prime}_2 p_1} \Psi(p_1 p_2 p_3)} + \nonumber\\[4pt]
&  & \,\, 
{{V}_{p^{\prime}_{3}p^{\prime}_{1}, p_3 p_1}
\delta_{p^{\prime}_2 p_2} \Psi(p_1 p_2 p_3)}
+\underset{p_2 \leftrightarrow p_3} 
 {{V}_{p^{\prime}_{3}p^{\prime}_{1}, p_1 p_2}
\delta_{p^{\prime}_2 p_3} \Psi(p_1 p_2 p_3)}\Big\}\nonumber\\[3pt]
&   & \,\, =  \cdots = \sum_{p_1 < p_3 } {V}_{p^{\prime}_{1}p^{\prime}_{3}, p_1
p_3}  \Psi(p_1 p^{\prime}_2 p_3), 
\end{eqnarray}
\newline
and,   the three remaining terms result in 
\newline
\begin{eqnarray} \label{eq:appendeqV2_2}
&  & \sum_{p_1 < p_2 <p_3} \!\! \! \Big\{ \underset{p_1 \leftrightarrow p_3}
{{V}_{p^{\prime}_{1}p^{\prime}_{2}, p_2 p_3}
\delta_{p^{\prime}_3 p_1} \Psi(p_1 p_2 p_3)} +\nonumber\\[3pt]
&  &  \underset{p_2 \leftrightarrow p_3}{{V}_{p^{\prime}_{1}p^{\prime}_{2}, p_3 p_1}
\delta_{p^{\prime}_3 p_2} \Psi(p_1 p_2 p_3)}
+ {{V}_{p^{\prime}_{1}p^{\prime}_{2}, p_1 p_2}
\delta_{p^{\prime}_3 p_3} \Psi(p_1 p_2 p_3)}\nonumber\\[3pt]
& &  = \cdots =  \sum_{p_1 < p_2 } {V}_{p^{\prime}_{1}p^{\prime}_{2}, p_1
p_2}  \Psi(p_1 p_2 p^{\prime}_3).
\end{eqnarray}
Taking into account Eqs.~(\ref{eq:appendeqT}-\ref{eq:appendeqV2_2}), the Schr\"odinger
equation  Eq.~(\ref{eq:appendSchro31})   in the three-body basis Eq.~(\ref{eq:append3basis}) can be written as 
\begin{eqnarray}\label{eq:appendSchro32}
 &  & \!\! \big({t}_{p^{\prime}_1 }  + {t}_{p^{\prime}_2 } +
 {t}_{ p^{\prime}_3}\big) \Psi({p^{\prime}_1 p^{\prime}_2
p^{\prime}_3}) +
\sum_{p_1 < p_2}  \!\!2 {V}_{{p^{\prime}_1 p^{\prime}_2, p_1 p_2}} \Psi({p_1 p_2
p^{\prime}_3})\nonumber\\[4pt]
&  &\!\! +\! \sum_{p_1 < p_3} \!\! 2 {V}_{{p^{\prime}_1 p^{\prime}_3, p_1 p_3}} \Psi({p_1
p^{\prime}_2
p_3}) + \!\!\sum_{p_2 < p_3} \!\! 2 {V}_{{p^{\prime}_2 p^{\prime}_3, p_2
p_3}} \Psi({p^{\prime}_1 p_2p_3})\nonumber\\[3pt]
&  & \,\, = E \Psi({p^{\prime}_1 p^{\prime}_2 p^{\prime}_3}),
\end{eqnarray}
which, as one expects, differs from the Schr\"odinger equation in the two-body
basis  Eq.~(\ref{eq:appendSchro2})  by the kinetic energy  of the third particle and the
two-body  interactions  between particles $1$-$3$ and $2$-$3$. Again, the factor of
$2$ in front of the potential vanishes when the incoming particles are
identical and the summations include all states $\tilde{p}_1, \tilde{p}_2$
 etc. 
For  illustration purposes,  let us  consider 
Eq.~(\ref{eq:appendSchro32})  in an explicit basis consisting  of four states, $|\{N
\Lambda \Lambda\}\rangle$, $|\{N
\Sigma \Sigma\}\rangle$, $|\{N \Lambda \Sigma\}\rangle$ and $|\{N
N \Xi\}\rangle$. The norm of the wave function 
in this four-particle-state basis can be calculated as follows
\begin{eqnarray} \label{eq:appendnorm3Body}
&  & \langle \Psi |  \Psi \rangle  = \int d^3{\tilde{p}_1} d^3 
\tilde{p}_2   d^3{\tilde{p}_3} \Big\{ |\frac{1}{\sqrt2}\Psi_{N \Lambda \Lambda }(\tilde{p}{_1} \tilde{p}_2  \tilde{p}_3)|^2\nonumber\\[3pt] 
&  &\qquad \quad + \,|\frac{1}{\sqrt2}\Psi_{N \Sigma \Sigma}(\tilde{p}_1 \tilde{p}_2 \tilde{p}_3)|^2
  +\, \, |\Psi_{N \Lambda \Sigma}(\tilde{p}_1 \tilde{p}_2 \tilde{p}_3)|^2\nonumber\\[3pt]
&  &  \qquad\quad + \, \,  |\frac{1}{\sqrt2}\Psi_{N N \Xi}(\tilde{p}_1 \tilde{p}_2 \tilde{p}_3)|^2\Big\}.
\end{eqnarray}
Based on   Eq.~(\ref{eq:appendnorm3Body}), we define a new set of  wavefunction components 
\begin{eqnarray}\label{eq:appendScale3}
&  &  \Phi_{N \Lambda \Lambda} = \frac{1}{\sqrt{2}} \Psi_{N \Lambda \Lambda}\,; \quad 
\Phi_{N \Sigma \Sigma} = \frac{1}{\sqrt{2}} \Psi_{N \Sigma \Sigma}\,; \quad\nonumber\\
&  & \Phi_{N \Lambda \Sigma} =  \Psi_{N \Lambda \Sigma }\,; \quad 
\Phi_{N N \Xi} =  \frac{1}{\sqrt2}\Psi_{N N \Xi }.\\
& & \nonumber
\end{eqnarray}
The   Schr\"odinger equation Eq.~(\ref{eq:appendSchro32}),  applying   
the   wave function  components in 
Eq.~(\ref{eq:appendScale3}), now has a symmetric form,

\begin{eqnarray} \label{eq:appendeqin3Bbasissys}
&  &  { (T + V )} \left( \begin{array}{c} \ \Phi_{N \Lambda
\Lambda} \\ \Phi_{N\Sigma \Sigma} \\ \Phi_{N \Lambda \Sigma} \\ 
 \Phi_{ N N \Xi } \end{array} \right)
  = E \left( \begin{array}{c}  \Phi_{N \Lambda
\Lambda} \\  \Phi_{N \Sigma \Sigma} \\ \Phi_{N \Lambda \Sigma}  \\
\Phi_{N N \Xi }  \end{array} \right),
\end{eqnarray}
\newline

with  ${T}$  being a diagonal  matrix

\begin{equation} \label{eq:appendsymTin3Bbasis}
\resizebox{.41 \textwidth}{!}{ 
 $\!\! {T}= \left( \begin{array}{c c c  c }
2 {t}_{\Lambda} + {t}_N  & 0 &  0 & 0 \\[4pt] 
0  & 2{t}_{ \Sigma} + {t}_{N} & 0 & 0 \\[4pt]
0 & 0 &  {t}_{\Lambda}  + {t}_{\Sigma} +{t}_N & 0 \\[4pt]
0 & 0 &  0 & 2 {t}_{N}  + {t}_{\Xi}   
\end{array} \right),
$}
\end{equation} 
\vspace{1cm}
and the symmetric  potential matrix \hfill
\begin{widetext}
\begin{equation}
\hspace{ 1.5cm}
\label{eq:appendsymHamilin3Bbasis}
\resizebox{.75 \textwidth}{!}{ 
${V}= \left( \begin{array}{c c c  c }
2 {\tilde{V}}_{N \Lambda, N \Lambda} + {\tilde{V}}_{\Lambda \Lambda, \Lambda \Lambda}  &
{\tilde{V}}_{\Lambda \Lambda, \Sigma \Sigma} &  {\tilde{V}}_{\Lambda
\Lambda, \Lambda \Sigma} + \sqrt2 {\tilde{V}}_{N \Lambda, N \Sigma} & 
\sqrt2{\tilde{V}}_{\Lambda \Lambda,N \Xi } \\[9pt]
 {\tilde{V}}_{\Sigma\Sigma, \Lambda \Lambda}  & 2{\tilde{V}}_{N \Sigma, N \Sigma} +
{\tilde{V}}_{\Sigma\Sigma, \Sigma \Sigma} & 
{\tilde{V}}_{\Sigma \Sigma, \Lambda \Sigma}  + \sqrt2{\tilde{V}}_{N \Sigma, N
\Lambda} &   \sqrt2{\tilde{V}}_{\Sigma
\Sigma, N \Xi } \\[9pt]
 {\tilde{V}}_{\Lambda\Sigma, \Lambda \Lambda}  + \sqrt2 {\tilde{V}}_{N \Sigma, N
\Lambda} & 
{\tilde{V}}_{\Lambda\Sigma, \Sigma \Sigma} + \sqrt2{\tilde{V}}_{N \Lambda, N \Sigma} & 
{\tilde{V}}_{N \Lambda, N \Lambda} + {\tilde{V}}_{N\Sigma, N \Sigma} +
{\tilde{V}}_{\Lambda \Sigma, \Lambda \Sigma}
& \sqrt2 {\tilde{V}}_{\Lambda\Sigma, N \Xi } \\[9pt]
\sqrt2 {\tilde{V}}_{N \Xi , \Lambda \Lambda} &  \sqrt2 {\tilde{V}}_{N \Xi , \Sigma \Sigma}
& \sqrt2 {\tilde{V}}_{N \Xi , \Lambda \Sigma} &  {\tilde{V}}_{NN,NN}+ 2
\tilde{{V}}_{N \Xi
, N \Xi } \end{array} \right) 
$}.
\end{equation} 
\end{widetext}

In the last step, we have expressed the potential 
matrix elements in terms of $\tilde V$ as given in 
Eq.~(\ref{eq:appendHamilinLLSXiNbasis2}). 
Eqs.~(\ref{eq:appendeqin3Bbasissys}-\ref{eq:appendsymHamilin3Bbasis})  
define the combinatorial factors of the  two-body potentials present in the three-body Hamiltonian. In following, we want to generalize this result to an $A$-body system. 

\begin{table*}[tbp]
 \renewcommand{\arraystretch}{1.8}
 \vskip 1 cm 
    \setlength{\tabcolsep}{0.09cm}
\centering
\begin{tabular}{|*{5}{c|}}
\cline{1-5}  
\diagbox[innerrightsep=0.3cm,innerwidth=2cm,height=0.87cm,trim=l]{transition}{  YN} &  $\tilde{V}_{N\Lambda,N\Lambda}$ &
$\tilde{V}_{N\Lambda, N\Sigma}$ & $\tilde{V}_{N\Sigma, N\Lambda}$ &
 $\tilde{V}_{N\Sigma, N\Sigma}$\\
\cline{1-5}
$\Lambda\Lambda \rightarrow \Lambda \Lambda$ & $2(A-2)$ & -  & - & \\
\cline{1-5}
$\Lambda\Lambda \rightarrow \Lambda \Sigma$ & - & $\sqrt2(A-2)$  & - & \\
\cline{1-5}
$\Lambda\Sigma \rightarrow \Lambda \Sigma$ & $A-2$ &  -  & - & $A-2$\\ 
\cline{1-5}
$\Lambda\Sigma \rightarrow \Sigma \Sigma$ &   - & $\sqrt2(A-2)$& - &- \\
\cline{1-5}
$\Sigma\Sigma \rightarrow \Lambda \Sigma$ & - &  -  & $ \sqrt2(A-2)$ & - \\
\cline{1-5}
$\Sigma\Sigma \rightarrow \Sigma \Sigma$ & - & -  & - & $2(A-2)$  \\
\cline{1-5}
\end{tabular}
\vspace{0.4 cm}
\caption{ Combinatorial factors of the two-body  YN  interactions embedded   in the
$A$-body space with strangeness  $S=-2$.}
\label{tab:appendt1}
\end{table*}

\begin{table*}[tbp]
  \renewcommand{\arraystretch}{0.9}
 \vskip 1 cm 
    \setlength{\tabcolsep}{0.04cm}
\centering
\begin{turn}{0} 
\begin{tabular}{|*{11}{c|}}
\cline{1-11} 
\diagbox[height=1cm, innerwidth=2cm, innerrightsep=0.3cm,trim=l]{transition}{ \\ YY}  &   { ${\tilde V}_{\Lambda \Lambda,
\Lambda \Lambda}$} & {$ {\tilde{V}}_{\Lambda \Lambda, \Lambda
\Sigma}$} &
{${\tilde V}_{\Lambda \Lambda, \Sigma \Sigma}$} &
{$ {\tilde  V}_{\Lambda
\Sigma, \Lambda \Sigma}$} &   {$ {\tilde  V}_{\Lambda \Sigma, \Sigma
\Sigma}$}  &
{${\tilde{V}}_{\Sigma \Sigma, \Sigma \Sigma}$} & 
{$ {\tilde V}_{\Lambda
\Lambda, N \Xi}$} &{$ { \tilde V}_{\Lambda \Sigma, N \Xi} $} & 
{${\tilde V}_{\Sigma \Sigma
, N \Xi}$} & { ${\tilde V}_{N\Xi, N \Xi }$} \\[3pt]
& & & & & & & & & & \\
\cline{1-11}
& & & & & & & & & & \\
$\Lambda \Lambda \rightarrow \Lambda \Lambda $ & 1 & - &- &- &- &- &-
& - &- &-\\[9pt]
\cline{1-11}
& & & & & & & & & & \\
$\Lambda \Lambda \rightarrow \Lambda \Sigma $ & - & 1 &- &- &- &- &
- & - &- &-\\[9pt]
\cline{1-11}
& & & & & & & & & & \\
$\Lambda \Lambda \rightarrow \Sigma \Sigma $ & - & - &1 &- &- &- &
- & - &- &-\\[9pt]
\cline{1-11}
& & & & & & & & & & \\
$\Lambda \Sigma \rightarrow \Lambda \Sigma $ & - & - &- &1 &- &- &
- & - &- &-\\[9pt]
\cline{1-11}
& & & & & & & & & & \\
$\Lambda \Sigma \rightarrow \Sigma \Sigma $ & - & - &- &- &1 &- &
- & - &- &-\\[9pt]
\cline{1-11}
& & & & & & & & & & \\
$\Sigma \Sigma \rightarrow \Sigma \Sigma $ & - & - &- &- &- &1 &
- & - &- &-\\[9pt]
\cline{1-11}
& & & & & & & & & & \\
$\Lambda \Lambda \rightarrow N \Xi$ & - & - &- &- &- & - & $\sqrt{A-1}$ &
 - &- &-\\[9pt]
\cline{1-11}
& & & & & & & & & & \\
$\Lambda \Sigma \rightarrow N \Xi$ & - & - &- &- &- & - & -&  $\sqrt{A-1}$ &
 - &-\\[9pt]
\cline{1-11}
& & & & & & & & & & \\
$\Sigma \Sigma \rightarrow N \Xi$ & - & - &- &- &- & - & -& -& $\sqrt{A-1}$ &
 -\\[9pt]
\cline{1-11}
& & & & & & & & & & \\
$N \Xi \rightarrow N \Xi$ & - & - &- &- &- & - & -& -&- & $ A-1$ \\[9pt]
\cline{1-11}
\end{tabular} 
\end{turn}
 \vskip 0.4 cm 
\caption{ Combinatorial factors of the two-body  $YY$  interactions  embedded in the
$A$-body space with strangeness $S=-2$. }
\label{tab:appendt2}
\end{table*}

\subsection{$A$-body Schrödinger equation}

With the preparation of the $A=3$ system, we are now able to generalize the 
combinatorial factors to arbitrary $A$. For the kinetic energy, the generalization is 
trivial and leads to the sum of the single particle kinetic energies since 
no particle  conversion can take place for this operator. Interactions are more  involved. 
To the general $A$-body matrix element $\langle \{p_1' \ldots p_A' \}   |  V |\{p_1 \ldots p_A \} \rangle$  
of the $n$-particle interaction 
\begin{equation}
 V = \frac{1}{n!} \sum_{\substack{   k_1,\ldots,k_n \\  k_1',\ldots,k_n' }} V_{ k_1'\ldots k_n', k_1 \ldots k_n } 
 \, a^{\dagger}_{k_1^{\prime}} a^{\dagger}_{k_2^{\prime}} \cdots  a^{\dagger}_{k_n^{\prime}} a_{k_n} \cdots  a_{k_2} a_{k_{1}}
\end{equation}
a total of 
\begin{equation}
  \frac{1}{n!} \, \binom{A}{n}   \, \binom{A}{n} \, n! \, n! (A-n)!  
\end{equation}
different permutations of $V_{ k_1'\ldots k_n', k_1 \ldots k_n }$ contribute. \break Therein, the first $\frac{1}{n!}$ 
is just from the definition of $V$. Following the same steps that lead to Eq.~(\ref{eq:appendeqV}), these 
terms can be rearranged such that the application to an arbitrary state $\Psi$ can be written as 
\begin{eqnarray}
\label{eq:psialla}
\langle \{p_1' \ldots p_A' \}   | \Psi' \rangle & = & \sum_{i_1 < i_2 \ldots <i_n}  \sum_{   p_{i_1}<\ldots<p_{i_n}} \nonumber \\
 & & \times n! \ V_{ p_{i_1}',\ldots,p_{i_n}', p_{i_1},\ldots,p_{i_n} }  \nonumber \\
 & & \times \langle \{p_1' \ldots p_{i_1} \ldots p_{i_n} \ldots  p_A' \}   | \Psi \rangle  \ . 
\end{eqnarray}
For this form, we assume that $\Psi'$ is represented using the ordered states $p_{1}'<\ldots<p_{A}'$. 
Then only one of the $(A-n)!$ different spectator permutations contributes. One of the $\binom{A}{n}$ 
terms is needed to make the sorting on the spectator particles and on the  interacting 
particles independent from each other as done in Eq.~(\ref{eq:appendeqV}). The other one 
is explicitly taken care of  by the sum over $i_1 < i_2 \ldots <i_n$.

If the interacting particles are (partly) identical, we will again replace the sum over   
$p_{i_1}<\ldots<p_{i_n}$ by (partly) full sums and add the appropriate  combinatorial factor (e.g. $1/2!$ in the case of two identical particles). Note that this factors depend on the kind of particles in the 
incoming $\Psi$ state. 

We again introduce rescaled wave functions by studying the norm of the states similar to Eq.~(\ref{eq:appendnorm3Body}). The appropriate factors for states with $p$ particles 
species and $n_1,\ldots, n_p$ particles of each species are $\sqrt{n_1! \ldots n_p!}$. 
The potential matrix element needs to be multiplied with (divided by) this factor for incoming states 
 (outgoing states) to reexpress Eq.~(\ref{eq:psialla}) in terms of $\Phi$ states. 
We note that the potential matrix in terms of these states is symmetrical. In summary, 
the potential matrix elements then reads 
\begin{equation}
\label{eq:sympotmat}
\frac{n!\sqrt{n_1! \ldots n_p!}}{\sqrt{n_1'! \ldots n_p'!}}  \ V_{ p_{i_1}',\ldots,p_{i_n}', p_{i_1},\ldots,p_{i_n}}  \ .    
\end{equation} 
Note that here the factor does not include the additional factor required when identical 
particles are involved in the sum of Eq.~(\ref{eq:psialla}). 

We then simplify the expressions by identifying $n$-particles  that contribute identically 
to Eq.~(\ref{eq:psialla}). The sum over $i_1 < i_2 \ldots <i_n$ can then be reduced and 
tuples of outgoing states involving the same kind of particles can be combined by the appropriate factor. 
Finally, we build the ratio of the factors for the $A$-body and $n$-body systems to find the 
correct combinatorial factors that enter our J-NCSM calculations.

As an example, we now consider some selected matrix elements of the $S=-2$ 2-body interaction for 
$(A-2)$-$\Lambda \Lambda $, 
$(A-2)$-$\Lambda \Sigma $,
$(A-2)$-$\Sigma \Sigma $, 
and $(A-1)$-$ \Xi $ states. 
For the diagonal matrix elements in particle space, the square root 
factors in Eq.~(\ref{eq:sympotmat}) cancel. In this case, the prefactor 
is just $2 \times$ the number of  pairs contributing in the outgoing channel $\times 1/2 $ if the active pair in the incoming channel 
consists of identical particles. Therefore, for our example, we find \hfill 
\begin{widetext}
\begin{eqnarray}
 \langle (A-2)-\Lambda \Lambda | V | (A-2)-\Lambda \Lambda \rangle & = & 2 \cdot 1  \cdot \frac{1}{2} 
 \cdot V_{ p_{\Lambda,1}'p_{\Lambda,2}', p_{\Lambda,1} p_{\Lambda,2}} = 
\tilde V_{ p_{\Lambda,1}' p_{\Lambda,2}', p_{\Lambda,1} p_{\Lambda,2}} \nonumber \\[5pt]
 \langle (A-2)-\Lambda \Sigma | V | (A-2)-\Lambda \Sigma \rangle & = & 2 \cdot 1  
 \cdot V_{ p_{\Lambda}' p_{\Sigma}', p_{\Lambda} p_{\Sigma}} = 
\tilde V_{ p_{\Lambda}'p_{\Sigma}', p_{\Lambda} p_{\Sigma}} \nonumber \\[5pt]
\langle (A-2)-\Sigma \Sigma | V | (A-2)-\Sigma \Sigma \rangle & = & 2 \cdot 1  \cdot \frac{1}{2} 
 \cdot V_{ p_{\Sigma,1}'p_{\Sigma,2}', p_{\Sigma,1} p_{\Sigma,2}} = 
\tilde V_{ p_{\Sigma,1}' p_{\Sigma,2}', p_{\Sigma,1} p_{\Sigma,2}} \nonumber \\[5pt]
\langle (A-1) - \Xi  | V | (A-1) - \Xi \rangle & = & 2 \cdot (A-1)  
 \cdot V_{ p_{N}' p_{\Xi}', p_{N} p_{\Xi}} = 
 ( A - 1 )   {\tilde V}_{ p_{N}'p_{\Xi}', p_{N} p_{\Xi}} \ . 
\end{eqnarray}
\end{widetext}
In the last step, we have exploited  the results of the two-body system that relate $V$ to 
 $\tilde V$ in Eq.~(\ref{eq:appendHamilinLLSXiNbasis2}).
The resulting combinatorial factors agree with the expectation that 
the interaction just has to be multiplied by the number of pairs contributing.  
More interesting is the case of transitions. Here, we first look at 
transitions between $\Sigma \Sigma$ and $\Lambda \Sigma$ states. Because 
the identity of the particles changes, we now have additionally the 
contribution of the square root factors. They are also important to 
guarantee that the interaction matrix is symmetric. 

The result for the two matrix elements is \hfill 
\begin{widetext}
\begin{eqnarray}
\langle (A-2)-\Sigma \Sigma | V | (A-2)-\Lambda \Sigma \rangle & = & 2 \cdot 1  \cdot 1 \cdot \frac{\sqrt{(A-2)!}}{\sqrt{2!(A-2)!}} \ 
 \cdot V_{ p_{\Sigma,1}'p_{\Sigma,2}', p_{\Lambda} p_{\Sigma}} = 
\tilde V_{ p_{\Sigma,1}' p_{\Sigma,2}', p_{\Lambda} p_{\Sigma}} \nonumber \\[5pt]
\langle (A-2)-\Lambda \Sigma | V | (A-2)-\Sigma \Sigma \rangle & = & 2 \cdot 1  \cdot \frac{1}{2} \cdot \frac{\sqrt{2!(A-2)!}}{\sqrt{(A-2)!}} \ 
 \cdot V_{ p_{\Lambda}'p_{\Sigma}', p_{\Sigma,1} p_{\Sigma,2}} = 
\tilde V_{ p_{\Lambda}' p_{\Sigma}', p_{\Sigma,1} p_{\Sigma,2}} 
\end{eqnarray}
\end{widetext}
and just reflects the number of $YY$ pairs in the $A$-body state. The final examples 
are transitions of $\Lambda \Lambda$ and $N \Xi$. For these matrix elements, 
several pairs contribute and the identity of the particles changes. It is reassuring that we also find in this case symmetry of the potential matrix elements \hfill 
\begin{widetext}
\begin{eqnarray}
\langle (A-2)-\Lambda \Lambda | V | (A-1)- \Xi \rangle & = & 2 \cdot 1  \cdot 1 \cdot \frac{\sqrt{(A-1)!}}{\sqrt{2!(A-2)!}} \cdot 
 V_{ p_{\Lambda,1}'p_{\Lambda,2}', p_{N} p_{\Xi}} = \sqrt{A-1} \ 
\tilde V_{ p_{\Lambda,1}' p_{\Lambda,2}', p_{N} p_{\Xi}} \nonumber \\[5pt]
\langle (A-1)-\Xi | V | (A-2)- \Lambda \Lambda \rangle & = & 2 \cdot ( A - 1 )  \cdot \frac{1}{2} \cdot \frac{\sqrt{2!(A-2)!}}{\sqrt{(A-1)!}} \cdot 
 V_{ p_{N}'p_{\Xi}', p_{\Lambda,1} p_{\Lambda,2}} = \sqrt{A-1} \ 
\tilde V_{ p_{N}' p_{\Xi}', p_{\Lambda,1} p_{\Lambda,2}} \ . 
\end{eqnarray}
\end{widetext}
In this way, it is straightforward to identify all relevant combinatorial 
factors for our calculations. For the $S=0$ interactions, where particle 
transitions do not occur, it is simply given by the number of NN pairs in 
the state, i.e. $\binom{A-2}{2}$ and $\binom{A-1}{2}$ for $(A-2)$-YY and 
$(A-1)$-$\Xi$ states, respectively. The factors for $S=-1$ ($S=-2$) interactions 
are summarized in Table~\ref{tab:appendt1} (Table~\ref{tab:appendt2}). 
To shorten the presentation, we only include particle transitions in one direction. The other one is  given by the symmetry of the potentials.

\bibliographystyle{unsrturl}

\bibliography{./bib/hyp-literatur.bib,./bib/ncsm.bib,./bib/hypernuclei.bib,./bib/nn-interactions.bib,./bib/yn-interactions.bib,./bib/srg.bib,./bib/double-strangeness.bib}

\end{document}